\documentclass[twocolumn,trackchanges]{aastex631}
\usepackage{amsmath}
\usepackage{epsf}
\usepackage{color}
\usepackage{color}
\usepackage{enumitem}
\usepackage{mathtools}
\usepackage[mathcal]{eucal}
 \let\mathscr\relax
\usepackage[scr]{rsfso}

\shorttitle{A CEERS Look at $z =$ 9--12}
\shortauthors{Finkelstein et al.}

\newcommand{\sol}{$_{\odot}$}

\def\arcs{\hbox{$^{\prime\prime}$}}

\begin{document}
\title{CEERS Key Paper I: \\ An Early Look into the First 500 Myr of Galaxy Formation with \emph{JWST}}


\author[0000-0001-8519-1130]{Steven L. Finkelstein}
\affiliation{Department of Astronomy, The University of Texas at Austin, Austin, TX, USA}
\email{stevenf@astro.as.utexas.edu}

\author[0000-0002-9921-9218]{Micaela B. Bagley}
\affiliation{Department of Astronomy, The University of Texas at Austin, Austin, TX, USA}

\author[0000-0001-7113-2738]{Henry C. Ferguson}
\affiliation{Space Telescope Science Institute, Baltimore, MD, USA}

\author[0000-0003-3903-6935]{Stephen M.~Wilkins}
\affiliation{Astronomy Centre, University of Sussex, Falmer, Brighton BN1 9QH, UK}
\affiliation{Institute of Space Sciences and Astronomy, University of Malta, Msida MSD 2080, Malta}

\author[0000-0001-9187-3605]{Jeyhan S. Kartaltepe}
\affiliation{Laboratory for Multiwavelength Astrophysics, School of Physics and Astronomy, Rochester Institute of Technology, 84 Lomb Memorial Drive, Rochester, NY 14623, USA}

\author[0000-0001-7503-8482]{Casey Papovich}
\affiliation{Department of Physics and Astronomy, Texas A\&M University, College Station, TX, 77843-4242 USA}
\affiliation{George P.\ and Cynthia Woods Mitchell Institute for Fundamental Physics and Astronomy, Texas A\&M University, College Station, TX, 77843-4242 USA}

\author[0000-0003-3466-035X]{L. Y. Aaron\ Yung}
\altaffiliation{NASA Postdoctoral Fellow}
\affiliation{Astrophysics Science Division, NASA Goddard Space Flight Center, 8800 Greenbelt Rd, Greenbelt, MD 20771, USA}

\author[0000-0002-7959-8783]{Pablo Arrabal Haro}
\affiliation{NSF's National Optical-Infrared Astronomy Research Laboratory, 950 N. Cherry Ave., Tucson, AZ 85719, USA}

\author[0000-0002-2517-6446]{Peter Behroozi}
\affiliation{Department of Astronomy and Steward Observatory, University of Arizona, Tucson, AZ 85721, USA}
\affiliation{Division of Science, National Astronomical Observatory of Japan, 2-21-1 Osawa, Mitaka, Tokyo 181-8588, Japan}

\author[0000-0001-5414-5131]{Mark Dickinson}
\affiliation{NSF's National Optical-Infrared Astronomy Research Laboratory, 950 N. Cherry Ave., Tucson, AZ 85719, USA}

\author[0000-0002-8360-3880]{Dale D. Kocevski}
\affiliation{Department of Physics and Astronomy, Colby College, Waterville, ME 04901, USA}

\author[0000-0002-6610-2048]{Anton M. Koekemoer}
\affiliation{Space Telescope Science Institute, 3700 San Martin Dr., Baltimore, MD 21218, USA}

\author[0000-0003-2366-8858]{Rebecca L. Larson}
\affiliation{NSF Graduate Fellow}
\affiliation{Department of Astronomy, The University of Texas at Austin, Austin, TX, USA}

\author[0000-0002-9466-2763]{Aur{\'e}lien Le Bail}
\affiliation{Universit{\'e} Paris-Saclay, Universit{\'e} Paris Cit{\'e}, CEA, CNRS, AIM, 91191, Gif-sur-Yvette, France}

\author[0000-0003-4965-0402]{Alexa M.\ Morales}
\affiliation{Department of Astronomy, The University of Texas at Austin, Austin, TX, USA}

\author[0000-0003-4528-5639]{Pablo G. P\'erez-Gonz\'alez}
\affiliation{Centro de Astrobiolog\'{\i}a (CAB), CSIC-INTA, Ctra. de Ajalvir km 4, Torrej\'on de Ardoz, E-28850, Madrid, Spain}

\author[0000-0002-4193-2539]{Denis Burgarella}
\affiliation{Aix Marseille Univ, CNRS, CNES, LAM Marseille, France}

\author[0000-0003-2842-9434]{Romeel Dav\'e}
\affiliation{Institute for Astronomy, University of Edinburgh, Blackford Hill, Edinburgh, EH9 3HJ UK}
\affiliation{Department of Physics and Astronomy, University of the Western Cape, Robert Sobukwe Rd, Bellville, Cape Town 7535, South Africa}

\author[0000-0002-3301-3321]{Michaela Hirschmann}
\affiliation{Institute of Physics, Laboratory of Galaxy Evolution, Ecole Polytechnique Fédérale de Lausanne (EPFL), Observatoire de Sauverny, 1290 Versoix, Switzerland}
\affiliation{INAF, Osservatorio Astronomico di Trieste, Via Tiepolo 11, 34131 Trieste, Italy}

\author[0000-0002-6748-6821]{Rachel S. Somerville}
\affiliation{Center for Computational Astrophysics, Flatiron Institute, 162 5th Avenue, New York, NY, 10010, USA}

\author[0000-0003-3735-1931]{Stijn Wuyts}
\affiliation{Department of Physics, University of Bath, Claverton Down, Bath BA2 7AY, UK}

\author[0000-0003-0212-2979]{Volker Bromm}
\affiliation{Department of Astronomy, The University of Texas at Austin, Austin, TX, USA}

\author[0000-0002-0930-6466]{Caitlin M. Casey}
\affiliation{Department of Astronomy, The University of Texas at Austin, Austin, TX, USA}

\author[0000-0003-3820-2823]{Adriano Fontana}
\affiliation{INAF - Osservatorio Astronomico di Roma, via di Frascati 33, 00078 Monte Porzio Catone, Italy}

\author[0000-0001-7201-5066]{Seiji Fujimoto}
\altaffiliation{NASA Hubble Fellow}
\affiliation{Department of Astronomy, The University of Texas at Austin, Austin, TX, USA}
\affiliation{Cosmic Dawn Center (DAWN), Jagtvej 128, DK2200 Copenhagen N, Denmark}
\affiliation{Niels Bohr Institute, University of Copenhagen, Lyngbyvej 2, DK2100 Copenhagen \O, Denmark}

\author[0000-0003-2098-9568]{Jonathan P. Gardner}
\affiliation{Astrophysics Science Division, Goddard Space Flight Center, Code 665, Greenbelt, MD 20771, USA}

\author[0000-0002-7831-8751]{Mauro Giavalisco}
\affiliation{University of Massachusetts Amherst, 710 North Pleasant Street, Amherst, MA 01003-9305, USA}

\author[0000-0002-5688-0663]{Andrea Grazian}
\affiliation{INAF--Osservatorio Astronomico di Padova, Vicolo dell'Osservatorio 5, I-35122, Padova, Italy}

\author[0000-0001-9440-8872]{Norman A. Grogin}
\affiliation{Space Telescope Science Institute, Baltimore, MD, USA}

\author[0000-0001-6145-5090]{Nimish P. Hathi}
\affiliation{Space Telescope Science Institute, Baltimore, MD, USA}

\author[0000-0001-6251-4988]{Taylor A. Hutchison}
\altaffiliation{NASA Postdoctoral Fellow}
\affiliation{Astrophysics Science Division, NASA Goddard Space Flight Center, 8800 Greenbelt Rd, Greenbelt, MD 20771, USA}

\author[0000-0001-8738-6011]{Saurabh W. Jha}
\affiliation{Department of Physics and Astronomy, Rutgers, the State University of New Jersey, Piscataway, NJ 08854, USA}

\author[0000-0002-1590-0568]{Shardha Jogee}
\affiliation{Department of Astronomy, The University of Texas at Austin, Austin, TX, USA}

\author[0000-0001-8152-3943]{Lisa J. Kewley}
\affiliation{Center for Astrophysics | Harvard \& Smithsonian, 60 Garden Street, Cambridge, MA 02138, USA}

\author[0000-0002-5537-8110]{Allison Kirkpatrick}
\affiliation{Department of Physics and Astronomy, University of Kansas, Lawrence, KS 66045, USA}

\author[0000-0002-7530-8857]{Arianna S. Long}
\altaffiliation{NASA Hubble Fellow}
\affiliation{Department of Astronomy, The University of Texas at Austin, Austin, TX, USA}

\author[0000-0003-3130-5643]{Jennifer M. Lotz}
\affiliation{Gemini Observatory/NSF's National Optical-Infrared Astronomy Research Laboratory, 950 N. Cherry Ave., Tucson, AZ 85719, USA}

\author[0000-0001-8940-6768]{Laura Pentericci}
\affiliation{INAF - Osservatorio Astronomico di Roma, via di Frascati 33, 00078 Monte Porzio Catone, Italy}

\author[0000-0002-2361-7201]{Justin~D.~R.~Pierel}
\affiliation{Space Telescope Science Institute, Baltimore, MD, USA}

\author[0000-0003-3382-5941]{Nor Pirzkal}
\affiliation{ESA/AURA}
\affiliation{Space Telescope Science Institute, Baltimore, MD, USA}

\author[0000-0002-5269-6527]{Swara Ravindranath}
\affiliation{Space Telescope Science Institute, Baltimore, MD, USA}

\author[0000-0003-0894-1588]{Russell E.\ Ryan Jr.}
\affiliation{Space Telescope Science Institute, Baltimore, MD, USA}

\author[0000-0002-1410-0470]{Jonathan R. Trump}
\affiliation{Department of Physics, 196 Auditorium Road, Unit 3046, University of Connecticut, Storrs, CT 06269, USA}

\author[0000-0001-8835-7722]{Guang Yang}
\affiliation{Kapteyn Astronomical Institute, University of Groningen, P.O. Box 800, 9700 AV Groningen, The Netherlands}
\affiliation{SRON Netherlands Institute for Space Research, Postbus 800, 9700 AV Groningen, The Netherlands}

\author[0000-0003-0883-2226]{Rachana Bhatawdekar}
\affiliation{European Space Agency, ESA/ESTEC, Keplerlaan 1, 2201 AZ Noordwijk, NL}

\author[0000-0003-0492-4924]{Laura Bisigello}
\affiliation{Dipartimento di Fisica e Astronomia "G.Galilei", Universit\'a di Padova, Via Marzolo 8, I-35131 Padova, Italy}
\affiliation{INAF--Osservatorio Astronomico di Padova, Vicolo dell'Osservatorio 5, I-35122, Padova, Italy}

\author[0000-0003-3441-903X]{V\'eronique Buat}
\affiliation{Aix Marseille Univ, CNRS, CNES, LAM Marseille, France}

\author[0000-0003-2536-1614]{Antonello Calabr{\`o}} 
\affiliation{INAF - Osservatorio Astronomico di Roma, via di Frascati 33, 00078 Monte Porzio Catone, Italy}

\author[0000-0001-9875-8263]{Marco Castellano}
\affiliation{INAF - Osservatorio Astronomico di Roma, via di Frascati 33, 00078 Monte Porzio Catone, Italy}

\author[0000-0001-7151-009X]{Nikko J. Cleri}
\affiliation{Department of Physics and Astronomy, Texas A\&M University, College Station, TX, 77843-4242 USA}
\affiliation{George P.\ and Cynthia Woods Mitchell Institute for Fundamental Physics and Astronomy, Texas A\&M University, College Station, TX, 77843-4242 USA}

\author[0000-0003-1371-6019]{M. C. Cooper}
\affiliation{Department of Physics \& Astronomy, University of California, Irvine, 4129 Reines Hall, Irvine, CA 92697, USA}

\author[0000-0002-5009-512X]{Darren Croton}
\affiliation{Centre for Astrophysics \& Supercomputing, Swinburne University of Technology, Hawthorn, VIC 3122, Australia}
\affiliation{ARC Centre of Excellence for All Sky Astrophysics in 3 Dimensions (ASTRO 3D)}

\author[0000-0002-3331-9590]{Emanuele Daddi}
\affiliation{Universit{\'e} Paris-Saclay, Universit{\'e} Paris Cit{\'e}, CEA, CNRS, AIM, 91191, Gif-sur-Yvette, France}

\author[0000-0003-4174-0374]{Avishai Dekel}
\affil{Racah Institute of Physics, The Hebrew University of Jerusalem, Jerusalem 91904, Israel}

\author[0000-0002-7631-647X]{David Elbaz}
\affiliation{Universit{\'e} Paris-Saclay, Universit{\'e} Paris Cit{\'e}, CEA, CNRS, AIM, 91191, Gif-sur-Yvette, France}

\author[0000-0002-3560-8599]{Maximilien Franco}
\affiliation{Department of Astronomy, The University of Texas at Austin, Austin, TX, USA}

\author[0000-0003-1530-8713]{Eric Gawiser}
\affiliation{Department of Physics and Astronomy, Rutgers, the State University of New Jersey, Piscataway, NJ 08854, USA}

\author[0000-0002-4884-6756]{Benne W. Holwerda}
\affil{Physics \& Astronomy Department, University of Louisville, 40292 KY, Louisville, USA}

\author[0000-0002-1416-8483]{Marc Huertas-Company}
\affil{Instituto de Astrof\'isica de Canarias, La Laguna, Tenerife, Spain}
\affil{Universidad de la Laguna, La Laguna, Tenerife, Spain}
\affil{Universit\'e Paris-Cit\'e, LERMA - Observatoire de Paris, PSL, Paris, France}

\author[0000-0002-6790-5125]{Anne E. Jaskot}
\affiliation{Department of Astronomy, Williams College, Williamstown, MA, 01267, USA}

\author[0000-0002-9393-6507]{Gene C. K. Leung}
\affiliation{Department of Astronomy, The University of Texas at Austin, Austin, TX, USA}

\author[0000-0003-1581-7825]{Ray A. Lucas}
\affiliation{Space Telescope Science Institute, 3700 San Martin Drive, Baltimore, MD 21218, USA}

\author[0000-0001-5846-4404]{Bahram Mobasher}
\affiliation{Department of Physics and Astronomy, University of California, 900 University Ave, Riverside, CA 92521, USA}

\author[0000-0002-2499-9205]{Viraj Pandya}
\altaffiliation{NASA Hubble Fellow}
\affiliation{Columbia Astrophysics Laboratory, Columbia University, 550 West 120th Street, New York, NY 10027, USA}

\author[0000-0002-8224-4505]{Sandro Tacchella}
\affiliation{Kavli Institute for Cosmology, University of Cambridge, Madingley Road, Cambridge, CB3 0HA, UK}\affiliation{Cavendish Laboratory, University of Cambridge, 19 JJ Thomson Avenue, Cambridge, CB3 0HE, UK}

\author[0000-0001-6065-7483]{Benjamin J. Weiner}
\affiliation{MMT/Steward Observatory, University of Arizona, 933 N. Cherry Ave., Tucson, AZ 85721, USA}

\author[0000-0002-7051-1100]{Jorge A. Zavala}
\affiliation{National Astronomical Observatory of Japan, 2-21-1 Osawa, Mitaka, Tokyo 181-8588, Japan}

\begin{abstract}
We present an investigation into the first 500 Myr of galaxy evolution from the Cosmic Evolution Early Release Science (CEERS) survey.  CEERS, one of 13 {\it JWST} ERS programs, targets galaxy formation $z \sim$ 0.5 to $z >$ 10 using several imaging and spectroscopic modes.  We make use of the first epoch of CEERS NIRCam imaging, spanning 35.5 sq.\ arcmin, to search for candidate galaxies at $z >$ 9.  Following a detailed data reduction process implementing several custom steps to produce high-quality reduced images, we perform multi-band photometry across seven NIRCam broad and medium-band (and six {\it Hubble} broadband) filters focusing on robust colors and accurate total fluxes.  We measure photometric redshifts and devise a robust set of selection criteria to identify a sample of 26 galaxy candidates at $z \sim$ 9--16.  These objects are compact with a median half-light radius of $\sim$0.5 kpc.  We present an early estimate of the $z \sim$ 11 rest-frame ultraviolet (UV) luminosity function, finding that the number density of galaxies at $M_{UV} \sim -$20 appears to evolve very little from $z \sim$ 9 to $z \sim$ 11.  We also find that the abundance (surface density  [arcmin$^{-2}$]) of our candidates exceeds nearly all theoretical predictions.  We explore potential implications, including that at $z >$ 10 star formation may be dominated by top-heavy initial mass functions, which would result in an increased ratio of UV light per unit halo mass, though a complete lack of dust attenuation and/or changing star-formation physics may also play a role.  While spectroscopic confirmation of these sources is urgently required, our results suggest that the deeper views to come with {\it JWST} should yield prolific samples of ultra-high-redshift galaxies with which to further explore these conclusions.
\end{abstract}


\section{Introduction}\label{sec:intro}

The epoch of reionization marks the period when energetic photons (presumably from massive stars in early galaxies; e.g., \citealt{stark16,finkelstein19,robertson21}) ionized the gas in the intergalactic medium (IGM).  Understanding when and how this process occurs is crucial to constraining both the earliest phases of galaxy formation (which kick-started this process), and how the evolution of the IGM temperature affects the star-formation efficiency in low-mass halos throughout (and after) this transition.

Advances in deep near-infrared (near-IR) imaging with the {\it Hubble Space Telescope} ({\it HST}) have pushed constraints on galaxy evolution into the first billion years after the Big Bang.  Studies of public blank-field surveys including the Hubble Ultra Deep Field [HUDF;  \citealt{beckwith06,oesch10,bouwens10a}], the Cosmic Assembly Near-infrared Deep Extragalactic Legacy Survey [CANDELS; \citealt{grogin11,koekemoer11}], the Hubble Frontier Fields [HFF; \citealt{lotz17}], and the Brightest of Reionizing Galaxy survey [BoRG; \citealt{trenti11}] have uncovered thousands of galaxies at $z >$ 6 \citep[e.g.][]{bouwens15,finkelstein15,ishigaki15,mcleod16,oesch18,morishita18,bridge19,rojasruiz20,finkelstein22,bouwens22,bagley22}.  The evolution of the rest-frame ultraviolet (UV) luminosity function has been well studied to $z \sim$ 8 \citep[e.g.][]{finkelstein15,bouwens15}, with some constraints placed at $z \sim$ 9 and 10 \citep[e.g.,][]{oesch18,bouwens19,finkelstein22,bagley22}. However, little was known about the $z >$ 10 universe prior to {\it JWST}, beyond the unexpected discovery of an exceptionally bright $z =$ 10.957 galaxy \citep{oesch16,jiang21}.  This knowledge gap is due to the modest light-gathering power of the 2.4m {\it HST} and the fact that at $z >$ 10 galaxies become one-band (F160W; {\it HST}/WFC3's reddest filter) detections.  Rest-UV emission from galaxies completely redshifts out of {\it HST} observability at $z \gtrsim$ 12.5.  

This has left a major gap in our knowledge of galaxy formation at early times.  Do galaxies form stars fairly inefficiently, like our own Milky Way, and build up slowly?  Or is star formation in the early Universe more rapid due to high gas densities and frequent interactions?  Equally exciting and unknown, does the initial mass function (IMF) begin to show signs of evolution?  Models predict top-heavy IMFs should dominate at very low metallicities \citep[e.g.][]{bromm04}, so observations should begin to see such signatures.  Answering these questions about the physical processes dominating the earliest star formation requires detailed observations of the earliest galaxies to form in our universe, and {\it JWST} was designed to push our cosmic horizons to the highest redshifts.  The 7$\times$ larger light-gathering power combined with the large field-of-view and near-infrared sensitivity of NIRCam \citep{rieke05} sets the stage for major advances in our ability to study early galaxy formation.  Cycle 1 of {\it JWST} includes several programs encompassing 100's of hours which all have the early Universe as their primary science goal.  

Indeed in just the days-to-weeks after the first science data were released, several papers were submitted discussing the detection of objects not only at the expected redshifts of $z \sim$ 10--11 \citep[e.g.,][]{castellano22,naidu22,adams22,whitler22,labbe22}, but with some candidates at $z \sim$ 12--17 \citep[e.g.][]{finkelstein22c,donnan22,harikane22}.  The existence of galaxies at such early times, and especially at such bright magnitudes for some sources, could potentially challenge early models of galaxy formation \citep[e.g.][]{mason22,ferrara22,mbk22}.  However, these studies originally relied on very early photometric calibration; subsequent calibration data shifted the photometric zeropoints significantly \citep{boyer22}.  Now that the flux calibration, and overall data reduction pipeline has stabilized, it is prudent to take a detailed look at what constraints we can place on this early epoch.

Here we use the first epoch of data from the Cosmic Evolution Early Release Science Survey (CEERS; survey description to come in Finkelstein et al., in prep).  CEERS was designed in part to provide our earliest detailed glimpse
into the $z >$ 10 universe, and these CEERS data were among the first Cycle 1 science exposures taken, included in the first publicly released data on 14 July 2022.  We search these data for $z \gtrsim$ 9 galaxy candidates, heretofore difficult (if not impossible) to find with {\it HST}.  We place an emphasis on building a robust sample via a detailed photometric cataloging process, coupled with stringent selection criteria, both backed by simulations.  Section 2 describes the observations and data reduction, while Section 3 describes our photometry and photometric redshift measurements, and Section 4 discusses our sample selection procedure. We describe our sample in Section 5, and present a comparison to other early samples in Section 6.  In Section 7 we present the $z \sim$ 11 UV luminosity function and the cumulative surface density of early galaxies, and discuss implications on the physics dominating galaxy formation at the earliest times in Section 8.  We summarize our results and present our conclusions in Section 9.  In this paper we assume the latest {\it Planck} flat $\Lambda$CDM cosmology with H$_{0}=\ $67.36 km s$^{-1}$ Mpc$^{-1}$, $\Omega_m=\ $0.3153, and $\Omega_{\Lambda}=\ $0.6847 \citep{planck20}.  All magnitudes are in the absolute bolometric system \citep[AB;][]{oke83}.

\section{Data}\label{sec:data}

CEERS is one of 13 early release science surveys designed to obtain data covering several scientific themes of astronomy early in Cycle 1, along with testing out multiple instrument modes and providing early public data to the community.  CEERS consists of a mosaic of 10 NIRCam pointings in the CANDELS Extended Groth Strip (EGS) field, with six obtained in parallel with prime NIRSpec observations, and four in parallel with prime MIRI observations (four of these pointings also include NIRCam wide-field slitless grism spectroscopy).  Here we use imaging data from NIRCam obtained during the first epoch of CEERS, during June 21-22, 2022.  This consists of short and long-wavelength images in both NIRCam A and B modules, taken over four pointings, labeled NIRCam1, NIRCam2, NIRCam3, and NIRCam6.  Each pointing was observed with seven filters: F115W, F150W, and F200W on the short-wavelength side, and F277W, F356W, F410M, and F444W on the long-wavelength side.  

\subsection{Data Reduction}

The NIRCam images used here are those released with the first CEERS public data release (Data Release 0.5), which are fully described in \citet{bagley22b}.  Here we briefly highlight the key features of the data reduction, directing the reader to \citet{bagley22b} for more details.

We reduce the raw NIRCam imaging through version 1.7.2 of the \textit{JWST} Calibration Pipeline, with custom modifications designed to correct for additional features in the data. We use the calibration reference file context\footnote{\url{jwst-crds.stsci.edu}, pmap 0989 corresponds to NIRCam instrument mapping 0232} pipeline mapping (pmap) 0989.
We begin by running Stage 1 of the calibration pipeline, which performs detector-level corrections and outputs a countrate image in units of counts/s, adopting all default parameters for this stage. We then perform custom corrections to flag and remove snowballs from all exposures, subtract off the large-scale wisps in F150W and F200W using wisp templates created by the NIRCam team, and measure and remove $1/f$ noise via a median measured (amplifier-by-amplifier) along rows and columns. The images are then flat fielded and flux calibrated using Stage 2 of the calibration pipeline, again adopting the default values, to produce images in units of MJy/sr. The pmap 0989 reference files include ground flats that have been corrected for illumination gradients measured with in-flight data, and improved but still preliminary photometric calibration reference files. We find that the flux calibration does a good job of synchronizing the zeropoints across the NIRCam detectors \citep[to within the 2-5\% level,][]{bagley22b}, but that an additional absolute flux calibration may be required at the few percent level (see \S 3.6). These flat and photometric calibration reference files will continue to be improved and updated throughout Cycle 1.

We align the images using a custom version of the TweakReg routine, which is designed to register image to an absolute WCS frame by matching sources detecting in each image with those in a reference catalog. Our modified version of the routine uses \textsc{Source Extractor} (hereafter \texttt{\textsc{SE}}; \citealt{bertin96}) to measure source centroids in each individual image. We then align each image to a reference catalog constructed from an \textit{HST} F160W mosaic with astrometry tied to Gaia EDR3 \citep[see Section~\ref{sec:hst} and][for details on the methods used for the mosaic construction]{koekemoer11}. We first align images from the same detector but different dithers to each other, allowing for shifts in $x$ and $y$ and achieving an RMS of $\sim 3-6$ mas per source for this relative alignment. We next align all images to the F160W reference image, allowing $xy$ shifts and rotations in the SW images and an additional scaling factor to account for large-scale distortions in the LW images. The RMS of this absolute alignment is $\sim 12-15$ mas and $\sim 5-10$ mas when comparing WFC3 to NIRCam and NIRCam to NIRCam, respectively. We note that we followed a slightly different procedure for NIRCam3, aligning F277W to {\it HST}/WFC3 F160W and then using F277W as the new reference for all other NIRCam filters. This altered procedure was required to address additional offsets registered in one portion of the F160W image in this region \citep[see][for details]{bagley22b}.

Finally, we create mosaics for each pointing in all filters in the following way. We subtract a pedestal value off of each individual image, and scale the readnoise variance maps such that they include an estimate of the robustly-measured sky variance. The mosaics are then created using the Resample routine in Stage 3 of the calibration pipeline, which uses the drizzle algorithm with an inverse variance weighting \citep{casertano00,fruchter02}. We drizzle the images to an output pixel scale of 0\farcs03/pixel and use the same tangent point as that of the \textit{HST} mosaics, such that the images in all filters, NIRCam and HST, are pixel-aligned. 

\begin{figure*}[!t]
\epsscale{1.2}
\plotone{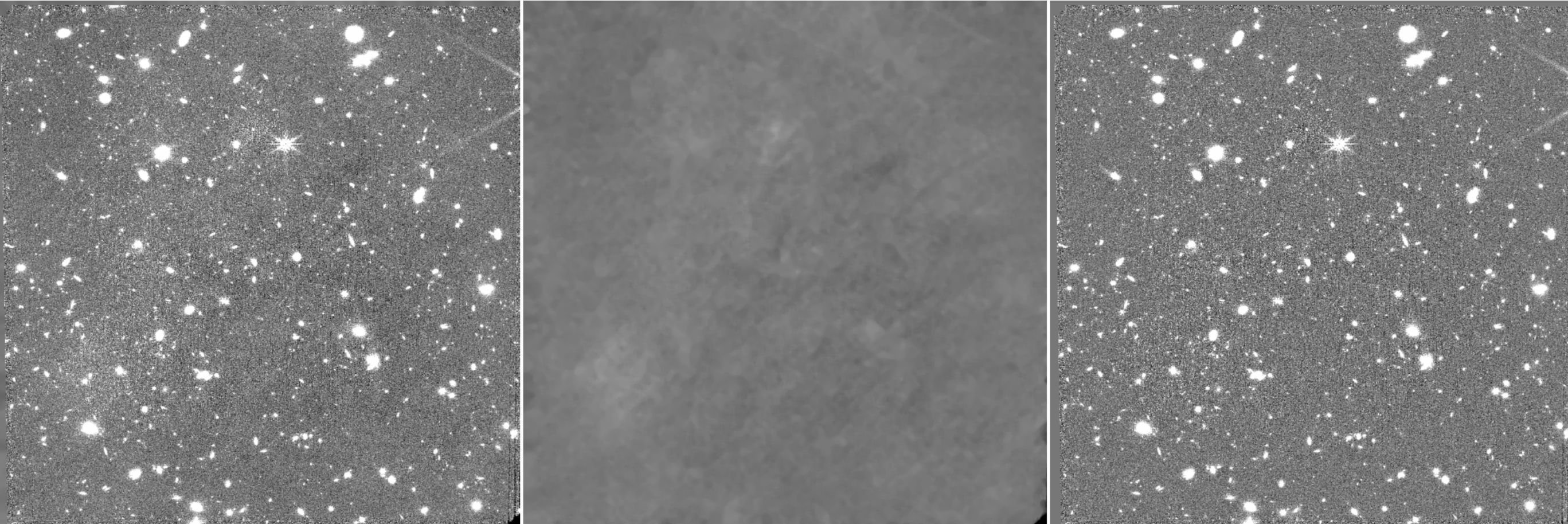}
\caption{An example of the results of our background subtraction procedure.  The left panel shows a zoom in on the science image of the ALONG module of our F444W mosaic in the CEERS1 pointing.  The middle panel is the derived background, and the right panel is the background-subtracted science image.  By progressively masking out objects in smaller tiers, this method is able to capture both small and large-scale fluctuations. Full details on this process are available in \citet{bagley22b}.}
\label{fig:background}
\end{figure*}

\subsection{HST Imaging Data}\label{sec:hst}
The EGS field has archival {\it HST} imaging from the All-wavelength Extended Groth Strip International Survey (AEGIS, \citep{davis07}), the Cosmic Assembly Deep Extragalactic Legacy Survey \citep[CANDELS, ][]{grogin11,koekemoer11}, 3D-HST \citep{momcheva16}, and various followup programs.  The entire CEERS field is covered by F606W, F814W, F125W, (shallow 800 s) F140W, and F160W; portions are covered by F105W.  For CEERS, we produced an updated version (v1.9) of the CANDELS EGS mosaics\footnote{https://ceers.github.io/releases.html\#hdr1} specifically aligning their astrometry onto {\it Gaia} DR3, and on the same 30 mas pixel scale as our NIRCam images.
In each of the {\it HST} mosaics, we create smaller cutouts to match the footprints of the drizzled NIRCam mosaics in each pointing. In this way we have pixel-aligned imaging in 12--13 filters per field (NIRCam1 does not include F105W coverage) from $\sim0.5-5$\micron.

\section{Photometry}

In this section we describe the creation of our $HST + JWST$ catalog.  As the focus is on high-redshift galaxies, this catalog is optimized for faint, compact sources.  To ensure accurate photometric redshifts, significant attention is paid to calculation of accurate colors and uncertainties.

\subsection{Background Subtraction}
The procedure adopted to remove background non-uniformities is described in detail by \citet{bagley22b}. Briefly, it involves several different tiers of source masking, aimed at removing large-scale structures -- including the wings of bright, extended galaxies -- while preserving small-scale structures (including the wings of faint galaxies). We have found this procedure to be more effective than the built-in background subtraction procedure in \texttt{\textsc{SE}} , and we therefore use these images as inputs to \texttt{\textsc{SE}} (\S 3.3) and disable its background-subtraction step.

\subsection{Point-Spread Function Matching}
The full-width at half-maximum (FWHM) of the NIRCam point-spread function (PSF) varies significantly across the wavelengths of the filters used.  As the selection of high-redshift galaxies depends on our ability to measure accurate colors, ensuring a similar fraction of light is measured in all bands is crucial.  In this work, we accomplish this by matching the point-spread functions in our images to the F444W filter, as it is the reddest filter and thus has the largest PSF (FWHM=0.161\arcs).

Our procedure for PSF matching follows \citet{finkelstein22}, which we summarize here.  In each filter, we create a preliminary photometric catalog made using Source Extractor v2.25.0
\citep{bertin96}, and identify potential PSF stars by searching for the stellar locus in a plane of half-light radius versus source magnitude (making custom cuts in both quantities for each filter).  We excluded objects with neighbors within 50 pixels with magnitudes brighter than the magnitude of the star in question minus one.  As stars were more difficult to identify in some bands (e.g., F115W, and the shallower F410M), we combined stars identified in the short-wavelength and long-wavelength channel bands to one list for each channel, which resulted in typically 10-20 stars per NIRCam pointing.  We then visually inspected each star in each image, removing stars near detector edges, other defects, or with close neighbors not excluded by the previous cut, to generate one star list per filter per pointing.

PSFs were then generated by stacking stars that passed this inspection.  As our observations in all four pointings utilized the same dither pattern and were taken at a similar time, we create one PSF per filter by stacking all stars over all four pointings, increasing the signal-to-noise ratio of our PSF.  For each star, we extract a 101x101 pixel box, upsample by a factor of 10, measure the centroid, and shift the star to be centered in this upsampled image.  We then downsample back to the native resolution,  rotate the star by a random position angle (to account for situations when the position angle of the observations was not identical), and normalize the star's peak flux to unity.  The final PSF was made by median-combining the individual stars.  The final PSFs have a centroiding accuracy of $\sim$0.05--0.1 pixels.

Kernels to match bluer PSFs to F444W were created with the \textsc{pypher} Python routine\footnote{https://pypher.readthedocs.io} \citep{boucaud16}, and these bluer images were then convolved with their respective kernels.  We included the {\it HST}/ACS F606W and F814W imaging in this PSF-matching process, as their PSF FWHMs are smaller than that of F444W.  However, the {\it HST}/WFC3 band PSF FWHMs are larger than F444W, so we do not convolve these images.  This will necessitate a correction accounting for the lower fraction-of-flux encompassed in a given aperture in these filters, which we discuss in \S 3.3.1.

We tested our PSF-matching process by measuring curves-of-growth of the PSF stars in the images.  We find that the median enclosed flux at an aperture diameter of 0.3\arcs\ was within 5\%\ (and often less) of the F444W value for all filters (while prior to this PSF matching process, the bluer filter measurements encompassed $\sim$20\% more light at this radius).  We provide the median FWHM values in each filter in Table~\ref{tab:tab1}.

\begin{deluxetable*}{cccccc}
\vspace{2mm}
\tablecaption{Imaging Data Summary}
\tablewidth{\textwidth}
\tablehead{\multicolumn{1}{c}{Camera} & \multicolumn{1}{c}{Filter} & \multicolumn{1}{c}{FWHM} & \multicolumn{1}{c}{PSF Enclosed} & \multicolumn{1}{c}{Point-Source Limiting} & \multicolumn{1}{c}{Zeropoint}\\
\multicolumn{1}{c}{$ $} & \multicolumn{1}{c}{$ $} & \multicolumn{1}{c}{$ $} & \multicolumn{1}{c}{Flux (d$=$0.2\arcs)} & \multicolumn{1}{c}{Magnitude (5$\sigma$)} & \multicolumn{1}{c}{Correction}}
\startdata
{\it JWST}/NIRCam SW&F115W&0.066$^{\prime\prime}$&0.80&29.2&1.07 $\pm$ 0.03\\
{\it JWST}/NIRCam SW&F150W&0.070$^{\prime\prime}$&0.80&29.0&1.05 $\pm$ 0.02\\
{\it JWST}/NIRCam SW&F200W&0.077$^{\prime\prime}$&0.76&29.2&1.03 $\pm$ 0.03\\
{\it JWST}/NIRCam LW&F277W&0.123$^{\prime\prime}$&0.64&29.2&1.00 $\pm$ 0.03\\
{\it JWST}/NIRCam LW&F356W&0.142$^{\prime\prime}$&0.58&29.2&1.01 $\pm$ 0.02\\
{\it JWST}/NIRCam LW&F410M&0.155$^{\prime\prime}$&0.56&28.4&1.00 $\pm$ 0.02\\
{\it JWST}/NIRCam LW&F444W&0.161$^{\prime\prime}$&0.52&28.6&0.99 $\pm$ 0.02\\
{\it HST}/ACS&F606W&0.118$^{\prime\prime}$&0.70&28.6&1.02 $\pm$ 0.02\\
{\it HST}/ACS&F814W&0.124$^{\prime\prime}$&0.63&28.3&0.96 $\pm$ 0.03\\
{\it HST}/WFC3&F105W&0.235$^{\prime\prime}$&0.35&27.1&0.97 $\pm$ 0.04\\
{\it HST}/WFC3&F125W&0.244$^{\prime\prime}$&0.33&27.3&0.95 $\pm$ 0.03\\
{\it HST}/WFC3&F140W&0.247$^{\prime\prime}$&0.32&26.7&0.95 $\pm$ 0.03\\
{\it HST}/WFC3&F160W&0.254$^{\prime\prime}$&0.30&27.4&0.95 $\pm$ 0.03\\
\enddata
\tablecomments{PSFs were created by stacking stars across all four pointings.  For our photometry, we PSF match all filters with FWHM smaller than the F444W PSF FWHM to the F444W PSF.  We note the {\it HST} imaging used is on the same 30mas pixel scale, which affects the FWHM of the PSF. The limiting magnitude is that measured in a 0.2\arcs\ diameter aperture on the unmatched images, corrected to total based on the PSF flux enclosed in that aperture size, averaged over the four fields.  The derived corrections to the photometric zeropoints for each filter were derived using best-fitting \textsc{EAZY} models to $\sim$900 galaxies with secure spectroscopic redshifts.  These corrections are due to a combination of residual differences between our estimated total fluxes and true total fluxes, differences between the model templates and true galaxies, and true photometric zeropoint inaccuracies (using the photometric reference files from pmap 0989 for NIRCam),   Because these corrections depend specifically on our photometry procedure, they may not be appropriate for other photometric catalogs.}
\label{tab:tab1}
\vspace{-8mm}
\end{deluxetable*}

\subsection{Source Extraction}
We use  \texttt{\textsc{SE}} in two-image mode to measure accurate photometry for each of our four pointings.
\texttt{\textsc{SE}} requires a detection image to identify sources.  We elect to use the inverse-variance-weighted sum of the PSF-matched F277W and F356W images as our detection image, to better detect faint sources.  We do not include F200W in this stack as the Ly$\alpha$ break enters this filter at $z =$ 13.4, and we do not wish to bias our catalog against extreme redshift galaxies (the blue edge of F277W corresponds to a Ly$\alpha$ break redshift of $z =$ 18.9), while the inclusion of F444W could have potentially begun to bias against very blue sources.

Using this detection image, we run \texttt{\textsc{SE}} cycling through the seven NIRCam images and six {\it HST} images as the measurement image.  The key source detection parameters were initially optimized using the CEERS simulated imaging\footnote{https://ceers.github.io/releases.html\#sdr3} \citep{bagley22b}, and further tweaked by inspecting their performance on the final mosaics.  The parameters we used best recovered faint sources while minimizing contamination by spurious objects.  These key parameters were: DETECT\_THRESH$=$1.3, DETECT\_MINAREA$=$5 pixels, and a top-hat convolution kernel with a width of 4 pixels.  

We forced \texttt{\textsc{SE}} to skip the background subtraction step as this was previously removed (\S 3.1).  We use MAP\_RMS for the source weighting.  As the pipeline-produced ERR images include Poisson noise, they are not appropriate for source detection.  We thus convert the weight map associated with the detection image into an effective rms map by taking 1/sqrt(WHT), and assign this to the detection image.  For the measurement image, we use the pipeline ERR image.

Following previous work \citep[e.g.,][]{finkelstein22} we measure colors in small elliptical apertures, which has been shown to accurately recover colors of distant galaxies.  In \texttt{\textsc{SE}} these apertures are defined by two parameters - a Kron factor, and a minimum radius.  We set these two quantities to (1.1, 1.6).  These are the same values found by \citet{finkelstein22} via optimization simulations, and we verified via our own simulations (\S 3.3.1) that the signal-to-noise ratio in these apertures was significantly better than larger parameters, and that gains in signal-to-noise ratio were negligible for smaller values.  We estimate an aperture correction to the total flux for these small apertures by performing a second run of \texttt{\textsc{SE}} on the F444W image with the Kron parameters set to the default ``MAG\_AUTO" parameters of (2.5, 3.5), deriving an aperture correction as the ratio between the flux in this larger aperture to that in the smaller aperture for each object.  The median aperture correction across all four fields was 1.5.  This aperture correction was then applied multiplicatively to the fluxes and uncertainties for all filters.

\subsubsection{Residual Aperture Correction}
While we use small Kron apertures to derive accurate colors, the aperture correction applied above should yield total fluxes close to the true value.  However, several previous studies have noted that the default Kron parameters we use for this aperture correction can miss light in the wings of the PSF \citep[e.g.,][]{bouwens15,finkelstein22}, yielding underestimates of the total fluxes at the 5-20\% level.  

We estimate these corrections using source-injection simulations, adding 3000 mock sources to our real images in each field.  We add sources from $m =$ 23--27 mag (to ensure a robust photometric measurement), with a log-normal half-light radius distribution peaking at $\sim$1.5 pixels ($\sim$0.2 kpc at $z =$ 10; compact but modestly resolved, comparable to high-redshift sources, see \S \ref{sec:sizes}), with a log-normal S\'ersic parameter distribution, peaking at 1.2.  These mock sources were generated with \textsc{galfit} \citep{peng02}, and added at random positions to the F277W, F356W and F444W images.  We combined the former two to create a detection image, and ran \texttt{\textsc{SE}} in the same way as on our real data to generate a F444W catalog (focusing on this one band as all images were PSF-matched to F444W).  Finally, we match sources in the \texttt{\textsc{SE}} catalog to their input values, and compare the ratio of input-to-recovered fluxes.  We find a median ratio of 1.08, measured between 25 $< m_{F444W} <$ 26.  There is a slight trend with magnitude of lower corrections for brighter sources, and higher for fainter sources, but only at the 1-2\% level.  We thus elect to use a single correction factor of 1.08 to all NIRCam fluxes and uncertainties.

For the {\it HST} fluxes, we elect to derive any residual aperture correction from comparison to the \citet{finkelstein22} photometric catalog, which performed similar simulations to derive total fluxes.  Matching sources in each of the six {\it HST} bands, we find a typical needed correction factor of $\sim$1.35 ($\pm$ 0.02).  These values are roughly consistent with the combination of the correction derived in \citet{finkelstein22} of 1.20 with the NIRCam correction derived here of 1.08.  We apply this same 1.35 correction to all {\it HST} bands, such that colors amongst these bands are not changed.  We note that in \S 3.6 below we test for the presence of any remaining photometric offsets in our catalog, and find these to be small ($\lesssim$5\%), indicating our procedure for deriving total fluxes in all 13 {\it HST} $+$ {\it JWST} bands is robust, especially for this nascent observatory.

\subsection{Noise Estimation}
While \texttt{\textsc{SE}} does provide an estimate of the noise, it is reliant on the accuracy of the provided error maps.  Given the nascent nature of the {\it JWST} reduction pipeline, we obtain estimates of our image noise directly from the images themselves.  We follow the methods of \citet{finkelstein22}, based on previous methodology outlined in \citet{papovich16}.  Our goal is to estimate the noise based on the number of pixels in an aperture.  We fit for the noise as a function of aperture size by measuring the fluxes in circular apertures with 30 different diameters, ranging from 0.1\arcs\ (3.33 pixels) to 3\arcs\ (100 pixels). 

When defining these positions, we restrict aperture placement to pixels with real non-zero values in the ERR image, and a zero value in the \texttt{\textsc{SE}} segmentation map within the aperture, avoiding real objects.  We also require these apertures to be non-overlapping to avoid correlating our noise estimation.  To improve statistics for smaller apertures, we placed apertures in two separate iterations -- ``small" (d $\leq$ 1.5\arcs) and ``large" (d $>$ 1.5\arcs).  We were able to place 5000 and 500 non-overlapping apertures in these two iterations, respectively.

We create a detection image setting the pixels at these positions to unity, with the rest of the image set to zero, and ran \texttt{\textsc{SE}} in two-image mode.  We measured fluxes at these positions in all 30 circular apertures with diameters ranging from 1 -- 200 pixels.  We calculate the 1$\sigma$ noise in each aperture size by measuring the median absolute deviation of the measured flux values (multiplying by 1.48 to convert to a Gaussian-like standard deviation).  Finally, we fit a curve to the noise in a given aperture as a function of pixels in that aperture, using this equation \citep{gawiser06a}:
\begin{equation}
\sigma_{N} = \sigma_1 \alpha N^{\beta}
\end{equation}
where $\sigma_N$ is the noise in an aperture containing N pixels, and $\sigma_1$ is the pixel-to-pixel noise measured in each image as the sigma-clipped standard deviation of all non-object pixels (see Figure 3 in \citealt{finkelstein22} for an example of this process). 

We fit the free parameters $\alpha$ and $\beta$ with an IDL implementation of \textsc{emcee} (see \citealt{finkelstein19} for details).  We used these functional form fits for each filter to calculate the photometric uncertainties for each object, using both the number of pixels in its Kron aperture (Area $=$ $\pi \times$ A\_IMAGE $\times$ B\_IMAGE $\times$ KRON\_RADIUS$^2$), as well as a value for a given circular aperture.  These values were scaled by the ratio of the error image value at the central position of a given source to the median error value of the whole map, thereby allowing the noise to be representative of the local noise level.

Finally, to account for variable image noise not captured by the error image value at the central pixel, for each object in our catalog we also calculate a local noise measurement.  This local noise was calculated at 0.2\arcs, 0.3\arcs, 0.4\arcs\ and 0.5\arcs-diameter apertures, by fitting a Gaussian to the negative side of the flux distribution in the 200 closest apertures from the above process.

\begin{figure*}[!t]
\epsscale{1.2}
\plotone{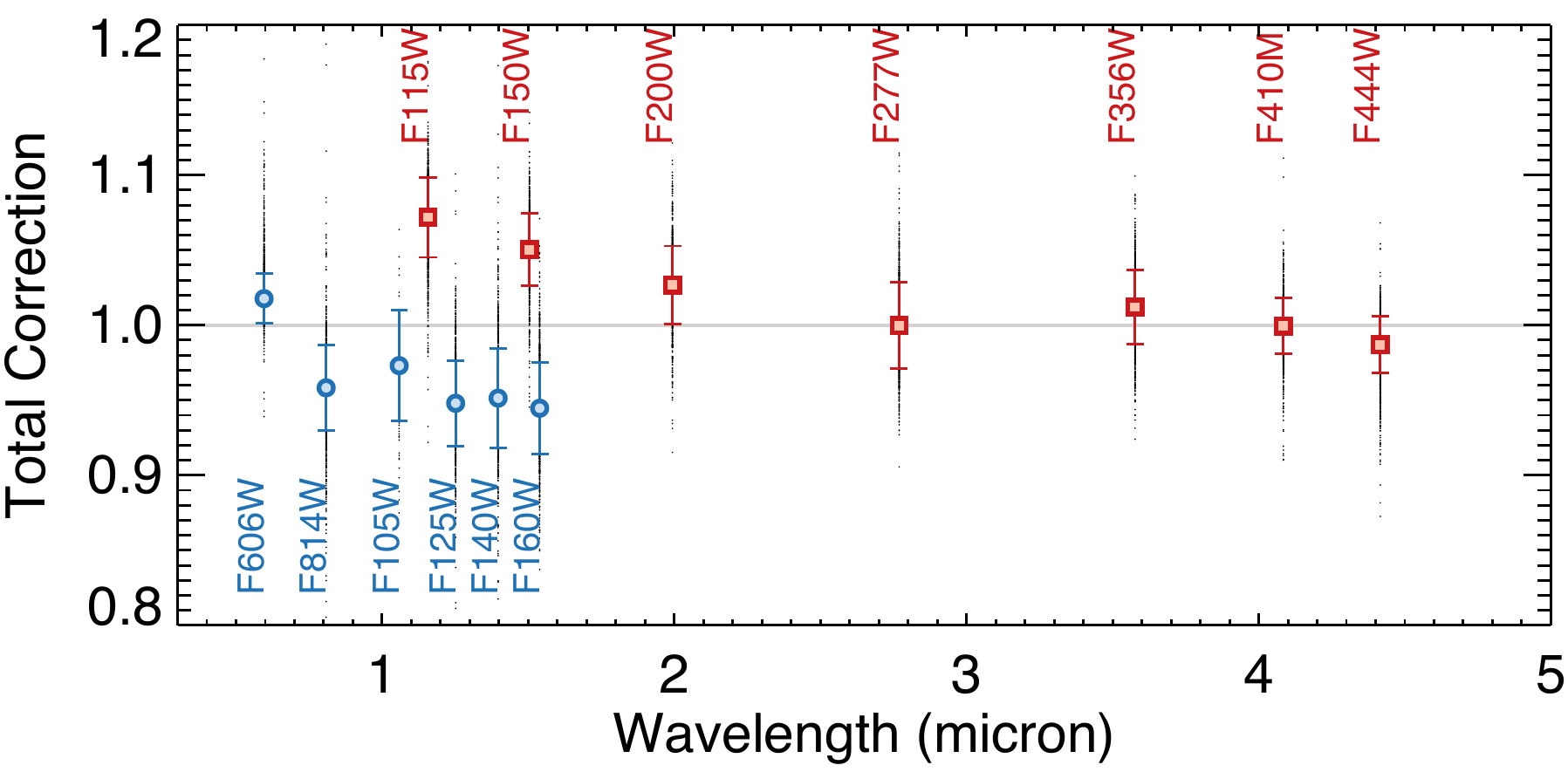}
\caption{The total corrections applied to our catalog to account for residual systematic offsets in our total flux estimations, mismatches between the spectral templates used for photometric redshift fitting, and any remaining photometric calibration corrections (shown as squares; small dots are individual galaxies).  These multiplicative offsets were derived by comparing the observed fluxes in each filter to the best-fitting \texttt{\texttt{\textsc{EAZY}}} model templates for $\sim$900 sources with robustly known spectroscopic redshifts in our field. The values are tabulated in Table 1.}
\label{fig:zpcorr}
\end{figure*}

\subsection{Multi-band Catalog}

We compose a multi-band catalog from the individual-filter catalogs created by \texttt{\textsc{SE}}.  As \texttt{\textsc{SE}} cannot parse the world-coordinate system in the {\it JWST} data model image headers, we use astropy.wcs \textsc{wcs\_pix2world} to derive celestial coordinates from the \texttt{\textsc{SE}} x, y positions.  We apply a photometric zeropoint to convert the image from MJy sr$^{-1}$ to erg s$^{-1}$ cm$^{-2}$ Hz$^{-1}$, and apply both aperture corrections derived above to all flux and flux error estimates.  We correct for Galactic extinction using an E(B-V) of 0.006 for the EGS field, and a \citet{cardelli89} Milky Way attenuation curve.

Both the MAG\_AUTO-derived aperture corrections and simulation-based (and {\it HST} catalog-based) residual aperture corrections were applied to all Kron-derived fluxes and uncertainties, maintaining our accurate colors while representing the total fluxes for each source.  In our final catalog, we include both these Kron-based fluxes, as well as fluxes measured in circular apertures with diameters ranging from 0.05\arcs\ to 2.0\arcs.  While the Kron fluxes will be used for most of the analysis, we do use the circular apertures (with a fiducial diameter of 0.2\arcs, or 6.67 pixels) as a measure of detection significance, as the sizes of Kron apertures can be affected by the proximity to bright sources.  These circular apertures are corrected for Galactic attenuation, but not corrected to total, as we will use them solely for detection significance.

For both the Kron and circular apertures, we calculate the noise per source following the method in \S 3.4, which is dependent on both the aperture area and the effective rms map value at the position of the source.  Finally, we flag any sources which had either a zero or NaN in any error column, replacing their flux error with 10$^{12}$ nJy (several orders of magnitude larger than any real source error).

In Table 1, we include an estimate of the limiting 5$\sigma$ magnitude for our catalog.  To calculate this, we use the noise functions described above to derive the flux density uncertainty in an aperture of diameter 0.2\arcs.  We then measure the enclosed flux at this radius from the stacked PSF.  We then divide the flux uncertainty by the enclosed fraction of flux to estimate the total noise for a point source.  Finally, we multiply this value by five, and convert to an AB magnitude.  Both the enclosed flux values and the limiting magnitudes are listed in Table 1.  While the depths were broadly as expected based on the pre-launch exposure-time calculator, the F115W image (with double the exposure time), was expected to be 0.3 mag deeper.  The very low background at that wavelength has led to those images being more read-noise dominated than expected, thus this image has a depth comparable to the bulk of the NIRCam filters.  The primary impact is that photometric redshifts will be slightly more uncertain at $m >$ 29 than originally planned.

\subsection{Photometric Calibration Validation}
While the photometric calibration of NIRCam has substantially improved over the first few months since science acquisition began, there is still some uncertainty, and many of the reference files still used in the pipeline are preliminary (to be refined during all of Cycle 1; see \citealt{boyer22} for more details on the NIRCam photometric calibration used here).  It is thus prudent to check the accuracy of the photometric calibration.  The {\it JWST} pipeline applies a photometric calibration taking into account the area of each pixel (using a {\it photom} reference file), resulting in image units of MJy sr$^{-1}$.  For the 30mas scale of our images, the conversion factor from MJy sr$^{-1}$ to erg s$^{-1}$ cm$^{-2}$ Hz$^{-1}$ is 2.1154 $\times$ 10$^{-31}$.

To check the accuracy of this calibration, we use a sample of objects with published spectroscopic redshifts in this field (compiled by N.\ Hathi, private communication; including DEEP2; \citealt{newman13}, MOSDEF; \citealt{kriek15}, and 3DHST; \citealt{momcheva16}, among others).  Applying a matching radius of 0.5\arcs, removing duplicates, keeping the two highest quality redshift flags, and restricting to F160W $<$ 24 mag, we find 988 matches which fall in our CEERS/NIRCam-covered area.  These objects span $z =$ 0--4, with a median $z_{spec} =$ 1.1.

We generate the expected fluxes of these sources in all 13 photometric bands using the  \texttt{\textsc{EAZY}} \citep{brammer08} photometric-redshift fitting software.  While in \S 4 below we will use  \texttt{\textsc{EAZY}} to measure photometric redshifts, here we run \texttt{\textsc{EAZY}} with the redshift fixed to the spectroscopic redshift value.  This thus obtains the best-fitting galaxy template to the observed photometry.  Assuming that this template set spans the color range of real galaxies (see \S 4.1), a comparison of the observed fluxes in a given filter to those predicted by \texttt{\textsc{EAZY}} can inform us on any systematic discrepancies in the fluxes in our catalog.  We measured the median offset for the subset of these sources with a measured signal-to-noise ratio of $>$ 5 and an  \texttt{\textsc{EAZY}} goodness-of-fit ($\chi^2$) $<$ 20.  This ensures that a well-fit model is found to a set of robust photometric measurements.  This resulted in typically $\sim$850--900 sources per filter (84 in F105W, which has significantly less coverage), from which we measured a sigma-clipped median and standard deviation.

We tabulate these derived corrections in Table 1, and plot the median values and full dispersion in Figure~\ref{fig:zpcorr}.  For NIRCam, the LW bands are all consistent with unity, which is a significant improvement over the state of the calibration in late summer 2022.  For the short-wavelength channels, we do find a needed correction on the 3--7\% level.  For \textit{HST}, we find that with the exception of F606W, the remaining bands are $\sim$3--5\% too bright.  As {\it HST} is extremely well calibrated, this implies that the 35\% correction we applied based on the comparison to the \citet{finkelstein22} catalog may have been slightly too large (we note that that study did not complete such a zeropoint-offset analysis).  Nonetheless, we applied these corrections.  To test their accuracy, we performed another iteration of this analysis after applying these offsets, and found that no significant residual correction was present.  We thus apply these corrections, listed in Table 1, to all fluxes and flux errors in our final photometric catalog. We reiterate that while photometric calibration was our motivation for this test, the resulting corrections are a combination of residual differences between our estimated total fluxes and true total fluxes, differences between the model templates and true galaxies, and true photometric zeropoint inaccuracies.

\section{Selection of Redshift $>$ 9 Galaxies}

Our analysis method produced a photometric catalog which contains our best estimates of the total fluxes in each of 13 filters spanning 0.6--5.0$\mu$m, with robust flux uncertainty values.  We also include fluxes measured in a range of circular apertures.  In this section we will use this catalog to select galaxies at $z >$ 8.5.  Below we will create identifiers for each object inclusive of the CEERS field the object was covered in, and the \texttt{\textsc{SE}} number within that field.  For example, ``Maisie's Galaxy", the previously identified $z \sim$ 12 galaxy candidate from \citet{finkelstein22c}, is referred to here as CEERS2\_5429.

\subsection{Photometric Redshift Estimation}

We measure photometric redshifts with  \texttt{\textsc{EAZY}} for our entire photometric catalog, which contains $\sim$40,000 sources across all four fields.   \texttt{\textsc{EAZY}} fits non-negative linear combinations of user-supplied templates to derive probability distribution functions
(PDFs) for the redshift, based on the quality of fit of the various
template combinations to the observed photometry for a given source.
The template set we use includes the ``tweak\_fsps\_QSF\_12\_v3" set
of 12 FSPS \citep{conroy10} templates recommended by the  \texttt{\textsc{EAZY}} documentation. 

As the population of $z >$ 9 galaxies is expected to exhibit fairly blue rest-frame UV colors, we follow Larson et al. (in prep) in adding six additional templates.  They derived these templates
by combining stellar population spectra from BPASS \citep{eldridge09} with (optional) nebular emission derived with \textsc{Cloudy} \citep{cloudy}.  They used models with low metallicities (5\% solar), young stellar populations (stellar ages of 10$^6$, 10$^{6.5}$, and 10$^7$
Myr), inclusive of binary stars, and with a high ionization parameter (log $U = -$2).  Larson et al.\ (in prep) showed that the inclusion of these templates significantly improved the recovery of photometric redshifts from a mock catalog derived by a semi-analytic model \citep{yung22}, due to the better match between galaxy and template colors.  These templates were also used by \citet{finkelstein22c} in their analysis of the $z\sim$ 12 ``Maisie's Galaxy", where they found the inclusion of these bluer templates significantly improved the goodness-of-fit between the data and the best-fitting model.

We assume a flat prior in luminosity; while bright galaxies will have a redshift distribution significantly tilted towards lower redshift, the bright end of the high-redshift luminosity function is poorly known, thus we do not wish to bias against the selection of true, bright distant galaxies.  We include a
systematic error of 5\% of the observed flux values, and fit to our measured total flux and flux error values.  In our fiducial  \texttt{\textsc{EAZY}} run the redshift can span 0--20.  We perform an additional ``Low-z" run with the maximum redshift set to $z_{max} =$ 7 such that we can visualize the best-fitting low-redshift solution.

\subsection{Sample Selection}

Here we describe our selection criteria we use to identify candidate $z >$ 8.5 galaxies.  Following our previous work \citep{finkelstein10,finkelstein15,finkelstein22,finkelstein22c}, we utilize a combination of flux detection significance values and quantities derived from the full photometric redshift PDF (denoted $\mathcal{P}[z]$) to select our galaxy sample.  As a part of this selection, we make use of the integral of the $\mathcal{P}(z)$ in $\Delta z =$ 1 bins centered on integer redshift values.  Specifically, we denote the unit redshift where the integral in a $z \pm 0.5$ bin is the maximum compared to all other redshifts as $\mathcal{S}_z$.  For example, a sources with $\mathcal{S}_z =$ 10 would have $\int_{9.5}^{10.5}$ $\mathcal{P}(z)$ $dz$ greater than the integrated $\mathcal{P}(z)$ in all other $\Delta z =$ 1 bins.

Our primary selection criteria are:

\begin{itemize}
    \item A signal-to-noise ratio, as measured in 0.2\arcs\ diameter apertures in the non-PSF-matched images, of $>$ 5.5 in \emph{at least two} of the F150W, F200W, F277W, F356W or F444W bands.  We required this to be true with both the global as well as local noise values.  This allows the selection of galaxies across a broad range of redshifts and rest-UV colors.  We note that we experimented with requiring a higher significance detection in just one band, but this significantly increased the spurious source fraction.
    
    \item Error map values $<$ 1000 in all of the F115W, F150W, F200W, and F277W images.  This includes only objects with a measurable (though not necessarily significant) flux in the bluest four filters, necessary for selection of galaxies at $z \sim$ 9--13.
    
    \item A signal-to-noise ratio of $<$ 3 in bands blue-ward of the Ly$\alpha$ break.  While studies occasionally use more stringent signal-to-noise cuts, we choose this value to both account for the fact that any positive flux in all dropout bands is already accounted for by  \texttt{\textsc{EAZY}}, and that $>$1$\sigma$ random fluctuations can coincide at the positions of real galaxies with non-Gaussian noise as is present in these images.  For this criterion, we include F606W and F814W for all redshifts considered here.  We add F115W for $\mathcal{S}_z =$ 11--12, and F150W for $\mathcal{S}_z =$ 13--17.  These redshift values were chosen to ensure that the Ly$\alpha$ break is fully red-ward of the $\Delta z =$ 1 range for a given filter.
    
    \item $\int \mathcal{P}(z > 7) \geq$ 0.7.  This requires less than 30\% of the integrated $\mathcal{P}(z)$ to be at $z <$ 7.

    \item Best-fitting photometric redshift $z_{best} >$ 8 (defined as ``za'' with EAZY; the redshift corresponding to the highest likelihood) with a goodness-of-fit $\chi^2 <$ 60.
    
    \item $\mathcal{S}_z$ $\geq$ 9 (selecting a sample of galaxies at $z \gtrsim$ 8.5).
    
    \item $\Delta \chi^2$ $>$ 4, calculated as the difference between the best-fitting $\chi^2$ value for the low-redshift restricted model to the fiducial model.  By requiring a value greater than four, we require the low-redshift model to be ruled out at $\geq$ 2$\sigma$ significance.  We note that \citet{harikane22} required a more conservative $\Delta \chi^2$ $>$ 9, based on results from the CEERS simulated imaging.  We explore the impacts of such a cut in \S 7.2.

\end{itemize}

\begin{figure*}[!t]
\epsscale{1.15}
\plotone{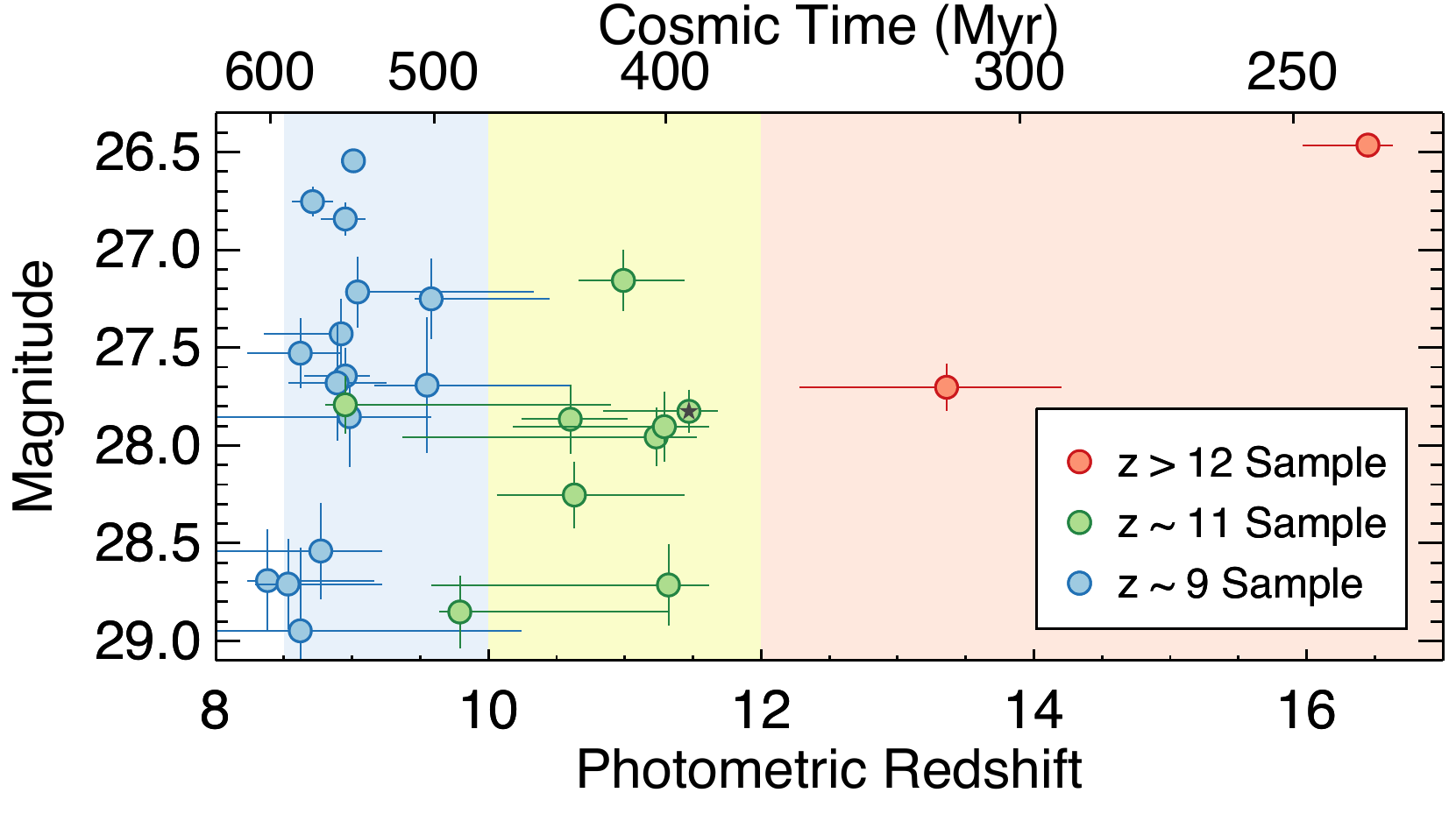}
\caption{The distribution of the best-fitting (minimized $\chi^2$) photometric redshift versus apparent magnitude.  Each object is color-coded by its redshift sample (with the background shading denoting the redshift range where the $\mathcal{P}(z)$ was integrated to determine the sample placement; the green circle far into in the blue region [CEERS6\_7641] has a double peaked $\mathcal{P}(z)$).  The magnitude plotted is F150W for $z \sim$ 9, F200W for $z \sim$ 11, and F277W for $z >$ 12 galaxies.  The small black star denotes ``Maisie's Galaxy" \citep{finkelstein22c}, known here as CEERS2\_5429.  The top axis shows ``Cosmic Time" (the time since the Big Bang) for our adopted cosmology.}
\label{fig:zmag}
\end{figure*}

\subsection{Sample Vetting}
This automated selection resulted in a sample of 64 candidate galaxies with $z \gtrsim$ 8.5.  To explore the impact of potential contamination by spurious (e.g., non-astrophysical) sources, we visually inspected each object in all images to search for obvious diffraction spikes, sources on image edges (as the dithering is different in different filters, the edges do not necessarily line up between different images), and un-flagged cosmic rays.

To perform this inspection, three authors (SF, JK, PAH) viewed ``bio" plots which showed small 1.5\arcs\ cutouts of the candidate in all images, with two different stretches, a 5\arcs\ cutout in the F200W and detection images, the fiducial and $z <$ 7 restricted $\mathcal{P}(z)$, and the spectral energy distribution plotted against both fiducial and Low-$z$ \texttt{\textsc{EAZY}} models. From this process we identified 36 spurious sources: six diffraction spikes, 11 un-flagged cosmic rays, and 19 sources on image edges.  We remove these 36 sources from our galaxy sample, and show images of all in Figure \ref{fig:spurious} in the Appendix, and list their positions in Table~\ref{tab:spurious}.  The cosmic rays were predominantly an issue in the long-wavelength image in pointings 3 and 6; these have longer exposure times than pointings 1 and 2, amplifying the impact of cosmic rays.  We note that future versions of our images and catalogs should be able to minimize these types of spurious sources by improving the cosmic-ray removal and using the {\it JWST} data model ``Context map" to identify pixels close to the image edge.

\begin{figure*}[!t]
\epsscale{1.15}
\plotone{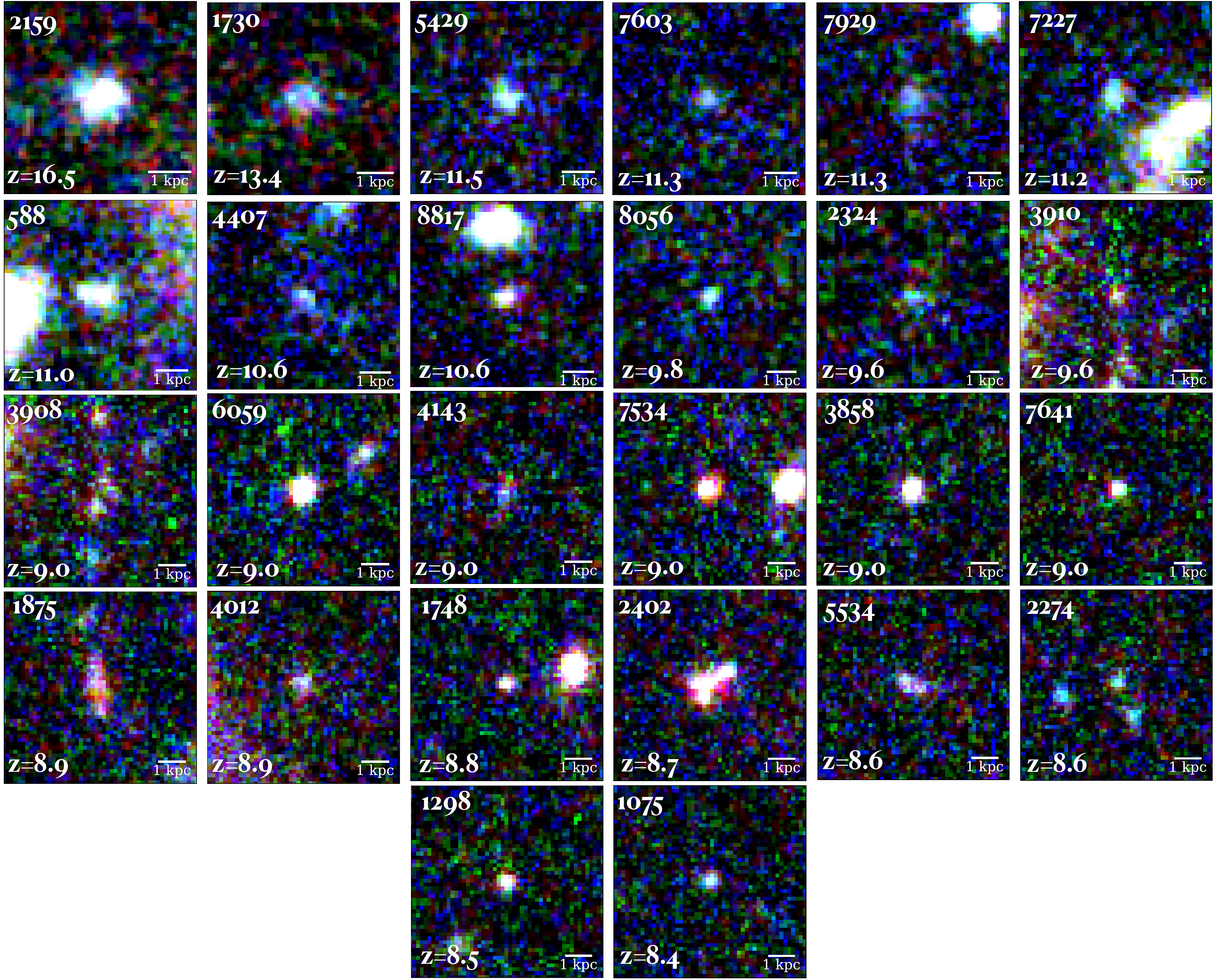}
\caption{A montage of images for the galaxy candidates identified here. For each galaxy the image shows as blue-green-red the three NIRCam bandpasses closest to the Ly$\alpha$ break, this includes: F150W, F200W, and F277W for galaxies at $z=9$; F200W, F277W and F356W for galaxies in the $z=10$--12 samples; F277W, F356W, and F444W for galaxies in the $z \geq 12$ samples.   In each case the images are 1.5\arcsec\ $\times$ 1.5\arcsec.  The inset scale bar shows 1 (physical) kpc.  We list the numerical IDs and the best-fitting photometric redshift values (the redshift uncertainties are listed in Table 2).}
\label{fig:colorims}
\end{figure*}

Upon visual inspection, we noticed that a small fraction of the sample (four objects) had Kron apertures which were drawn too large due to the presence of bright, nearby galaxies.  To explore the impact of potential light from neighboring galaxies affecting the colors, we performed an additional run of \texttt{\textsc{EAZY}} using the colors measured in 0.3\arcs\ diameter apertures to more accurately measure the colors of the candidate galaxies.  For this run we used the PSF-matched photometry, but did not apply any aperture correction as the photometric redshift depends only on colors and not luminosity.  We found that two sources, CEERS3\_1537 (initial $z_{phot} =$ 10.5) and CEERS6\_7478 (initial $z_{phot} =$ 8.6) had their photometric redshifts shift to lower redshift with these smaller apertures; these were the same two sources where the Kron aperture was most significantly stretched.  We thus remove these two objects from our sample, though we show their SEDs and image stamps in Figure~\ref{fig:largekron} in the Appendix.

\begin{figure*}[!t]
\epsscale{1.15}
\plotone{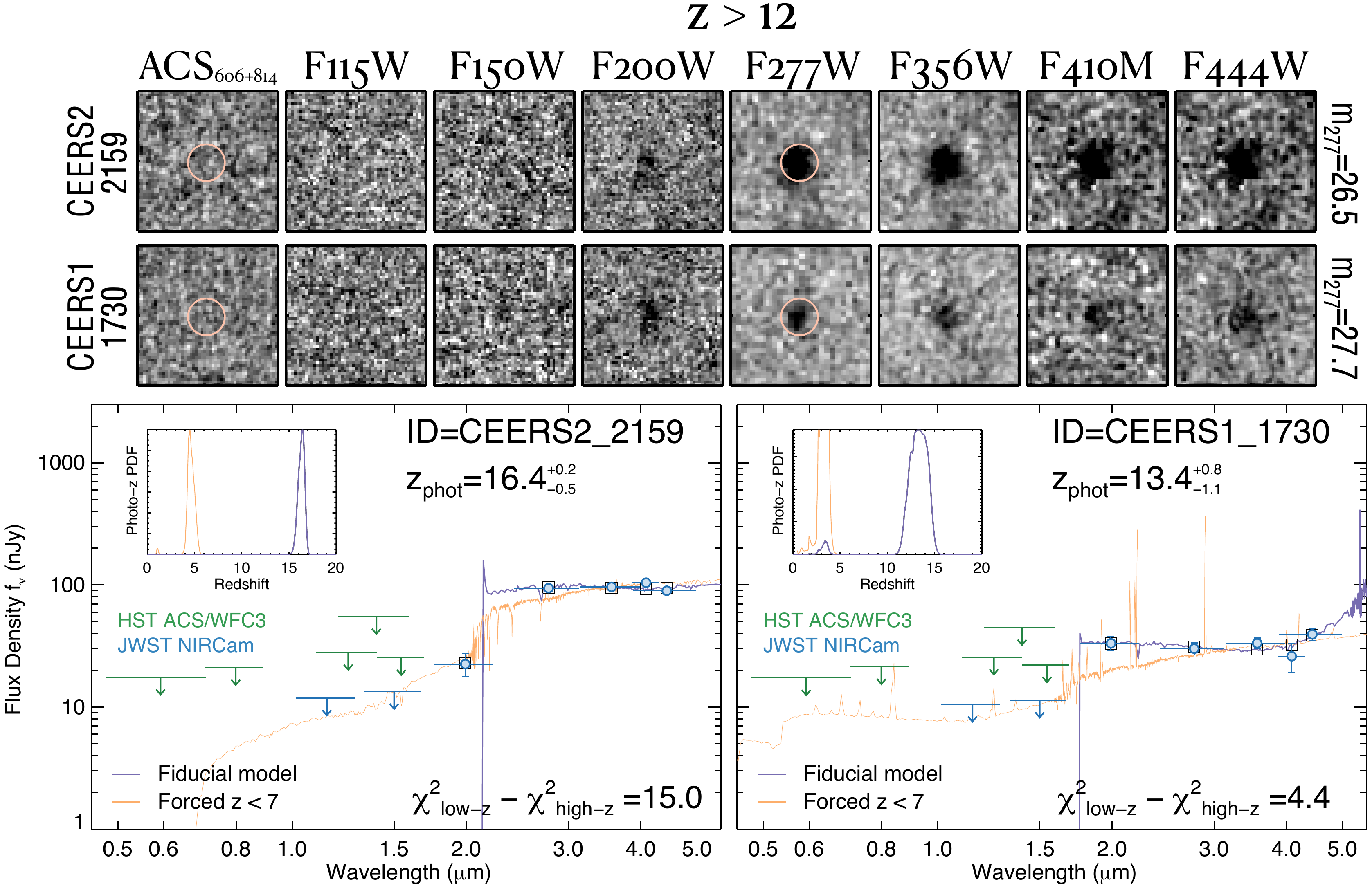}
\caption{Top) Stamp images, 1.5\arcs\ on a side, of the two galaxies that are best-fit with $z >$ 12 (shown from the non-PSF-matched images). The red circle denotes a 0.4\arcs\ diameter region around the source (which we show in only two bands for clarity). CEERS1\_1730 has a best-fit photo-z of 13.4, albeit with a wide 68\% confidence range of 12.3--14.2.  CEERS2\_2159 is best-fit by $z =$ 16.5 (16.0--16.6), and was first identified by \citet{donnan22}. The bottom panels show the best-fitting \texttt{\textsc{EAZY}} models (both overall, and constrained to $z <$ 7), the model bandpass fluxes (open squares), alongside the observed photometry (circles; upper limits are 2$\sigma$).  The inset panels show the $\mathcal{P}(z)$ distributions.  Both sources exhibit well-constrained Ly$\alpha$ breaks, implying the redshifts are $z >$ 11. CEERS1\_1730 does show a small low-redshift solution, and its primary $\mathcal{P}(z)$ peak extends to $z \sim$ 10.5.}
\label{fig:z12}
\end{figure*}

The remaining two sources (CEERS1\_3908 and CEERS1\_3910) had no significant change in their photometric redshift when measured with these small apertures.  For these two objects we correct their fluxes by a single factor to account for potential excess brightness from the neighbor.  We do this by deriving a flux correction as the ratio between the median total flux for objects with similar 0.3\arcs\ diameter aperture fluxes as a given source to the actual Kron-based (aperture-corrected) total flux for those comparison sources, and divided this quantity by the same quantity for the sources in question.  We apply this correction to all filters such that the colors do not change.  For CEERS1\_3908 this correction factor was 0.6; for CEERS1\_3910 it was close to unity, so we applied no correction.

\begin{deluxetable*}{cccccccccc}
\vspace{2mm}
\tablecaption{Summary of $z >$ 9 Candidate Galaxies}
\tablewidth{\textwidth}
\tablehead{\multicolumn{1}{c}{ID} & \multicolumn{1}{c}{RA} & \multicolumn{1}{c}{Dec} & \multicolumn{1}{c}{m$_{F277W}$} & \multicolumn{1}{c}{$\int_7^{20} \mathcal{P}(z)$} & \multicolumn{1}{c}{Sample} & \multicolumn{1}{c}{Photometric} & \multicolumn{1}{c}{M$_{1500}$} & \multicolumn{1}{c}{R$_{h}$} & \multicolumn{1}{c}{$\Delta \chi^2$}\\
\multicolumn{1}{c}{$ $} & \multicolumn{1}{c}{(J2000)} & \multicolumn{1}{c}{(J2000)} & \multicolumn{1}{c}{(mag)} & \multicolumn{1}{c}{$ $} & \multicolumn{1}{c}{$\mathcal{S}_z$} & \multicolumn{1}{c}{Redshift} & \multicolumn{1}{c}{(mag)} & \multicolumn{1}{c}{(pix)} & \multicolumn{1}{c}{$ $}}

\startdata
CEERS2\_2159&214.914548&52.943023&26.5&1.00&$>$12&16.45$^{+0.18}_{-0.48}$&$-$21.7$^{+0.1}_{-0.1}$&4.4 $\pm$ 6.0&4.4\\
CEERS1\_1730&215.010022&53.013641&27.7&0.97&$>$12&13.36$^{+0.84}_{-1.08}$&$-$20.4$^{+0.2}_{-0.2}$&5.3 $\pm$ 1.0&15.0\\
\hline
CEERS2\_588&214.906633&52.945504&26.9&0.98&11&10.99$^{+0.45}_{-0.33}$&$-$20.6$^{+0.2}_{-0.2}$&4.0$^{\ast}$&7.4\\
CEERS1\_8817&215.043999&52.994302&28.1&1.00&11&10.60$^{+0.42}_{-0.36}$&$-$19.7$^{+0.2}_{-0.2}$&2.4 $\pm$ 0.7&12.1\\
CEERS2\_7929&214.922783&52.911528&28.1&0.99&11&11.29$^{+0.33}_{-1.11}$&$-$19.8$^{+0.4}_{-0.1}$&4.1 $\pm$ 1.0&9.1\\
CEERS6\_7641&214.897232&52.843854&28.1&0.99&11&8.95$^{+1.95}_{-0.15}$&$-$19.3$^{+0.1}_{-0.5}$&2.0 $\pm$ 0.6&8.8\\
CEERS2\_5429$^{a}$&214.943148&52.942442&28.3&1.00&11&11.47$^{+0.21}_{-0.63}$&$-$19.9$^{+0.2}_{-0.0}$&4.6 $\pm$ 1.2&25.8\\
CEERS1\_7227&215.037504&52.999394&28.3&0.85&11&11.23$^{+0.30}_{-1.86}$&$-$19.7$^{+0.3}_{-0.1}$&6.4 $\pm$ 2.6&4.0\\
CEERS6\_7603&214.901252&52.846997&28.9&0.98&11&11.32$^{+0.30}_{-1.74}$&$-$18.8$^{+0.4}_{-0.2}$&2.1 $\pm$ 0.8&7.1\\
CEERS6\_4407&214.869661&52.843646&29.0&0.99&11&10.63$^{+0.81}_{-0.57}$&$-$19.1$^{+0.2}_{-0.3}$&4.2 $\pm$ 1.2&9.3\\
CEERS6\_8056&214.850129&52.808052&29.1&0.99&11&9.79$^{+1.53}_{-0.15}$&$-$18.4$^{+0.1}_{-0.4}$&2.0 $\pm$ 0.7&7.2\\
\hline
CEERS2\_2402&214.844766&52.892104&26.7&1.00&9&8.71$^{+0.15}_{-0.15}$&$-$20.5$^{+0.1}_{-0.0}$&4.9 $\pm$ 0.7&63.9\\
CEERS1\_6059&215.011706&52.988303&27.0&1.00&9&9.01$^{+0.06}_{-0.06}$&$-$20.6$^{+0.0}_{-0.1}$&2.6 $\pm$ 0.4&48.0\\
CEERS1\_1875&214.951936&52.971742&27.1&0.92&9&8.92$^{+0.06}_{-0.57}$&$-$19.9$^{+0.3}_{-0.1}$&8.4 $\pm$ 1.3&5.4\\
CEERS1\_3858&214.994402&52.989379&27.2&0.99&9&8.95$^{+0.15}_{-0.18}$&$-$20.4$^{+0.1}_{-0.1}$&2.1 $\pm$ 0.5&7.6\\
CEERS2\_7534&214.876144&52.880826&27.4&1.00&9&8.95$^{+0.18}_{-0.30}$&$-$19.5$^{+0.2}_{-0.1}$&1.4 $\pm$ 0.3&39.6\\
CEERS1\_3908&215.005189&52.996580&27.3&0.96&9&9.04$^{+1.29}_{-0.06}$&$-$20.1$^{+0.1}_{-0.3}$&10.0$^{\ast}$&5.3\\
CEERS6\_4012&214.888127&52.858987&27.6&0.92&9&8.89$^{+0.36}_{-0.36}$&$-$20.0$^{+0.2}_{-0.2}$&5.5 $\pm$ 2.0&4.8\\
CEERS2\_2324&214.861602&52.904603&27.6&1.00&9&9.58$^{+0.87}_{-0.12}$&$-$20.1$^{+0.2}_{-0.2}$&2.0 $\pm$ 0.8&10.7\\
CEERS1\_3910&215.005365&52.996697&28.0&0.96&9&9.55$^{+1.05}_{-0.39}$&$-$19.5$^{+0.5}_{-0.2}$&8.6 $\pm$ 1.4&5.7\\
CEERS1\_5534&214.950078&52.949267&27.9&1.00&9&8.62$^{+0.30}_{-0.39}$&$-$19.5$^{+0.2}_{-0.2}$&3.6 $\pm$ 0.7&12.6\\
CEERS1\_4143&214.966717&52.968286&28.1&0.84&9&8.98$^{+0.60}_{-1.23}$&$-$19.5$^{+0.3}_{-0.2}$&4.8 $\pm$ 1.5&4.0\\
CEERS3\_1748&214.830685&52.887771&28.5&0.93&9&8.77$^{+0.45}_{-1.08}$&$-$18.3$^{+1.0}_{-0.2}$&1.8 $\pm$ 0.7&4.2\\
CEERS2\_1298&214.902236&52.939370&28.6&1.00&9&8.53$^{+0.69}_{-0.24}$&$-$18.4$^{+0.2}_{-0.4}$&2.2 $\pm$ 0.8&12.7\\
CEERS2\_2274&214.846173&52.894001&29.1&0.85&9&8.62$^{+1.62}_{-1.41}$&$-$18.6$^{+0.3}_{-0.4}$&1.8 $\pm$ 0.8&4.2\\
CEERS2\_1075&214.907627&52.944611&29.1&1.00&9&8.38$^{+0.78}_{-0.15}$&$-$18.4$^{+0.2}_{-0.3}$&1.6 $\pm$ 0.7&12.7
\enddata
\tablecomments{The horizontal lines divide our three redshift samples (given by the sixth column). The photometric redshift is ``za'' from \texttt{\textsc{EAZY}}, which is the redshift where the $\chi^2$ is minimized.  The $\Delta \chi^2$ in the final column compares the best-fitting low-redshift (0.5 $< z <$ 7) model to the best-fitting high-redshift model; a value of $\geq$ 4 was required for selection.  $^{a}$Previously published as Maisie's Galaxy by \citet{finkelstein22c}.  The half-light radii are listed in pixels; our pixel scale is 30 mas.  $^{\ast}$Galfit did not converge for these two sources, so we list their \textsc{SE} half-light radii.}
\end{deluxetable*} \label{tab:sample}

\section{An Early Redshift $>$ 9 {\it JWST} Galaxy Sample}

\begin{figure*}[!t]
\epsscale{1.05}
\plotone{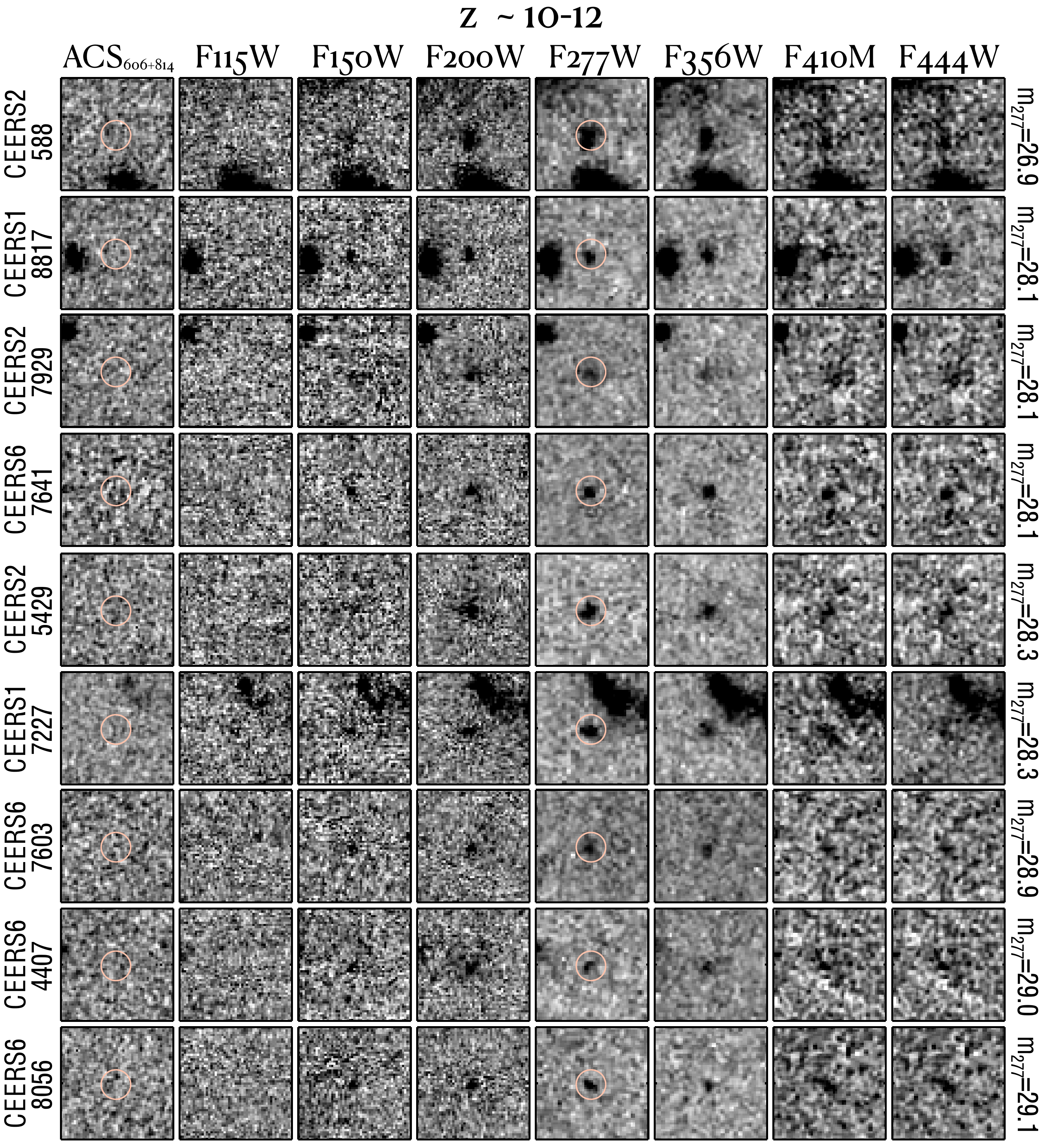}
\caption{Similar to Figure~\ref{fig:z12}, for the nine galaxies in the $z\sim$ 11 sample.}
\label{fig:z11stamps}
\end{figure*}

\begin{figure*}[!t]
\epsscale{1.15}
\plotone{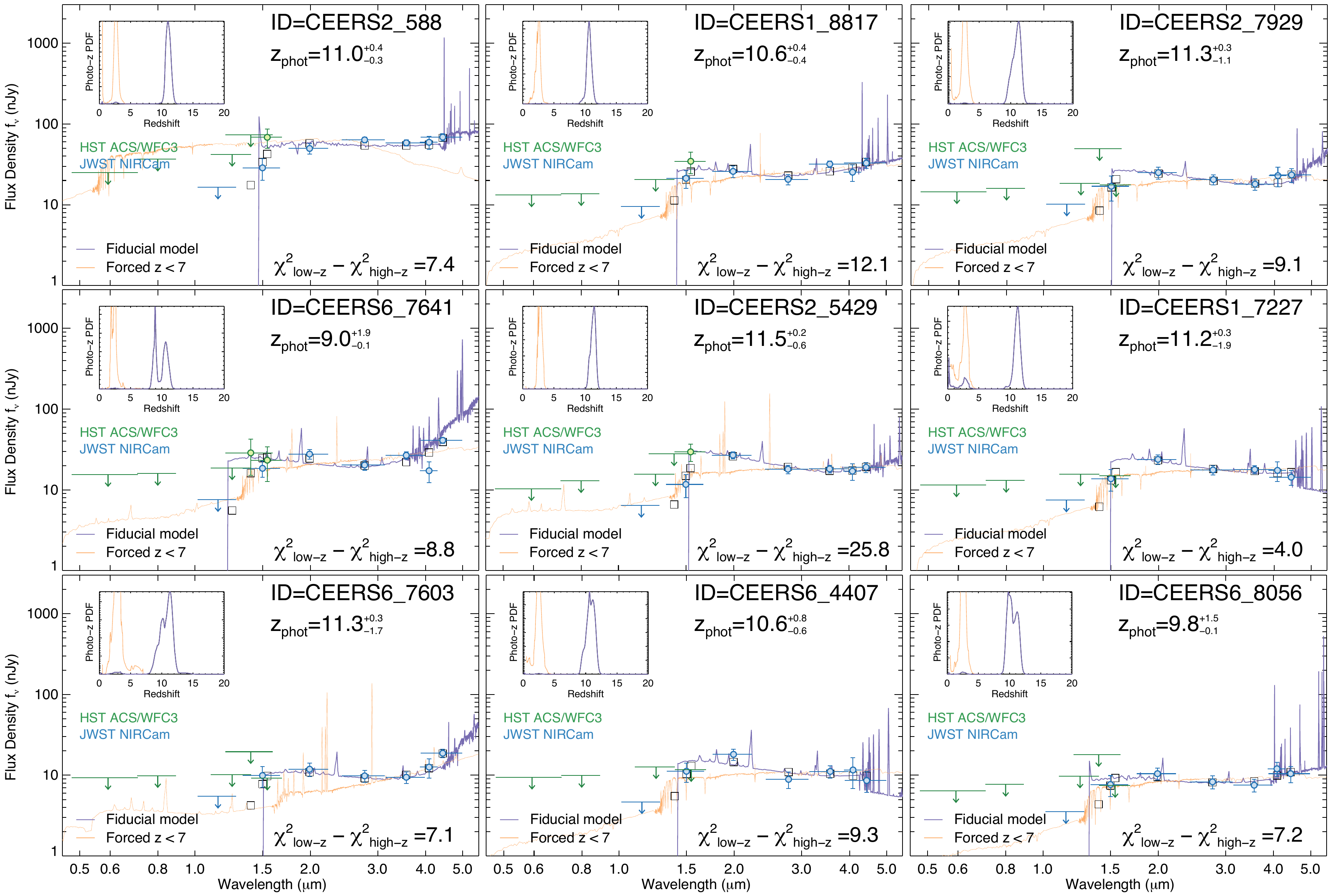}
\caption{The SEDs and $\mathcal{P}(z)$'s for the $z \sim$ 10--12 sample, with lines and symbols the same as Figure~\ref{fig:z12}.}
\label{fig:z11sed}
\end{figure*}

Our final sample consists of 26 galaxies.  We divide them into redshift bins for analysis.  The high redshifts now accessible with {\it JWST} coupled with the broad filter transmission functions make it harder to place galaxies in often-used $\Delta z =$ 1 redshift bins, and such bins span progressively smaller epochs of cosmic time (for example, a $\Delta z =$ 1 bin centered at $z =$ 12 would cover the Ly$\alpha$ break over only $\sim$40 Myr of cosmic time, compared to 200 Myr for one centered at $z =$ 6).  We thus split our sample into three redshift bins: a ``$z \sim$ 9" sample from $z =$ 8.5--10, a ``$z \sim$ 11" sample from $z =$ 10--12, and a ``$z >$ 12" sample from $z =$ 12--17.  We place candidate galaxies in the sample bin based on where their integrated $\mathcal{P}(z)$ across the bin is the largest.  These three bins cover roughly similar ranges of cosmic time (116, 105, 122 Myr, respectively).  We show the distribution of photometric redshifts and apparent magnitudes of our sample in Figure~\ref{fig:zmag}, and we tabulate the sample in Table \ref{tab:sample}.  We show cutout images of all candidates in Figures~\ref{fig:z12}, \ref{fig:z11stamps},
\ref{fig:z9stamps-bright} and \ref{fig:z9stamps-faint}, and SEDs in Figure~\ref{fig:z12}, \ref{fig:z11sed} and \ref{fig:z9sed}.  We also present a rest-UV color montage of our sample in descending redshift order in Figure~\ref{fig:colorims}.

As a check on our fiducial \texttt{\textsc{EAZY}} photometric redshifts, we also measured photometric redshifts with \texttt{\textsc{Cigale}} \citep{burgarella05}.  We find excellent agreement, with a median difference of 0.2 (with \texttt{\textsc{EAZY}} preferring the slightly higher redshifts).  Only the source CEERS1\_4143 has a difference in best-fitting redshift of $>$2, as \texttt{\textsc{Cigale}} prefers $z =$ 5.4 for this source, whereas \texttt{\textsc{EAZY}} finds $z =$ 9.0, though this source lies right at the boundary of our $\Delta \chi^2$ selection criterion.  Finally, we inspected the spatial distribution of our full sample of galaxies across the full CEERS field, and do not visually see any evidence of strong clustering (which would not be expected with such a small sample covering a broad redshift range).

\subsection{The $z >$ 12 Sample}

This highest-redshift sample contains two galaxies: CEERS2\_2159 and CEERS1\_1730, with 68\% confidence ranges on their photometric redshifts of 16.0--16.6 and 12.3--14.2, and F277W magnitudes of 26.5 and 27.7, respectively.  We will compare in detail our sample to those previously presented in the literature in \S6 below, but we note here that CEERS2\_2159 was first presented as a robust ultra-high-redshift candidate in \citet{donnan22}, with some caveats discussed in \citet[][a marginal 2.6$\sigma$ dust emission detection]{zavala22} and \citet[][environmental evidence]{naidu22b}.  CEERS1\_1730 is presented here for the first time.

The cutout images for these two objects are shown in the top panel of Figure~\ref{fig:z12}.  It is apparent that there is no significant flux at the positions of both candidates in the stacked ACS F606W$+$F814W images, nor in the F115W image.  For CEERS2\_2159 no flux is also seen in F150W, and the F200W flux is noticeably fainter than in F277W, consistent with its photometric redshift of $z \sim$ 16--16.6 (where the Ly$\alpha$ break would be in the red half of the F200W bandpass).  For CEERS1\_1730, faint flux is seen in F150W.  However, this is measured at just 1$\sigma$ significance in the Kron aperture.  This leads to a fairly broad $\mathcal{P}(z)$, spanning $z \sim$ 12--14.  If this flux is real, then $z \sim$ 12 is more likely.  If it is not significant, then $z \sim$ 14 is possible.  The SEDs, shown in the bottom panel of Figure~\ref{fig:z12}, show the strong observed Ly$\alpha$ break in both objects, with fairly blue UV spectral slopes redward of the break.  For both objects, the high-redshift model is a significantly better fit to the photometry than the best-fitting low-redshift model, though for CEERS1\_1730, $\Delta \chi^2$ is only 4.4; this low-redshift solution is visible as a small peak in the fiducial $\mathcal{P}(z)$.  CEERS2\_2159 shows no significant lower-redshift peak in its $\mathcal{P}(z)$ due to its brighter magnitude leading to more robust constraints on the full shape of the SED.

\subsection{The $z \sim$ 11 Sample}

The $z \sim$ 11 sample contains nine galaxies.  As shown by their cutout images in the bottom panel of Figure~\ref{fig:z11stamps}, all show no significant flux in the ACS stack and F115W images, consistent with $z >$ 9.5.  Many exhibit a red F150W$-$F200W color, suggesting the Ly$\alpha$ break is in F150W.  The SEDs and $\mathcal{P}(z)$s for all nine are shown in Figure~\ref{fig:z11sed}.  First looking at the $\mathcal{P}(z)$s in the upper-left of each panel, the amplitude of the detected Ly$\alpha$ break is strong enough to either eliminate, or leave a very small low-redshift solution.  As shown in Table 2, the integrated $\mathcal{P}(z>7)$ is $\geq$0.98 for 8/9 of these sources, with the remaining source (CEERS1\_7227) having a value of 0.85, significantly greater than our sample limit of 0.7 (this object also just barely satisfies our $\Delta \chi^2$ criterion, as only the F200W flux is discrepant with the best-fitting lower-redshift model, and it has a 2.2$\sigma$ significance detection in F115W, which could indicate a redshift closer to $z \sim$ 9). 

CEERS6\_7641 shows a double-peaked high-redshift $\mathcal{P}(z)$, with peaks at $z \sim$ 9 and $z \sim$ 11.  As its integrated $\mathcal{P}(z)$ at $z =$ 10--12 is larger than at $z =$ 8.5--10, this object was placed in the $z \sim$ 11 sample (this is the green $z \sim$ 11 symbol in Figure~\ref{fig:zmag} that is in the $z \sim$ 9 sample region), though clearly it has a near-equal probability of being at slightly lower redshift.  Finally, we note that similar to the $z >$ 12 galaxies, the UV spectral slopes for these nine sources all appear fairly blue.  Though a detailed analysis of this quantity is beyond the scope of this paper, it is clear that these objects all appear fairly low in dust attenuation.  We acknowledge that Ly$\alpha$-break selection can be biased against red sources \citep[e.g.][]{dunlop12}, thus a full quantitative analysis on the colors of these galaxies is reserved to future work.

\subsection{The $z \sim$ 9 Sample}

The remaining 15 candidate high-redshift galaxies fall into our $z \sim$ 9 sample.  We show the cutout images in Figures~\ref{fig:z9stamps-bright} and \ref{fig:z9stamps-faint} in the Appendix.  At $z <$ 9.5, the Ly$\alpha$ break is in the F115W band, thus we expect to see signal in that image for all but the highest-redshift sources in this subsample, though the F115W flux should be fainter than the F150W flux, which is apparent in the images.  Likewise, we see no significant flux in the ACS filters as expected.  Examining their SEDs and $\mathcal{P}(z)$s in Figure~\ref{fig:z9sed} in the Appendix, we see that nearly all show a tight high-redshift peak with very little probability of being at $z <$ 7.  Similar to the $z \sim$ 11 sample, the lowest integrated $\mathcal{P}(z>7)$ is 0.84, with 13/15 galaxies having integrated $\mathcal{P}(z>7) >$ 0.9, and 8/15 having integrated $\mathcal{P}(z>7) \geq$ 0.99.  

The object with the most complex $\mathcal{P}(z)$ constraints is CEERS2\_2274, which shows small peaks at $z \sim$ 2 and 6, a dominant peak at $z \sim$ 9, and a modest peak at $z \sim$ 10.5, with this uncertainty due to the lower signal-to-noise ratio on this object's faint fluxes.  While the majority of these $z \sim$ 9 galaxies show blue rest-UV spectral slopes, the lower redshift means that the reddest filters probe the 4000 \AA\ breaks, and it is clear that some objects have redder colors in these reddest bands.  This may be due to significant stellar mass in (somewhat) older stellar populations \citep[e.g.][]{labbe22}, though could also be due to the presence of very strong rest-frame optical nebular emission lines, which early {\it JWST} results indicate are ubiquitous in high-redshift star-forming galaxies \citep{endsley22,papovich22}.

Finally, we discuss the interesting pair of objects CEERS1\_3908 and CEERS1\_3910.  As discussed in \S 4.3, the presence of a bright neighbor skewed their Kron apertures, but even after correcting for this they appear as robust $z \sim$ 9 candidates.  In inspecting their cutout images, it is apparent that these two candidate galaxies are very close together with each consisting of a small knot of emission with centroids 0.6\arcs\ ($\sim$2.5 kpc) apart.  It is possible that these are two star-forming regions of the same host galaxy.  This hypothesis could be supported by the apparent very faint emission between the two clumps.  Deeper imaging could resolve this, though as they are resolved in our catalog we do not merge them here.  

As one additional check, we explored the impact of our photometric correction factors (\S 3.6) which we had applied to our catalog.  We re-ran  \texttt{\textsc{EAZY}} on our final sample of 26 high-redshift galaxy candidates removing this factor. We find that this makes effectively no change in the high-redshift solution, with best-fitting redshifts unchanged to more than the 1--2\% level, and all candidates continuing to satisfy our $\mathcal{P}(z)$ selection criteria.  We did notice a slight change in the best-fitting low-redshift model, leading to the median value of $\Delta \chi^2$ being reduced by 5\%.  This impacts sources which were close to our $\Delta \chi^2$ threshold, with CEERS1\_1730 ($\Delta \chi^2 =$ 4.4 $\rightarrow$ 3.5), CEERS1\_4143 ($\Delta \chi^2 =$ 4.0 $\rightarrow$ 3.7) and CEERS1\_7227 ($\Delta \chi^2 =$ 4.0 $\rightarrow$ 3.7) falling below our threshold of $\Delta \chi^2 >$ 4.  While these sources remain in the sample following our fiducial selection, their presence near this threshold leaves their inclusion more sensitive to these small correction factors.

\subsection{Galaxy Sizes} \label{sec:sizes}

We derive the sizes of all 26 galaxies in the sample by performing parametric fits on the F200W NIRCam images using \texttt{\textsc{Galfit}} \footnote{\href{https://users.obs.carnegiescience.edu/peng/work/galfit/galfit.html}{https://users.obs.carnegiescience.edu/peng/work/galfit/galfit.html}}  \citep{peng02,peng10}.  \texttt{\sc{Galfit}} finds the optimum S\'ersic fit to a galaxy's light profile using a least-squares fitting algorithm. As input, we use a 100x100 pixel cutout of the F200W image for each source, the corresponding error array (the `ERR' extension) as the input sigma image, and the empirically derived PSFs. We use the source location, magnitude, size, position angle, and axis ratios from the \texttt{\textsc{SE}} catalog as initial guesses. We allow the S\'ersic index to vary between 0.01 and 8, the magnitude of the galaxy between 0 and 45, the half-light radius (r$_{h}$) between 0.3 and 100 pixels, and the axis ratio between 0.0001 and 1. We also allow \texttt{\textsc{Galfit}} to oversample the PSF by a factor 9. We then visually inspected the best-fit model and image residual for each source to ensure that the fits were reasonable and that minimal flux remained in the residual. We also performed a fit on the PSF image itself in order to determine the smallest resolvable size. This value is 1.18$\pm$0.01 pixels. Two sources (CEERS1\_3908, a member of the pair discussed above, and CEERS2\_588) failed to converge on a fit.  

\begin{figure}[!t]
\epsscale{1.15}
\plotone{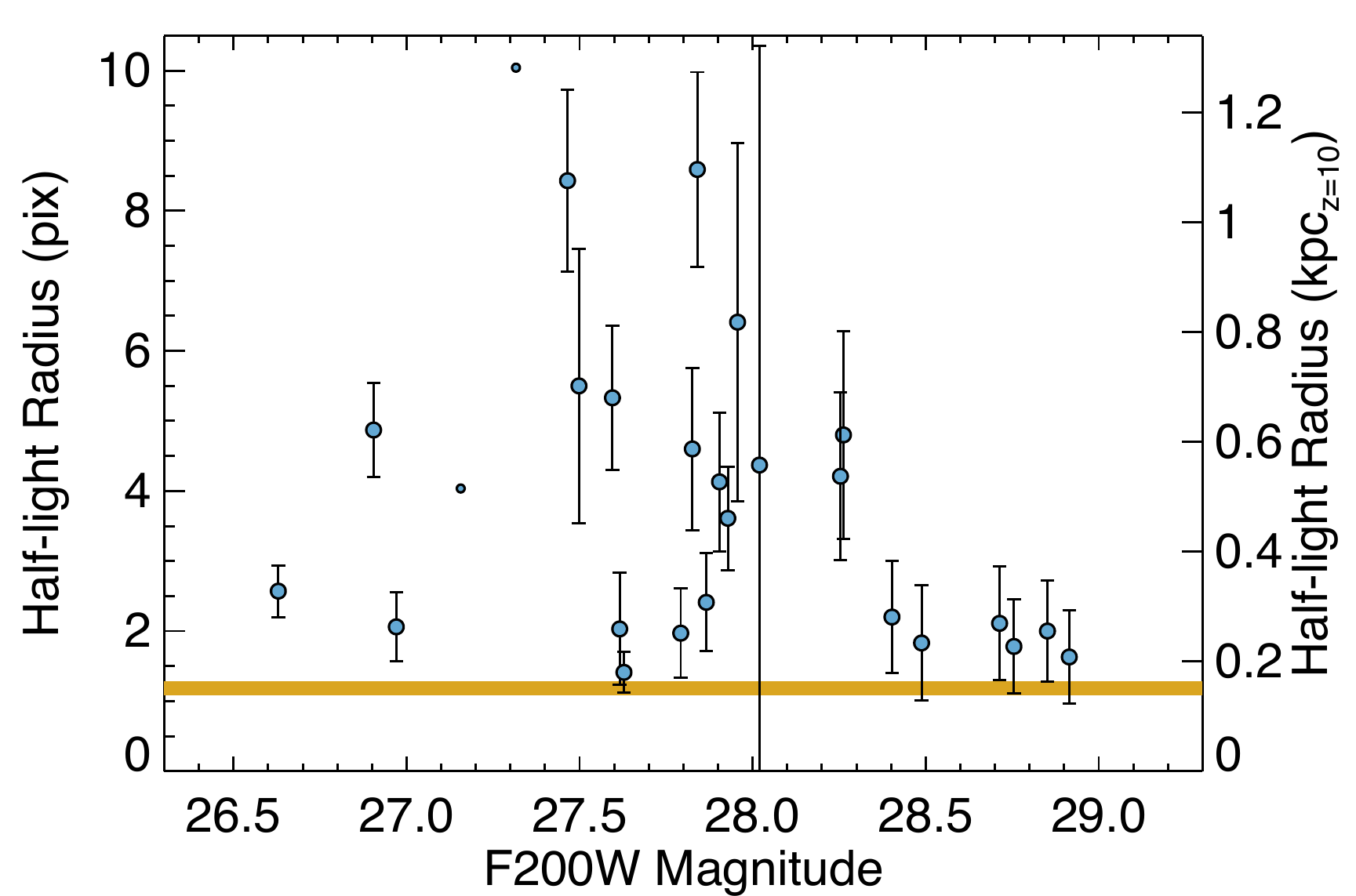}
\caption{The F200W half-light radii of our high-redshift galaxy candidates, measured with \textsc{Galfit}.  The shaded region denotes the half-light radius for the F200W PSF, of 1.18 $\pm$ 0.01 pixels (our pixel scale is 30 mas).  While the galaxies are compact, all but the faintest appear to be resolved.  Our candidate galaxies have a median half-light radius of 0.46 kpc.  The small circles denote the two objects where \textsc{Galfit} did not converge, thus we show the \texttt{\textsc{SE}} half-light radius.} 
\label{fig:rh}
\end{figure}

In Figure~\ref{fig:rh} we show these measured sizes, highlighting that the majority of the sample is significantly spatially resolved.  The measured r$_{h}$ values range from 0.41 to 8.59 pixels, with a median value of 3.6 pixels (0.11\arcs or 0.46 kpc).  These sizes are consistent with the rest-UV sizes found in the GLASS survey by \citet{yang22}, who found a median half-light radius of 0.45 $\pm$ 0.13 kpc for galaxies at 7 $< z <$ 15.
 
\subsection{Stellar Contamination Screening}

It is important to analyze whether any objects in our sample could potentially be low-mass stars or brown-dwarfs, as they can have similar colors as high-redshift galaxies when observed in broadband filters
\citep[e.g.,][]{Yan03,Ryan2005,Caballero2008,Wilkins2014}.  We first analyze the \texttt{\textsc{galfit}}-produced half-light radii of our candidates, as any resolved objects cannot be stellar in origin.  Figure~\ref{fig:rh} shows the \texttt{\textsc{Galfit}}-measured radii for the 24/26 of our candidate high-redshift galaxies where \texttt{\textsc{Galfit}} converged (showing the \texttt{\textsc{SE}} values for the remaining two).  We find that the majority of the sample is obviously resolved, with only five of the faintest sources having sizes within the 1$\sigma$ error bars that allow for a point-source.

For these five unresolved sources, we then follow the methodology of \citet{finkelstein22,finkelstein22c} to explore whether the colors of any of the few unresolved galaxy candidates could be consistent with stars.  We derive a grid of models for the
colors of low-mass stars and brown dwarfs (spectral types of M4--T8)
in the NIRCam filters, by integrating the IRTF SpEX brown dwarf
templates \citep{burgasser14}.  As these spectra end at 2.5$\mu$m, we
use the tabulated 2MASS photometry to link each SpeX model with {\it
  Spitzer}/IRAC photometry from \citet{patten06}.  Following \citet{finkelstein22c} we assume we can map IRAC
3.6$\mu$m onto F356W and 4.5$\mu$m onto F444W, though future spectroscopic observations of
brown dwarfs with {\it JWST} at $\lambda\gtrsim2.5~\mu$m will improve this methodology.  To explore which model is preferred, we use the Bayesian Information Criterion, which includes the goodness-of-fit ($\chi^2$), the number of photometric constraints (five for stars, 12 for galaxies [due to the use of {\it HST} for the latter]), and the number of free parameters (one for stars, 19 [18 templates fit simultaneously, plus the redshift] for galaxies).  We find that two of these five sources have BIC values which indicate a preference for the best-fitting stellar model. 

Although these two objects (CEERS2\_2274 and CEERS2\_1075) are formally unresolved and have optical/near-infrared colors that consistent with brown dwarfs \citep{patten06}, we nevertheless doubt this interpretation as they are very faint (m$_{F277W} =$ 29.1 mag).  The colors imply a very late spectral type of $\gtrsim$T5, which corresponds to an absolute magnitude of $K$=16.1 AB mag \citep{patten06}.  Such a brown dwarf would therefore be at a heliocentric distance of 3.6 kpc, or 3.1 kpc above the Galactic plane, which would place this object approximately nine scale heights out of the thin disk \citep[e.g.,][]{ryan11,holwerda14}.  These objects are therefore highly unlikely to be part of the Population I thin disk stars, but could be a member of the thick disk or Galactic halo \citep{ryan16}.  At present, there are very few constraints on the density of brown dwarfs in these more distant Galactic components at these extremely low effective temperatures, but based on more massive main sequence stars, these are expected to have many orders-of-magnitude fewer stars than the thin disk \citep[e.g.,][]{juric08}.  A rigorous Bayesian model of the halo brown dwarfs would provide the strongest  statistical evidence and prediction for or against this source, but this is beyond the scope of the present work.  We also note that CEERS2\_2274 appears to have a very nearby neighbor with a similar SED, which could indicate it is a merging system at high-redshift. We do note that both sources are below the brightness limit ($m <$ 28.5) we apply when analyzing our sample in \S 7, so their inclusion (or exclusion) does not affect any of our conclusions.

\subsection{Examining Ancillary Multi-wavelength Observations}

We have examined the positions of our 26 candidate $z \gtrsim$ 9 galaxies at both shorter and longer wavelengths, and find no significant detections, increasing confidence in the very high-redshift interpretation.  While the depth of these data are not necessarily sufficient to completely rule out low-redshift solutions, these non-detections do increase our confidence in the fidelity of our sample.  First we searched for X-ray emission coincident with our candidate positions using \emph{Chandra} imaging from the AEGIS-XD survey \citep{nandra15}, which has a flux limit of $1.5\times10^{-16}$ erg cm$^{-2}$ s$^{-1}$ in the 0.5-10 keV band.  There was no emission detected with a Poisson false probability less than $4\times10^{-6}$ in the soft (0.5-2 keV), hard (2-7 keV), full (0.5-7 keV) and ultrahard (4-7 keV) energy bands.

We then investigated possible far-infrared (FIR) emission at the  position of our high redshift galaxy candidates, using the super-deblending catalog technique from \citet{liu18} and \citet{jin18}, which was adapted for the EGS field.  To summarize, this multi-wavelength fitting technique is meant to optimize the number of priors fitted at each band to extract the deepest possible information. We use data from {\it Spitzer} (24$\mu$m from FIDEL; \citealt{dickinson07}), {\it Herschel} 100$\mu$m and 160$\mu$m from PEP \citep{lutz11} and 250$\mu$m, 350$\mu$m, 500$\mu$m from HerMES \citep{oliver12}, JCMT/SCUBA2, including 850$\mu$m from S2CLS \citep{geach17} and 450$\mu$m and 850$\mu$m from \citet{zavala17} and AzTEC 1.1mm from \citet{aretxaga15}. The key is to obtain an adaptive balance as a function of wavelength/dataset between the number of priors fitted, the quality of the fit, and the achievable deblending given the PSF sizes. We start with the deepest images and fit each band from deeper to shallower images.

In general, the high-redshift galaxy candidates are not already contained in our prior-lists, with the exception of two sources, for which all measurements are already contained in the forthcoming catalog (A. Le Bail et al., in preparation). For the other candidates we consider them one at a time and add a specific prior at its position, that we fit together with the rest of priors that are relevant for each band. Extensive Monte-Carlo simulations ensure that the uncertainties associated to the flux measurements are “quasi-Gaussians” \citep[see][A. Le Bail et al. in preparation]{liu18,jin18}.
None of the candidates are significantly detected at any band between 24$\mu$m and 1.1mm.  We do note that for one galaxy, CEERS1\_1730, this method implies a $\sim$ 3$\sigma$ detection at 500 $\mu$m.  However, inspecting the images, it is clear at 160, 250 and 350 $\mu$m there is significant emission from a bright neighbor to the south, which overlaps our source's position at 500$\mu$m due to the worsening PSF.  We conclude it is highly unlikely that this emission is associated with our candidate. 

We do also note that our object CEERS2\_2159, first published as ID$=$93316 by \citet{donnan22}, does have a formal 2.6$\sigma$ detection in the SCUBA-2 850$\mu$m data, as noted by \citet{zavala22}.  However, that marginal 850$\mu$m detection could plausibly be associated with other nearby $z\sim5$ galaxies within the beam as discussed in \citet{zavala22}; higher resolution mm interferometry is being carried out on this system to gain further insight (Fujimoto et al., in prep).

\section{Previously Identified $z \gtrsim$ 9 Candidates in CEERS}

\subsection{Candidates Identified with {\it HST}}

\begin{figure*}[!t]
\epsscale{1.15}
\plotone{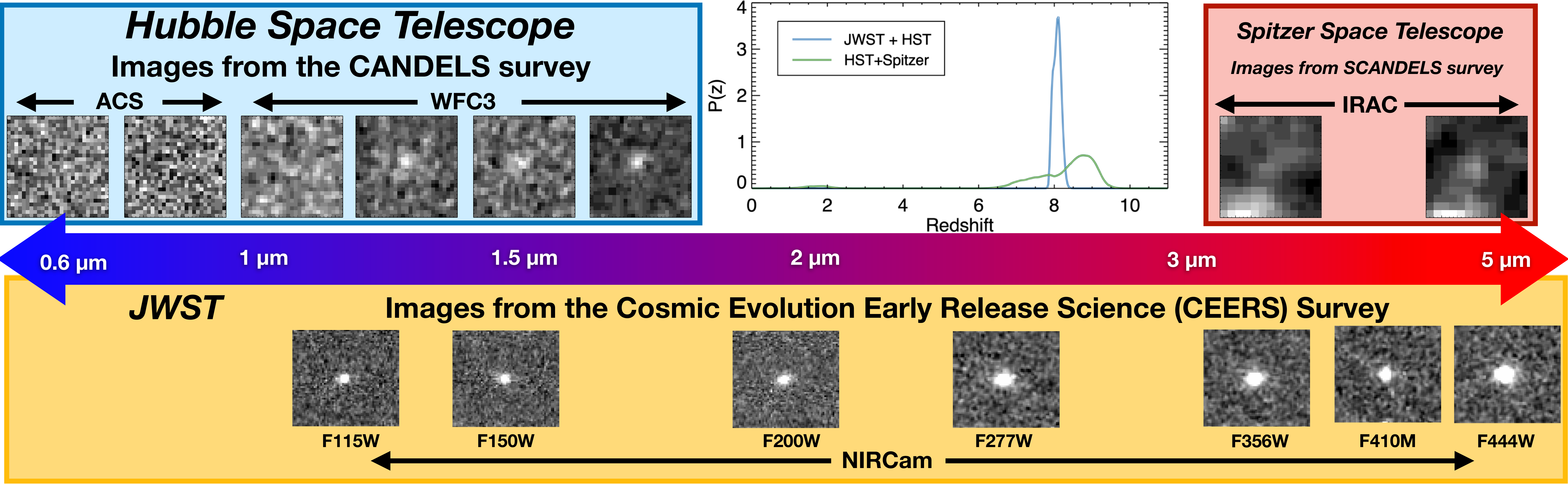}
\caption{The candidate galaxy EGS\_z910\_40898, first published by \citet{finkelstein22} as a $z \sim$ 9 galaxy candidate.  This source was detectable by {\it HST}, albeit in only three filters and at only modest significance.  Even in 50 hr of {\it Spitzer}/IRAC imaging, it was marginally detected.  In just $<$3000 sec with {\it JWST}/CEERS, this source is extremely well-detected in all seven NIRCam filters, highlighting the power of {\it JWST} to probe the very early universe.  In the top-middle panel we compare the previous $\mathcal{P}(z)$ to that now possible with {\it JWST}, finding a much sharper peak, centered at $z \sim$ 8.1 (the F115W$-$F150W color is only marginally red, implying a Ly$\alpha$ break at the blue edge of F115W).}
\label{fig:candelsz9}
\end{figure*}

The area presently covered by CEERS covers two $z \sim$ 9--10 galaxies previously identified by \citet{finkelstein22}, EGS\_z910\_40898 and EGS\_z910\_65860 (none of the candidate $z \gtrsim$ 9 galaxies from \citealt{bouwens19} fall in this first epoch of CEERS NIRCam imaging).  In Figure~\ref{fig:candelsz9}, we show a comparison of the first object in NIRCam versus {\it HST} and {\it Spitzer} imaging.  While the object was only detected at modest significance in the $\sim$few-orbit {\it HST} imaging, and barely at all in the 50 hr depth {\it Spitzer}/IRAC imaging, it is extremely well detected (signal-to-noise ratio $>$30) in $<$\,3000 seconds in all seven NIRCam filters.  

In this figure we also compare constraints on the photometric redshift from the previous {\it HST} $+$ {\it Spitzer} photometry of EGS\_z910\_40898 to the present {\it JWST} $+$ {\it HST} photometry.  While this object was previously selected as a $z \sim$ 9 candidate as the majority of the $\mathcal{P}(z)$ was at $z \gtrsim$ 9, the primary peak extended down to $z <$ 7, and there was a non-negligible secondary peak at $z \sim$ 2.  With the significant improvement in photometric constraints, there is now a single narrow peak at $z \sim$ 8.1.  The best-fitting redshift is modestly lower than the previous value due to the only marginally red F115W$-$F150W color, implying the Ly$\alpha$ break is towards the blue end of the F115W filter.  Observations of this object highlight the capabilities of {\it JWST} to significantly improve constraints on photometric redshifts compared to the pre-{\it JWST} era.

\begin{figure*}[!t]
\epsscale{1.1}
\plotone{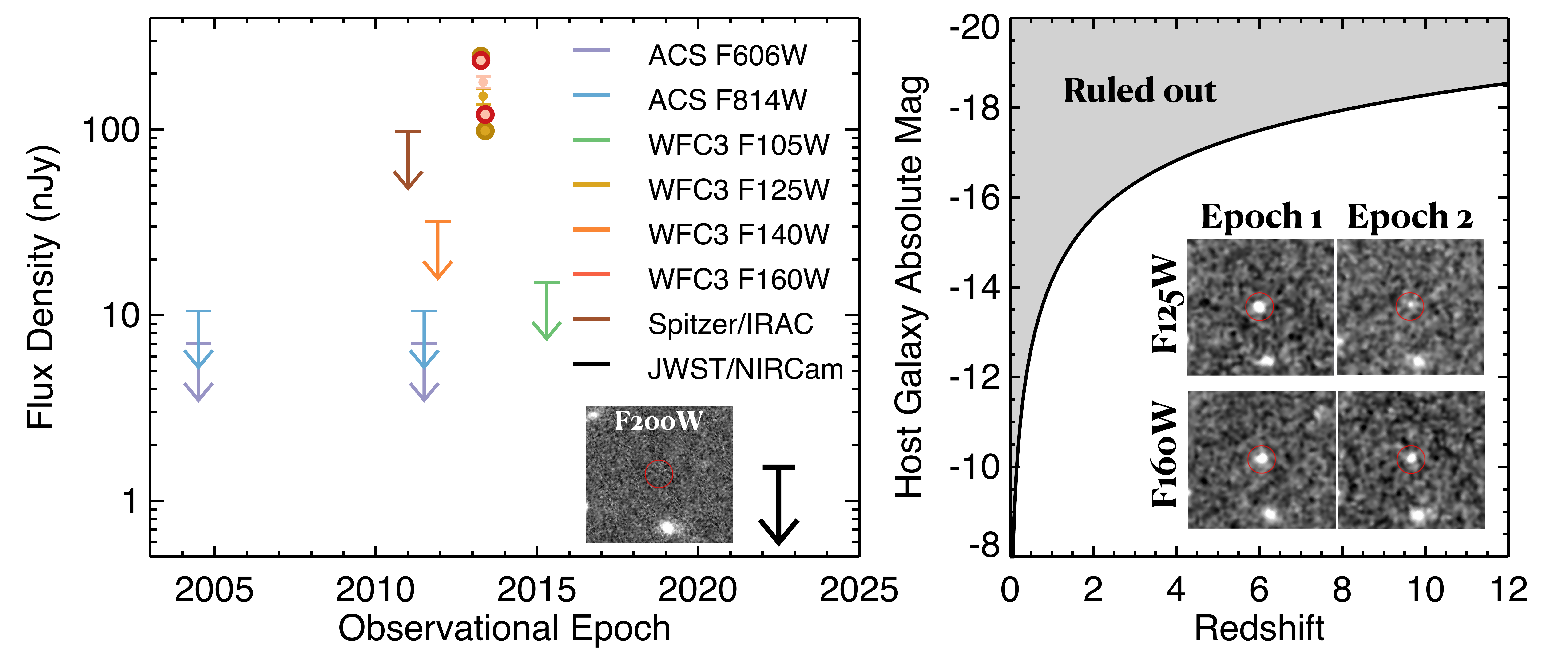}
\vspace{-2mm}
\caption{Left) Flux versus observational epoch for the object EGS\_z910\_68560, first published by \citet{finkelstein22} as a $z \sim$ 9 galaxy candidate.  The colors denote the filter.  The small circles denote the integrated CANDELS F125W and F160W fluxes, while the larger circles denote the fluxes measured in each epoch, 50 days apart.  The object is detected in both epochs in both bands, fading by a factor of $\sim$2.5 and 2.0 in F125W and F160W, respectively, across this time interval.  The object is not detected in any of the CEERS imaging; the inset image shows this position in the CEERS F200W image.  This implies the object is $>$150$\times$ fainter now than it was when it was detected.  Right) The limits on the host galaxy absolute magnitude, taken by applying the cosmological distance modulus to the CEERS upper limit.  With the exception of the extremely low-redshift Universe, a wide range of host galaxy absolute magnitudes are plausible, making it difficult to constrain the redshift of this supernova, though analysis of the light curve implies a likely redshift of $z \sim$ 0.2--1.2.  The inset panels show the images of this transient source in both CANDELS epochs in F125W and F160W.}
\label{fig:sne}
\end{figure*}

The second object, EGS\_z910\_65860, however shows no significant flux at all at the expected position (Figure~\ref{fig:sne}).  This implies that the source identified in the {\it HST} imaging was either spurious, or a transient phenomenon.  \citet{finkelstein22} performed a detailed vetting process to remove all forms of spurious sources, including persistence, thus this seems like an unlikely solution.  We therefore consider whether the observations are consistent with a transient source.  This object showed significant detections in {\it HST} F125W and F160W imaging, with no detections in F606W, F814W, F105W nor F140W.  While 3D-HST F140W pre-imaging was shallower than the CANDELS imaging, it was curious that this object showed no detection as its SED (anchored by F125W and F160W) should have been detectable.

We thus investigate the date of each of these observations.  The F125W and F160W images were taken in 2013, with two images taken $\sim$50 days apart (2 April, and 24 May).  We made updated {\it HST} mosaics (following the procedure in \citealt{koekemoer11}) around the position of this source in each epoch separately, and we do see a clear detection in both bands in both epochs.  This further refutes the hypothesis that this disappearing source was spurious.  Adding to the likelihood of a transient explanation is that the source is \emph{fading} between the two epochs, with the ratio of fluxes from the first to second epoch being 2.5 and 2.0 for F125W and F160W, respectively (fluxes were measured with \texttt{\textsc{SE}} on each epoch separately, using MAG\_AUTO to approximate total fluxes).  The non-detection of this source in F140W is easily explained, in the context of a transient interpretation, by the acquisition date of those images, which was 1.5 years earlier than F125W and F160W (2 Dec 2011).

While the apparent Ly$\alpha$ break between F105W and F125W could imply that this object is a $z \sim$ 9 transient, the F105W imaging was obtained much later, on 1 April, 2015.  Likewise, the F606W images were taken in 2004 and 2011, while the F814W were taken in 2004, 2011, and 2013.  The SCANDELS {\it Spitzer}/IRAC imaging was obtained over many years from 2003--2012, but the bulk was obtained between 2010--2012 \citep{ashby15}.  We therefore do not have the contemporaneous photometry needed for a reliable redshift of this source.  What we do know is that the host galaxy is fainter than the limit of our NIRCam imaging.  Taking the limit of our deepest image in Table 2, we find that the observed flux in the first F160W epoch (23.6 nJy) is $\sim$150 times brighter than the 1$\sigma$ upper limit of our NIRCam imaging.   The compact morphology of this source, along with the lack of apparent proper motion between the two CANDELS epochs leaves a supernova as the most likely explanation (a proper motion of less than the F160W PSF in 50 days indicates a distance of more than one parsec, ruling out a Solar System object).  We show the observed fluxes/limits versus time in the left panel of Figure~\ref{fig:sne}, while in the right panel we show constraints on host galaxy absolute magnitude.  

While nearly any redshift is plausible from a luminosity standpoint (from a low-redshift dwarf galaxy to a very high-redshift $\sim$L$^{\ast}$ galaxy), the significant fading over a 50 day time period argues against a very high redshift due to time dilation of the SNe decay curve.  From the observed (sparse) light curve, we can only place loose constraints on the nature of the transient, following the methods of \citet{2014AJ....148...13R}. The data favor classification as a core-collapse supernova with a redshift ranging from $z \simeq$ 0.2–1.2, with maximum likelihood at $z \simeq 1.1$ and a slight preference for SN Ib/c over SN II. Nevertheless, the data are also compatible with a SN Ia at $z \simeq 1.45$.

\subsection{Candidates Identified with CEERS}

At the time of this writing, there have been five previously submitted articles which have identified $z \gtrsim$ 9 candidates from these CEERS imaging data \citep{donnan22,finkelstein22c,harikane22,labbe22,whitler22}.  In this section we compare our current analysis to these previous results, noting that differences in selection techniques may bias the redshift ranges that a particular study is sensitive to.  We also emphasize that while the selection of a candidate by more than one study does increase the confidence in the object's fidelity, the presence or lack of a galaxy in a given sample is often easily attributable to differences in data reduction, photometric methodology, and selection criteria.

First, \citet{finkelstein22c} published a paper on a single galaxy, ``Maisie's Galaxy", which they identified as a $z \sim$ 12 galaxy candidate.  While they used the same full CEERS imaging dataset as we use here (albeit with an earlier version of the data reduction), they employed highly conservative selection criteria to identify a single extremely robust galaxy candidate due to the nascent state of the data reduction pipeline and photometric calibration c.\ July 2022.  This object is in our galaxy sample, here known as CEERS2\_5429.  Our photometric redshift constraints for this source are $z_{phot} =$ 11.5$^{+0.2}_{-0.6}$.  This is consistent at the $\sim$1$\sigma$ level with the value of $z_{phot} =$ 11.8$^{+0.3}_{-0.2}$ from \citet{finkelstein22c}, with the small differences in the redshift attributable to small changes in photometry due to the updated photometric calibration now available.

\begin{figure}[!t]
\epsscale{1.15}
\plotone{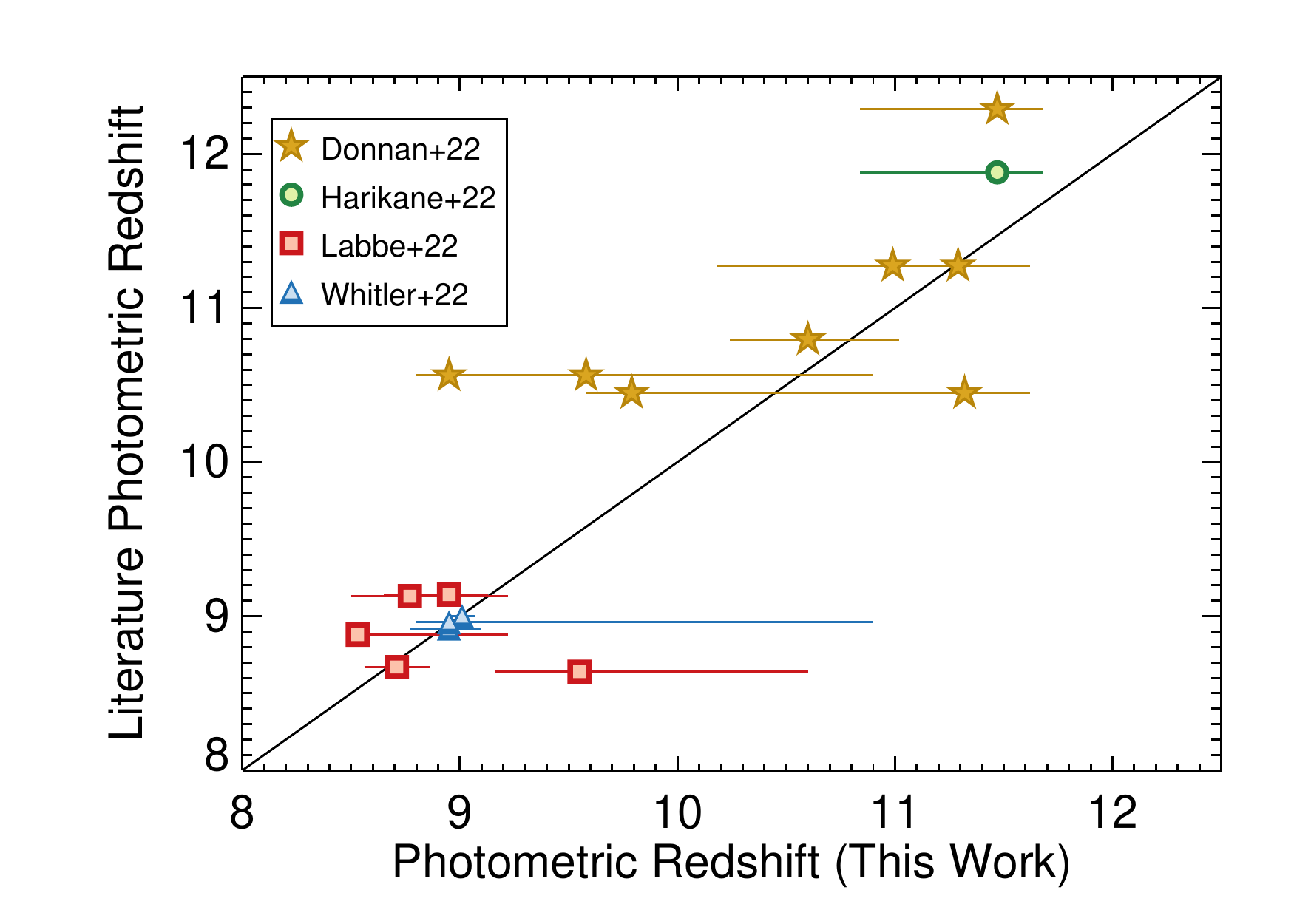}
\caption{A comparison of our photometric redshifts to those for objects in our sample which were previously published as $z >$ 8.5 galaxy candidates from CEERS data in the literature, showing good agreement (not shown is the $z \sim$ 16 candidate, where the agreement is good between the three studies who have published it).  As discussed in \S 6.2, for literature sources not present in our sample, we find in our catalog that they typically miss our detection significance and/or $\Delta \chi^2$ criteria, though often the $\mathcal{P}(z>10)$ is non-zero.}
\label{fig:zcomp}
\end{figure}

The largest previous sample of high-redshift galaxies in the CEERS field comes from \citet[][hereafter D22]{donnan22}, who selected 19 high-redshift galaxies over the CEERS field.  They required dropouts in the F115W (or redder) bands to be selected, biasing the sample to $z >$ 9.5 (where Ly$\alpha$ redshifts out of F115W).  Here we compare to their revised sample (c. Oct 2022).  We find nine galaxies in common between our two samples.  Of the 17 candidates in our sample but not in D22, 14 are at $z \lesssim$ 9.5, thus their absence from the D22 sample is expected via their requirement of no significant flux in F115W.  The remaining three candidates in our sample and not in D22, CEERS1\_1730, CEERS1\_7227 and CEER6\_4407 have $z_{best} >$ 10, thus in principle could have been selected by D22.  Of the nine sources in common, the median difference in photometric redshifts is 0.3 $\pm$ 0.7, with all redshifts agreeing within $\Delta z <$ 1 (Figure~\ref{fig:zcomp}), with the exception of our source CEERS6\_7641 (D22 ID 30585), which we find has $z_{best} =$ 9.0 compared to their value of $z_{best} =$ 10.6.  However, as shown in Figure~\ref{fig:z11sed}, the $\mathcal{P}(z)$ for this source is quite broad, and contains a second peak at $z \sim$ 10.4, thus our results are fully consistent with those of D22.  

Of note is that our sample contains D22 object 93316, as object CEERS2\_2159, and our photometric redshift is nearly precisely equal to the D22 value of $z =$ 16.4.  This object is remarkably bright ($m_{F277W} =$ 26.5), and some evidence is present both from sub-mm imaging and environmental studies that indicate $z \sim$ 5 \citep{zavala22,naidu22b}. The {\it JWST} photometry alone very strongly prefer this ultra-high redshift.  Followup spectroscopy and millimeter interferometry will soon reveal the true nature of this potentially record-breaking system.

We next explore the properties of the 10 galaxies selected by D22 that are not in our final high-redshift galaxy sample.  Of these 10, we find that five are faint enough that they did not meet our source detection significance criteria.  Some of these are faint enough in our catalog that their photometric redshifts are not well constrained, while others do show peaks at $z >$ 10.  Four more sources do show primary peaks at $z >$ 10, but have secondary peaks at $z <$ 4 that are large enough for our selection criteria to remove these sources (these D22 IDs 1434\_2, 26409\_4, 5628\_2 and 6647 all have $\Delta \chi^2 <$ 4 in our catalog).  For only one source, D22 ID 61486, do we find a strong low-redshift solution.  This object exhibits a fairly flat SED in our catalog, though it does exhibit a red F115W$-$F150W color, which could be indicative of a true $z >$ 9 Ly$\alpha$ break.  Overall, we find strong consistency between objects in common in both our sample and that of D22, and for the D22 sources not in our sample, a high-redshift nature is possible given our photometry.

We next compare to \citet{harikane22}.  At the time of this writing only the original sample, pre-photometric calibration update, was available for comparison.  In the CEERS field, they selected galaxies as F150W or F200W dropouts, restricting their sample to $z \gtrsim$ 12, selecting six high-redshift candidates.  Of these six, two are in common with our sample: CEERS2\_5429 (``Maisie's Galaxy''; Harikane ID CR2-z12-1) and CEERS2\_2159 (D22's 93316; Harikane ID CR2-z17-1), with Harikane et al.\ having $z_{best} =$ 11.88 and 16.45 for these two objects, respectively.  Of the four sources in their sample we do not recover, for CR2-z12-3, CR3-z12-1 and CR6-z12-1, our $\mathcal{P}(z)$ does show a primary peak at $z >$ 10, but the low-redshift peak is significant, with $\Delta \chi^2 <$ 4 for all three.  All three sources are also faint, and do not meet our detection significance criteria.  For their source CR2-z12-1, we find $z <$ 6, driven by a $\sim$6$\sigma$ detection in F115W, and a blue F115W$-$F150W color.

\citet{labbe22} selected candidate massive galaxies at 7 $< z <$ 11 as those with detectable Ly$\alpha$ and Balmer breaks in their photometry.  For this comparison, we use an updated sample made available following improvements in photometric calibration (I.\ Labbe, private communication).  Of their 13 sources, they report photometric redshifts of $>$ 8.5 for five of them.  Promisingly, all five of these sources are in our sample, with our (Labbe et al.) IDs of CEERS1\_3910 (39575), CEERS2\_1298 (21834), CEERS2\_2402 (16624), CEERS2\_7534 (14924) and CEERS3\_1748 (35300), with best-fitting photometric redshifts agreeing to $\Delta z <$1 (and $\lesssim$0.3 for 4/5).

Finally, \citet{whitler22} studied the stellar populations of bright galaxies at high-redshift in CEERS.  We compare to an updated sample, again following improved photometric calibration (L.\ Whitler, private communication).  In this sample, they have six galaxies with $z_{best} >$ 8.5.  Three of these sources are in our sample (CEERS1\_3858, CEERS1\_6059 and CEERS6\_7641, with Whilter et al.\ IDs of 37135, 37400 and 9711, respectively).  The photometric redshifts for these three agree extremely well ($\Delta z <$ 0.05) with our estimates.  Of the three sources we do not recover, we do find a strong $z \sim$ 10 peak for their ID 7860; this source narrowly misses our sample with $\Delta \chi^2 =$ 3.7.  ID 14506 does show a red F115W$-$F150W color, but our analysis prefers a 4000 \AA\ break rather than a Ly$\alpha$ break, though a very small peak is present at $z \sim$ 11.  For ID 34362, a $z \sim$ 9 peak is dominant in our measurements, but $\sim$50\% of the integrated $\mathcal{P}(z)$ is contained in a $z \sim$ 2 peak.

In Figure~\ref{fig:zcomp} we compare the photometric redshifts from our study to these previous works for sources in common.  As discussed above, especially considering the mix of photometric procedures and photometric redshift techniques, the agreement is generally good.

As one final comparison, we compare our estimated total magnitudes with the published magnitudes for those candidates in common (where, using the information available, we compare F200W for D22 and \citealt{whitler22}, F356W for \citealt{harikane22}, and F444W for \citealt{labbe22}).  We find the largest discrepancy with D22, where for the nine sources in common, the median magnitude differences between our cataloged total magnitudes and those of D22 is $-$0.3 mag (meaning our fluxes are brighter); however there is significant scatter, with one source being as much as 1 mag brighter in our catalog, and another being 0.4 mag fainter in our catalog.  Further exploring this discrepancy requires a deeper comparison between the procedures adopted to estimate total fluxes.  Comparing to the other studies, we find results more in agreement.  For the four sources in common with \citet{whitler22}, the median magnitude offset is $-$0.1, with individual objects ranging from $-$0.2 to 0.  For the two objects in common with \citet{harikane22}, our magnitudes are \emph{fainter} by 0.1 and 0.3 mag.  Finally for the five sources in common with \citet{labbe22} our magnitudes are all fainter by 0.05--0.2 mag, with a median of 0.15 mag.  We conclude that while there is significant scatter in the photometry due to the various processes used to reduce and analyze the data, we find no evidence that our photometry is significantly systematically shifted compared to other studies, especially by the large amount needed to explain the observed luminosity function excess (\S 7.1).  Clearly improving photometric agreement between different photometric catalogs is a key goal for the near future.

\begin{figure*}[!t]
\epsscale{1.0}
\plotone{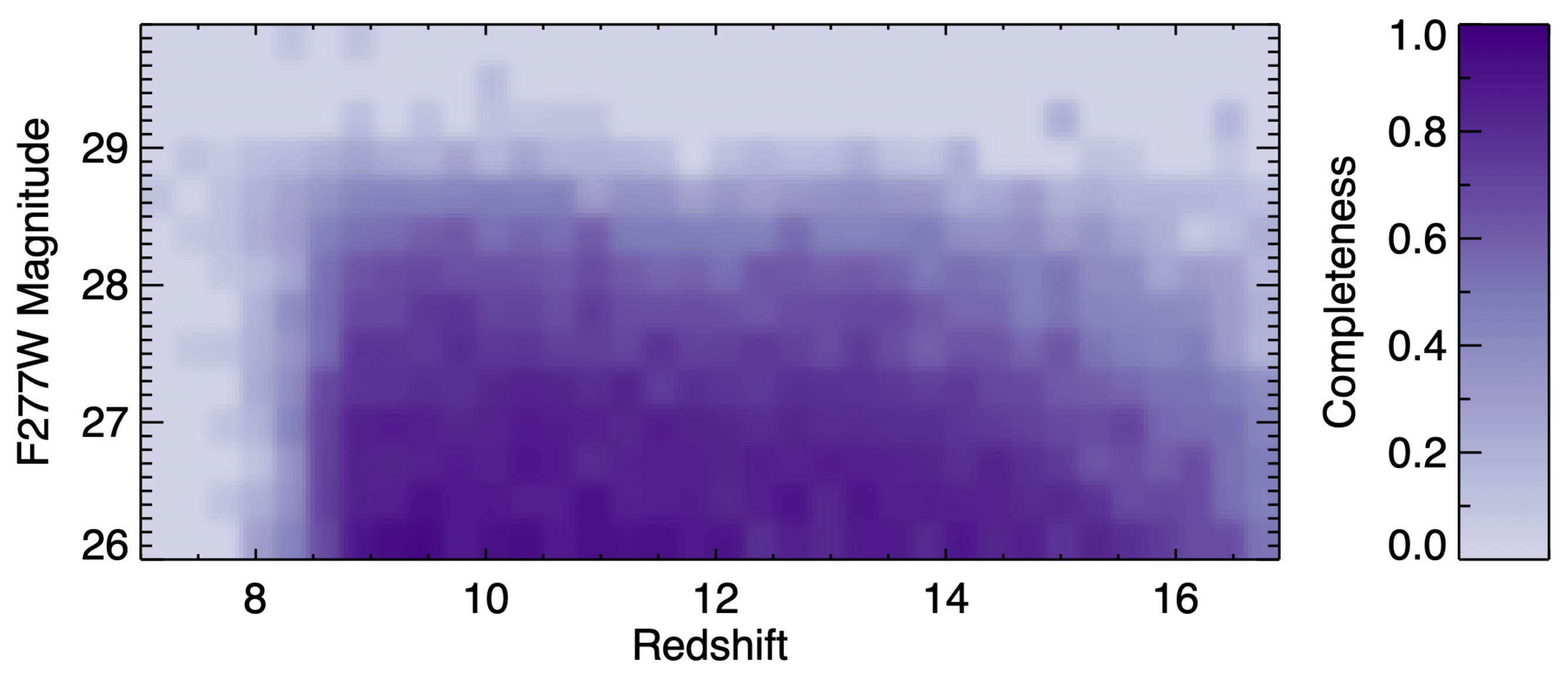}
\caption{The total completeness as a function of input redshift and F277W apparent magnitude.  These values were derived by simulations where we inject compact sources with realistic (modestly blue) SEDs into our imaging, and attempt to recover them with our analysis pipeline.  The shading denotes the fraction of sources in each bin of input redshift and magnitude which were both detected by \texttt{\textsc{SE}} and satisfied our sample selection. The incompleteness at $z <$ 8.5 is expected via our sample selection.  We see that the CEERS imaging allow recovery of galaxies across the entire redshift range of interest to very high completeness at $m_{F277W} <$ 28.0, with the completeness dropping steadily from $m =$ 28 to $m= $ 29.}
\label{fig:completeness}
\end{figure*}

\section{The Evolution of Galaxies in the First 500 Myr}

Here we investigate what constraints we can place on galaxy evolution with our sample of 26 candidate $z \sim$ 8.5--16.5 galaxies.  While it is early in the {\it JWST} mission and the sample size is modest, the fact that our sample contains a large number of galaxies at $z >$ 9 allows us to investigate what constraints are possible.  We first measure the rest-frame UV luminosity function at $z \sim$ 11 in comparison to both previous results as well as expectations from empirical extrapolations.  We then compare the cumulative number of galaxies to predictions from a suite of theoretical models, to explore the accuracy with which these models predicted galaxy abundances in this early epoch.

\subsection{The $z \sim$ 11 Rest-UV Luminosity Function}

The UV luminosity distribution function is one of the key observational diagnostics of galaxy evolution in the early universe.  With each technological leap leading to a new redshift era being observable, this quantity is always of immense interest \citep[e.g.][]{bouwens04,mclure13,finkelstein15} as it is directly comparable to simulation predictions, helping to constrain the physical mechanisms regulating galaxy evolution.  While the extensive amount of deep-field imaging data available from the full Cycle 1 dataset will lead to excellent constraints on this quantity, here we gain a first look by exploring constraints placed by the CEERS data.

For this first-look luminosity function we focus on the specific redshift range of $z \sim$ 9.5--12.  We exclude $z =$ 8.5--9.5 for two reasons.  First, there is a likely overdensity at $z =$ 8.7 \citep[e.g.][]{finkelstein22,larson22} which would bias this quantity high.  Second, at $z >$ 9.5, the Ly$\alpha$ break is fully redward of F115W, providing a true \textit{JWST} dropout sample that is not dependent on the shallower {\it HST} imaging.  We choose the upper bound of $z \sim$ 12 as only two galaxies in our sample have higher photometric redshifts.

To measure the UV luminosity function we require an estimate of the effective volume over which we are sensitive to galaxies.   This is a function of both redshift and source brightness, and accounts for incompleteness due to both photometric measurements and sample selection.  Following \citet{finkelstein22} we estimate the completeness by injecting mock sources of known brightness into our images.  We do this separately for each of the four fields, injecting 10$^3$ sources per iteration, with $\sim$30 iterations per field, to avoid crowding. 

We generate source morphologies with \textsc{galfit} \citep{peng02}.  While the completeness can depend sensitively on source size due to surface brightness dimming, our sources are quite compact (Figure~\ref{fig:rh}), thus we choose a log-normal half-light radius distribution such that the size distribution of the recovered sources matches well that of our galaxy sample (noting that if there were highly extended sources not present in our sample, we would be underestimating the incompleteness). While we choose S\'ersic profiles (with a log-normal distribution peaked at $n =$ 1), our size assumption results in a peak of unresolved sources, with a tail towards modestly resolved sources.  We generate galaxy SEDs assuming a log-normal distribution in magnitude and a flat distribution in redshift, using \citet{bruzual03} templates with stellar population properties tuned to generate fairly blue galaxies (this results in median UV spectral slope ($\beta$) of $-$2.3, with a tail to $-$3 and $-$1).

\begin{figure*}[!t]
\epsscale{1.1}
\plotone{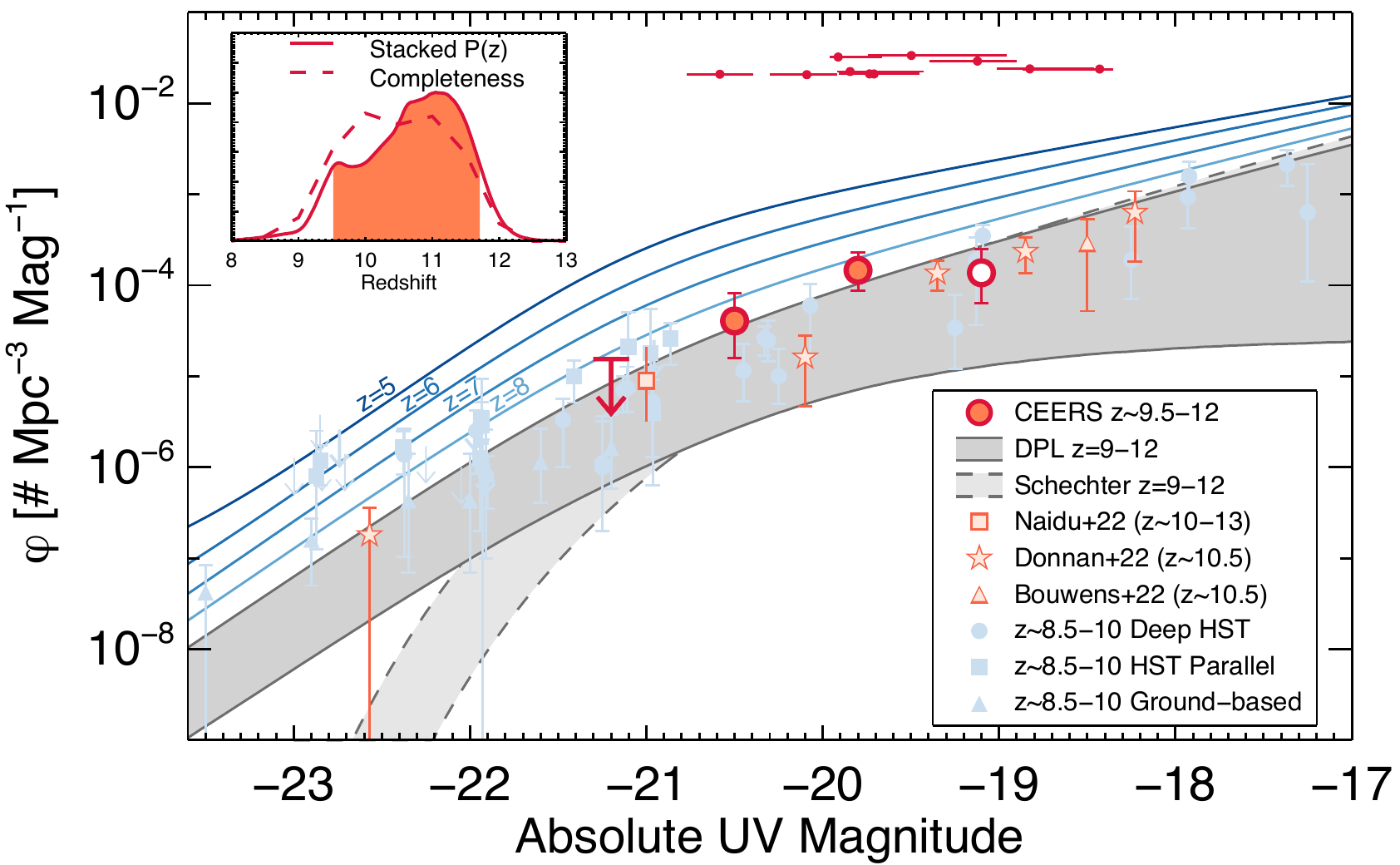}
\caption{The rest-frame UV luminosity function at $z \sim$ 11, shown as the red circles (the open circle denotes our faintest bin, where we are $<$30\% complete).   Each galaxy's magnitude and magnitude uncertainty is denoted by a small circle and line at the top of the figure.  
The light red symbols show literature constraints from {\it JWST} data, from GLASS \citep{naidu22}, CEERS$+$GLASS \citep[][, who also used UltraVISTA]{donnan22}, and the HUDF \citep{bouwens22b}.  The light blue points show a compilation of data from the literature at $z \sim$ 9--10.  Circles denote results from studies which used (modestly) deep imaging from surveys such as CANDELS and the Hubble Frontier fields, including \citet{mcleod15}, \citet{oesch18}, and \citet{bouwens19,bouwens21}.  The squares denote studies making use of {\it HST} pure parallel surveys, including \citet{bernard16}, \citet{morishita18} and \citet{rojasruiz20}.  The triangles denote results from wide-area ground-based studies of \citet{stefanon19} and \citet{bowler20}.  The blue lines show the evolving double power-law (DPL) luminosity functions from \citet{finkelstein22b} at $z =$ 4--8 (this model was fit to data at $z =$ 3--9).  The darker shaded gray region show the predictions from these DPL fits extrapolated to $z =$ 9 (upper bound) -- 12 (lower bound); the lighter gray (outlined with dashed lines) shows a similar extrapolation from the evolving Schechter function fits from \citet[][; this model was fit to data at $z =$ 4--8]{finkelstein16}.  The inset shows the stacked $\mathcal{P}(z)$ of the galaxies used in this luminosity function, as well as the redshift distribution estimated from the completeness simluations at $M_{UV} = -$20.
The observed $z \sim$ 11 luminosity function is consistent with the top end of both smooth extrapolations, implying that the observed smoothly UV luminosity function evolution from $z =$ 4 to $z =$ 9 may be slowing at $z \sim$ 11.}
\label{fig:lf}
\end{figure*}

After sources are added to the images, the images are processed through our entire analysis pipeline in the same way as our real data, measuring photometry with \texttt{\textsc{SE}}, applying all aperture corrections (though we did \emph{not} apply the small zeropoint offsets as those may be due to real instrumental offsets not captured in these simulations), measuring photometric redshifts with  \texttt{\textsc{EAZY}}, and applying our sample selection.  In Figure~\ref{fig:completeness} we show our completeness as a function of input redshift and F277W magnitude.  This highlights that our sample selection accurately begins to select galaxies at $z \gtrsim$ 8.5.  Our completeness remains high to $m \sim$ 28.0, beginning to fall off at $m >$ 28.5, consistent with pre-launch expectations.  

To calculate rest-UV absolute magnitudes for both our real and simulated sources, we follow \citet{finkelstein15} performing a basic round of SED fitting, measuring $M_{1500}$ from the bandpass-averaged flux over a top-hat bandpass spanning 1450--1550 \AA\ in the rest-frame.  We use 10$^3$ Monte Carlo simulations to obtain uncertainties on these magnitudes.  
For the completeness for our luminosity function estimate, we require the simulated sources to have best-fit photometric redshifts spanning 9.5--12, to match those of our galaxy sub-sample.  We then calculate our completeness as a function of absolute magnitude, and measure effective volumes in bins of absolute magnitude by integrating over the co-moving volume element multiplied by the completeness for a given magnitude and redshift, over our current survey area of 35.5 arcmin$^2$.  We list our effective volumes in Table 3.

We measure our luminosity function following the methodology of \citet{finkelstein15}, adopting a bin size of 0.7 magnitudes.  The number density in each bin is estimated via MCMC (see details in \citealt{finkelstein15}), sampling galaxies absolute magnitude posterior distributions, such that galaxies can fractionally span multiple magnitude bins.  We note that at $M_{UV} > -$19.5 our completeness falls below 30\% of its maximum value, thus our faintest bin is shaded in white to indicate that the value is dominated by the completeness correction.

In Figure~\ref{fig:lf} we show our luminosity function results.  In the top-left panel, we show the stacked $\mathcal{P}(z)$ of the 10 sources in our sample at 9.5 $< z_{best} <$ 12.0.  The FWHM of this normalized distribution spans $z =$ 9.5--11.7, with a peak at $z \sim$ 11.  The dashed line shows the completeness from our simulations as a function of redshift at the median absolute magnitude of our sample, which probes a broadly similar redshift range, albeit with a flatter distribution.
In this figure we also plot recent $z \sim$ 11 results from \citet{donnan22}, \citet{naidu22} and \citet{bouwens22}.  While numbers in all studies are fairly small, the agreement between all studies is encouraging.  We also compare to a wealth of studies in the literature from {\it HST} at $z =$ 8.5--11.  

Broadly speaking, our results at $z \sim$ 11 do not show significant evolution from these modestly lower redshifts.  This is consistent with the conclusion from \citet{bowler20} who noted that the bright-end of the UV luminosity function ($M_{UV} = -$22 to $-$23) shows little evolution from $z =$ 8--10.  However, here we are beginning to probe fainter, yet we still see little evolution.

Finally, we comment on both the shape of the UV luminosity function, and the overall evolution.  In Figure~\ref{fig:lf} we plot two empirical extrapolations.  The first is a Schechter function from \citet{finkelstein16}, who measured the evolution of Schechter function parameters as a linear function of (1$+z$) from $z =$ 4--8; we plot this function evolved to $z =$ 9 and 12, shaded by the light gray color.  The darker gray shading is a similar empirical evolution, this time using data from $z =$ 3--9, and assuming a double power-law (DPL) form, from \citet{finkelstein22b}.  Over the magnitude range of our observed sources, these two functions agree, and our observations are consistent with, albeit at the high end of, these empirical extrapolations, implying that pur observed $z \sim$ 11 UV luminosity function is similar to the $z =$ 9 DPL luminosity functional fit from \citet{finkelstein22b}.

These results suggest that the evolution of the UV luminosity function, which had been smoothly declining from $z \sim$ 4 to 8, begins to slow by $z \sim$ 11.  The luminosity function decline has been debated in the literature prior to the {\it JWST} era, notably by \citet{oesch18} and \citet{bouwens19}, who found evidence for an accelerated decline in the UV luminosity function at $z >$ 8.  While our small sample cannot conclusively distinguish between these two scenarios, should future luminosity function efforts validate our observed number densities, it provides further evidence that there is significant star formation in our universe at $z >$ 10.  Previous efforts to recover the star formation histories of galaxies detected at $z \sim 9$ to 11 with \textit{HST} and \textit{Spitzer}/IRAC \citep{tacchella22} and analysis of early {\it JWST} luminosity functions from \citet{donnan22} also arrived at a similar conclusion.

\begin{deluxetable}{ccccc}
\vspace{2mm}
\tablecaption{$z \sim$ 11 UV Luminosity Function}
\tablewidth{\textwidth}
\tablehead{\multicolumn{1}{c}{M$_{1500}$} & \multicolumn{1}{c}{Number of} & \multicolumn{1}{c}{Effective} & \multicolumn{1}{c}{$\phi \times$ 10$^{-5}$} & \multicolumn{1}{c}{V$_{eff}$}\\
\multicolumn{1}{c}{(mag)} & \multicolumn{1}{c}{Galaxies} & \multicolumn{1}{c}{Number} & \multicolumn{1}{c}{(Mpc$^{-3}$ mag$^{-1}$)} & \multicolumn{1}{c}{(Mpc$^3$)}}
\startdata
$-$21.2&0&0.0$\pm$0.2&$<$2.8&92200\\
$-$20.5&1&1.4$\pm$0.6&4.0$_{-2.5}^{+4.2}$&74300\\
$-$19.8&6&4.8$\pm$1.0&14.6$_{-5.9}^{+8.2}$&51900\\
$-$19.1$^{a}$&2&2.4$\pm$1.0&13.7$_{-7.4}^{+11.5}$&30600
\enddata
\tablecomments{The number densities are derived via a MCMC method which includes photometric uncertainties, thus galaxies can contribute to the number density in more than one bin.  The effective number column lists the mean and standard deviation of the number of galaxies per bin from these MCMC simulations, while the number column gives the actual value based on the measured magnitude. $^{a}$ Our data is $<$30\% complete in this faintest bin, so we do not consider this value reliable; it is indicated as an open symbol in Figure~\ref{fig:lf}.}
\end{deluxetable} \label{tab:lf}

\begin{figure*}[!t]
\epsscale{1.0}
\plotone{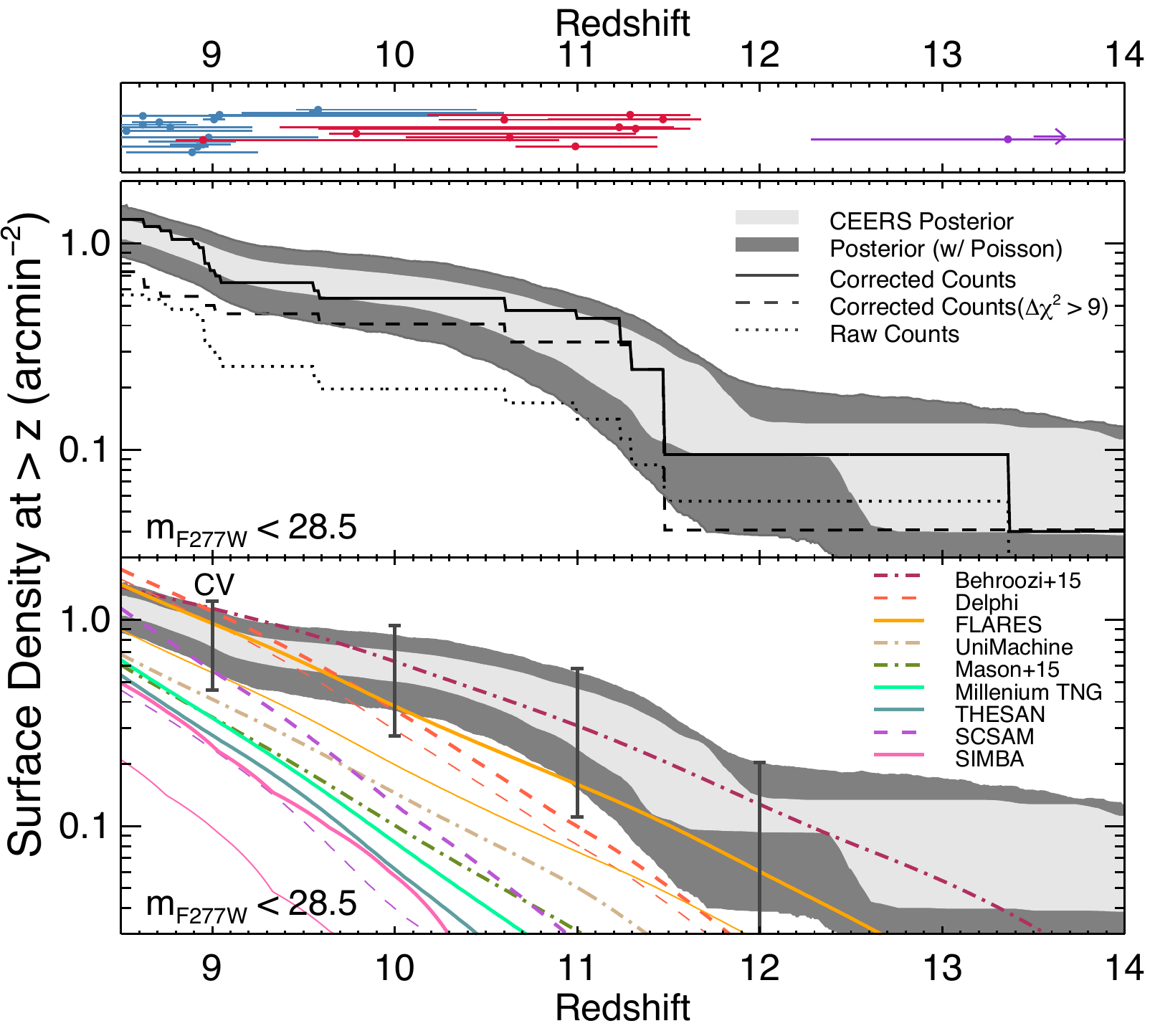}
\caption{The cumulative surface density of sources with $m_{F277W} <$ 28.5 at redshift greater than a given x-axis value, starting at $z \geq$ 8.5.  The top panel shows the redshifts of individual objects (with blue, red and purple denoting the $z \sim$ 9, 11 and $>$12 samples).  In the middle panel the solid line shows the observed surface density, after applying a correction for incompleteness; the dotted line shows the un-corrected (incomplete) values.  The light shaded region shows the posterior on the distribution of the completeness-corrected surface density derived from Monte Carlo simulations marginalizing over the uncertainties in magnitude and photometric redshift; the dark shading includes Poisson uncertainty in this marginalization.  The dashed line shows the completeness-corrected surface density if we had applied a more conservative sample selection criterion of $\Delta \chi^2_{low-z - high-z} >$ 9. In the bottom panel, we show the same shaded regions, now comparing several recent model predictions, shown by the various colored lines (with solid, dot-dashed and dashed denoting predictions from hydrodynamical, semi-empirical, and semi-analytic models, respectively).  For four models we plot thicker lines for predictions with no dust attenuation, and thinner lines for attenuated predictions.  In this panel we also show an estimate of the cosmic variance uncertainty using the method from \citet{bhowmick20}, though this is likely an upper limit on this quantity.  Even including all sources of uncertainty, our observed surface densities are higher than nearly all predictions, with the exception of the \citet{behroozi15} semi-empirical model. This apparent excess of high-redshift galaxies holds true at all $z =$ 8.5--14, regardless of the $\Delta \chi^2$ cut, and is even true of the raw, non-corrected counts at $z >$ 10.5.  These results strongly imply that these predictions lack the full complement of physics describing star formation in the early universe, which we discuss in \S 8.2.}
\label{fig:sdens}
\end{figure*}

\subsection{The Cumulative Surface Density of Galaxies at $z >$ 8.5}

As another view into the evolution of galaxies in the first 500 Myr, in Figure~\ref{fig:sdens} we plot the cumulative surface density of sources in our sample.  This plot shows the integrated surface density for sources at redshift greater than $z$, limiting to $m_{F277W} <$ 28.5 to avoid fainter luminosities where we are highly incomplete.  We correct for incompleteness by counting each galaxy as one divided by the estimated completeness at the redshift and magnitude of a given galaxy.   The solid black line shows the completeness corrected value from our sample, while the dotted line shows the results with no correction.  We note that across the redshift range considered here, the completeness correction is typically a factor of $\sim$2, which is typical of most analyses of modestly faint galaxies.  This, of course, means that the accuracy of completeness corrections remains an important systematic, and one we will explore in more detail in future works.  However, we note that with the exception of the lowest redshifts studied here, even the uncorrected counts exceed most of the model predictions, and therefore this is a reasonable lower limit on the true surface density.

To encompass the uncertainty in both flux and photometric redshift, we run Monte Carlo simulations, sampling both the F277W flux and photometric redshift posterior distributions, and plotting the 68\% confidence range on the surface density as the 68\% spread in values from these 10$^3$ simulations as the light gray shaded region. We also run a set of Monte Carlo simulations additionally sampling the Poisson uncertainty, shown by the wider dark gray region.  We find that at $z >$ 8.5 and $m_{F277W} <$ 28.5, our results suggest $>$1 galaxy per arcmin$^2$.  This trends downward when integrating from higher redshifts, albeit slowly, with this quantity not dropping to 0.1 arcmin$^{-2}$ until $z >$ 11.5--12.  We note that these surface densities are only slightly reduced when removing sources with 4 $< \Delta \chi^2 <$ 9 (dashed line).

We also show the cosmic variance calculated based on the \textsc{bluetides} simulation \citep{bhowmick20}. While these uncertainties appear comparable to the combined redshift, photometric and Poisson uncertainties, these cosmic variance uncertainties are likely an upper limit.  As they are simulation based, they rely on the predicted abundance of galaxies, and as we show here, most simulations under-predict the abundance of $z >$ 9 galaxies.  As the relation between UV luminosity and halo mass in this epoch is very uncertain (\S 8.2), this leads to uncertainty in these calculations. We show this upper limit on the true cosmic variance uncertainty to highlight that it implies our results are still significantly above the predictions.

In the bottom panel of this plot, we also compare to nine recent model-based predictions, including from the Santa Cruz \citep{yung19a, yung20b} and \textsc{delphi} \citep{dayal17} semi-analytic models (SAMs), empirical models by \citet{behroozi15}, \citet{mason15}, and \textsc{UniverseMachine} \citep{behroozi19, behroozi20}, and cosmological hydrodynamic simulations FLARES \citep{wilkins22a, wilkins22b}, THESAN \citep{kannan22}, MillenniumTNG \citep{kannan22a}, and {\sc Simba} \citep{dave19}.  Interestingly our results are noticeably higher than most predictions from physics based models.  At $z <$ 9.5, our observations are consistent with both the \citet{behroozi15} and \textsc{delphi} models (and the FLARES predictions with no dust attenuation), while at $z >$ 10 our results lie significantly above all predictions with the exception of \citet{behroozi15}. We include results with and without dust attenuation for the Santa Cruz SAM, \textsc{delphi}, FLARES, and \textsc{Simba}. We note that while there is likely an overdensity at $z \sim$ 8.7 in the observed field, this has no impact on the surface density measured at $z >$ 9, and the excess of our observed galaxy numbers over nearly all theoretical predictions persists at all redshifts.  We discuss possible explanations for this discrepancy in \S 8.2.

It is important to note that the compilation of theory results included in this comparison is made utilizing several different modeling approaches and with some different modeling assumptions (see \citet{somerville15a} for a thorough review). 
For instance, cosmological hydrodynamic simulations (e.g. MilleniumTNG, THESAN, {\sc Simba}) are carried out by solving the equations of hydrodynamics, thermodynamics, and gravity for large numbers (typically billions) of dark matter, gas, and star particles. These simulations are capable of self-consistently tracking various properties (e.g. stellar and gas mass) and morphology, as well as their correlation with the large-scale environment. 
Cosmological zoom-in simulations (e.g. FLARES, based on the EAGLE model) track galaxy evolution at higher resolution by \textit{zooming} into regions of interest in a cosmological simulation and re-simulate at higher mass resolution and in some cases with more fundamental treatments of baryonic processes. 

All numerical cosmological hydrodynamic simulations require the use of ``subgrid'' prescriptions to represent physical processes that operate at physical scales below the resolution limit of these simulations (e.g. star formation, stellar feedback, black hole growth and feedback, etc.).
Hydrodynamic simulations are subject to tension between the simulated volume and resolution. Large volumes are required to capture the rare over-dense regions that host massive galaxies (the ones of interest in this comparison) and high mass resolution is required to properly resolve the evolution and assembly history of these galaxies. 
Thus it is extremely challenging to simultaneously capture the rare peaks where high redshift galaxies form, while simultaneously resolving these low-mass halos with sufficient numbers of particles. 

On the other hand, the semi-analytic modeling technique uses phenomenological recipes to track the formation and evolution of galaxies in dark matter halo merger trees and is able to predict a wide variety of physical and observable properties of galaxies with a relatively low computational cost. 
These models are typically capable of simulating galaxies over a wider mass range and are less susceptible to the volume-resolution tension faced by hydrodynamic simulations (though even obtaining adequate dynamic range for dark matter only simulations and halo merger trees is challenging; see Section \ref{sec:theory_implication} for a more detailed discussion).
For both the semi-analytic models and the hydrodynamic simulations described above, the predictive power of these models relies on the assumption that the physical recipes within, often derived from or calibrated to nearby observations, accurately represent the processes that drive galaxy formation in the high-redshift universe.

A different class of models are (semi-)empirical methods, also known as subhalo abundance matching (SHAM) or halo occupation distribution (HOD), which efficiently map the properties of a large ensemble of galaxies onto the properties of dark matter halos guided by a set of observed quantities and scaling relations (see \citealt{wechsler18} for detailed review). This approach is independent of specific galaxy formation models and is guaranteed to match the observational constrains to which these models are calibrated. However, extrapolating these models to redshifts different from those where they were calibrated has less physical grounding.

This is only a broad overview of the simulation methods that produced the results presented in Fig.~\ref{fig:sdens}, with the aim of emphasizing that these results are produced with different methods, each having their own advantages and disadvantages. 
We refer the reader to these works for a full description of the design of these simulations and their performance against observational constraints.

\section{Discussion}

\subsection{Observational Effects}
\label{sec:obs_implications}

In the previous subsections we showed that the evolution of the UV luminosity function appears to be ``slowing" at $z >$ 10, and that the abundance of $z >$ 9 galaxies significantly exceeds predictions by most physically-based theoretical models.  Here we explore several possible reasons for these unexpected results.  At this early time with {\it JWST}, we must first acknowledge that the purity of the sample is uncertain.  Should the majority of our galaxy sample turn out to be low-redshift interlopers, it would put our results more in line with the theoretical predictions.  While this is unlikely given our detailed photometry and selection process, these strong claims require strong evidence, thus spectroscopic confirmation is a must.  Without this, it remains possible that heretofore unexpected contaminants could be dominating our sample.  
The CEERS spectroscopic program will observe a number of these sources, and thus we may soon be able to gain further confidence in our galaxy sample.  We do note that non-detections in ALMA dust-continuum followup of three $z \gtrsim$ 11 sources in other early {\it JWST} fields strengthens the high-redshift solutions in all four cases \citep[e.g.][]{fujimoto22,bakx22,yoon22}.  

One other observational systematic effect which could affect these results is that of aperture corrections.  In the above sections we described our multi-step approach to deriving total fluxes, accounting both for what can be detected in the images, and then using simulations to correct for any missing flux from the wings of the PSF.  As one additional check, in \S 6.2 we explored whether our fluxes are significantly systematically brighter (or fainter) than other objects published in the literature, finding that while there was significant scatter, there was no evidence that our fluxes were systematically brighter (especially by the large amount needed to explain the luminosity function excess). Nonetheless, future improvements in photometry can increase confidence in these results.

\subsection{Theoretical Implications}
\label{sec:theory_implication}
Here we speculate on potential physical causes for the observed abundances should future observations validate our results.  One potential explanation would be a complete absence of dust attenuation \citep[e.g.][]{ferrara22,mason22}.  As shown in Figure~\ref{fig:sdens} for those models where we plot the (un) attenuated predictions as thicker (thinner) lines (when available), higher number densities are predicted.  While a full analysis of the colors of these galaxies is beyond the scope of this paper, their SEDs do appear quite blue, and thus it is possible that they have little dust attenuation.  Of note is that bright $z \sim$ 9--10 galaxies exhibit only marginally redder spectral slopes.  For example, \citet{tacchella22} found a median UV spectral slope of $\sim -$2 for $M_{UV} = -$21, implying that a small amount of dust attenuation was present.  Simulations (FLARES, {\sc Simba}) predict a large $\sim 3\times$ reduction in surface densities due to dust, while the {\sc Delphi} SAM predicts much less.  Should a lack of dust attenuation at $z \sim$ 10--12 be responsible for our observed evolution, one may expect to see the luminosity function at higher redshifts continue to decline. 

Most models we have compared to employ some kind of ``Kennicutt-Schmidt" (KS) star-formation law, relating the star-formation rate surface density to the dense gas surface density, assuming a constant efficiency per free fall time.  While models such as the SC SAM \citep{yung19a, yung19b} adopt a KS law that becomes steeper at higher gas surface densities, leading to higher star formation efficiencies at early times, when the typical gas densities are higher, these models still imply a fairly long gas depletion time compared to the age of the Universe at these extreme high redshifts. 
For example, in the local universe, the depletion time for dense (molecular) 
gas is $\sim$1.5 Gyr.  This timescale does appear to be shorter at higher redshift, as short as $\sim$0.7 Gyr at $z =$ 2, and perhaps even shorter at higher redshifts \citep[e.g.][]{sommovigo22}.
The \citet{behroozi15} model would require stars to form at the same rate that gas is funneled into halos, effectively assuming a negligibly short gas depletion time. 
It is also possible that the star formation efficiency could be higher in low metallicity gas, though the opposite trend has been proposed \citep[e.g.][]{krumholz12}, or that stellar driven winds are weaker. Additionally, given the extremely high gas densities at these high redshifts, the cold neutral medium itself may also participate in star formation in addition to the molecular phase typically tracked by some simulations \citep{bialy19}.

Another explanation for the poor match between the predictions and observations in SAMs is the mass and temporal resolution of halo merger trees.  The halo merger and assembly histories, as an important component of semi-analytic models and some empirical models, can either be constructed with the extended Press-Schechter (EPS) formalism \citep[e.g.][]{somerville99b} or extracted from $N$-body cosmological simulations \citep[e.g.][]{behroozi13e}.
However, even the current generation state-of-the-art cosmological simulations only saved a few dozen snapshots at $z>6$ and among which only a handful at $z\gtrsim13$, which is insufficient to properly capture the merger histories of these early-forming halos. Furthermore, the bulk of these high-redshift halos have masses near the resolution limit of these cosmological simulations, which further limit their ability to resolve halo merger trees at these extreme redshifts. 
On the other hand, while EPS-based merger trees have been compared to and shown to be in good statistical agreement with $n$-body merger trees, they are untested in the mass and redshift ranges that are explored here (see \citet{yung20b} for a detailed discussion).

Conversely, the poor resolution of the gas content in cosmological hydrodynamic simulations that does not resolve the multi-phase nature of the ISM  could  contribute to the discrepancy between predictions and observations. It has been repeatedly shown that increasing the resolution typically leads to more dense gas and thus, burstier star formation (without adjusting any free parameters; although this depends on the nature of the sub-grid recipes used for star formation and stellar feedback). Thus, solely increasing the resolution in cosmological hydro simulations, such as in the Lyra cosmological zoom simulation set \citep{gutcke22}, and/or improving the sub-grid recipes for the ISM, star formation and stellar feedback, may partly alleviate the poor match between simulations and these observations.

\begin{figure*}[!t]
\epsscale{1.0}
\plotone{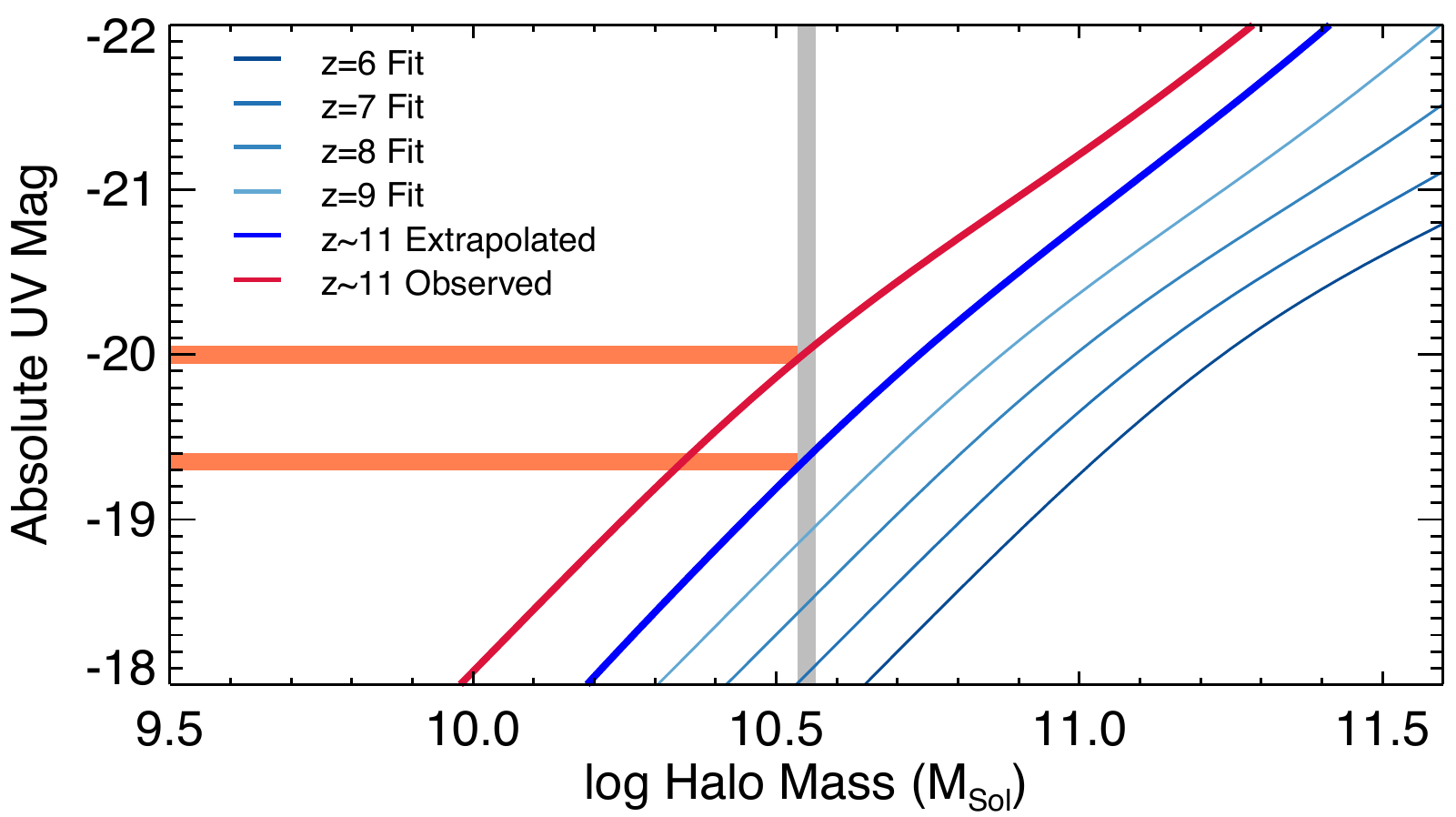}
\vspace{-2mm}
\caption{The relation between rest-UV absolute magnitude and halo mass, obtained via abundance matching the observed UV luminosity functions assuming the DPL fits from \citet{finkelstein22b} for $z =$ 4--8, and extrapolated to $z \sim$ 11 (the thick blue curve is equivalent to the average of the darker gray shaded region in Figure~\ref{fig:lf}).  For our observed $z \sim$ 11 UV luminosity function we use the $z =$ 9 DPL fit from \citet{finkelstein22b} as it is very consistent with our observations (Figure~\ref{fig:lf}).  This relation implies that our observed galaxies at $M_{UV} = -$20 (where our observations place the tightest constraints) have a host halo mass of log(M$_{h}$/M\sol) $=$ 10.55.  At this  halo mass, the expected UV luminosity based on the expected UV luminosity function is 0.6--0.7 mag fainter.  This implies that based on the observed abundances, our observed galaxies are 1.8$\times$ more UV luminous than expected from extrapolation of \textit{HST} results.  While larger sample sizes and spectroscopic confirmations are needed to have greater confidence in the $z \sim$ 11 luminosity function, should these results be confirmed it implies that $z >$ 10 galaxies are more luminous in the rest-UV than expected.  One possible explanation could be an increasing prevalence at higher redshifts of a top-heavy initial mass function, which is predicted to dominate at very low metallicities.}
\label{fig:mh}
\end{figure*}

Finally, and perhaps most interestingly, would be an evolution in the initial mass function (IMF).  It has long been predicted that the first stars would have a top-heavy IMF \citep[e.g.][]{bromm01,clarke03}, not only due to extremely low metallicities \citep[e.g.,][]{chon21,sharda22}, but also the rising cosmic microwave background temperture floor \citep[e.g.,][]{larson98}.  While the relatively massive systems we see here are unlikely to be dominated by metal-free stars, it is possible their metallicities are low enough to affect the characteristic mass of their IMF.  In fact, these observations now probe so close to the Big Bang, that it would be surprising if the IMF did not begin to evolve at some point.

We therefore explore what excess UV luminosity is needed to match our observation.  We first do a simple test by exploring how much we would have to shift our observed luminosity function data in Figure~\ref{fig:lf} in UV magnitude such that they would match the DPL luminosity function fits from lower redshift extrapolated to $z \sim$ 11. We find our data would need to shift $\sim$1 magnitude fainter to match these empirical predictions.  Therefore if an evolving IMF was the \emph{sole} explanation for the higher-than expected luminosities, the UV luminosity would need to be boosted by about a factor of $\sim 2.5$.

As a more complex version of this analysis we estimate the total masses for the host halos of galaxies across the luminosity function via abundance matching.  We follow the procedures of \citet{reddick13}, assuming a 0.2 dex scatter in UV luminosity at fixed halo mass, showing the observed relations between halo mass and UV absolute magnitude in Figure~\ref{fig:mh}; qualitatively similar findings applied for scatter ranging from $0-0.4$ dex.  For $z \leq$ 9, we use the DPL UV luminosity functions from \citet{finkelstein22b}.  For the expected evolution to $z \sim$ 11, we extrapolate their fits to $z \sim$ 12, and show the volume-averaged extrapolated LF for $z=9.5-12$ with the bright blue line.  For our actual observed values, as we have not fit a functional form here due to our limited dynamic range in luminosity, we assume the $z =$ 9 DPL fit from \citet{finkelstein22b} as this is most consistent with our observed number densities in Figure~\ref{fig:lf}. We show in red the M$_{halo}$ -- M$_{UV}$ relation when abundance matching the average of the $z \sim$ 9.5--12 halo mass functions to this assumed observed luminosity function.  In this figure, we highlight $M_{UV} = -$20, which is where our $z \sim$ 11 observations are most constraining.   Our abundance matching analysis predicts that based on our observations, these galaxies are hosted in halos with log(M$_{h}$/M\sol) $=$ 10.55 (upper light-red bar).  If we had instead used the extrapolated $z \sim$ 11 results, these halos should host galaxies with $M_{UV} =$ $-$19.3 to $-$19.4  (lower light-red bar).  This implies that our observed galaxies are $\sim -$0.6--0.7 mag brighter (1.8$\times$ in UV luminosity) than expected for a galaxy in their host halo.

A UV-luminosity boost of $\sim$1.8--2.5 is not unexpected in the context of a top-heavy IMF.  \citet{raiter10} explore the impact of differing IMFs on the UV luminosity (specifically investigating the conversion from UV luminosity to SFR).  They find that for the top-heavy IMFs proposed by \citet{tumlinson06}, which peak at $\sim$10--40 M\sol, yield a UV luminosity $\sim$0.4 dex higher than Salpeter at a stellar metallicity of log (Z/Z\sol) $= -$2.  Such a brightness increase is exactly what we find would be needed to bring our observed luminosity function in line with expectations.  This interpretation was also considered by \citet{harikane22}, who found that some top-heavy IMF models can boost the UV luminosity by $\sim$3--4$\times$ \citep[e.g.][]{zackrisson11}.  The top end of the expected UV luminosity boost was discussed by \citet{pawlik11}, who predicted up to a factor of 10 increase in UV luminosity for a top-heavy IMF in a zero metallicity system.

If the interpretation of invoking a top-heavy IMF were correct, there would be a number
of additional effects and consequences that may be detectable, thus providing an independent
cross-check. One such diagnostic is the predicted boost to nebular line emission, including
the occurrence of strong \ion{He}{2} 1640 \AA\ features \citep[e.g.][]{bromm01,raiter10}, accessible via our upcoming CEERS spectroscopic follow-up. A qualitatively different signature of a top-heavy IMF would be a supernova rate that may be enhanced by up to an order of magnitude, per unit stellar mass, resulting in strong SN-driven galactic winds \citep[e.g.,][]{dayal18,jaacks19}.
 
We note that the high observed galaxy number densities do not represent a fundamental problem for hierarchical structure formation models.  For instance, using a {\sc Simba} simulation with all explicit feedback processes turned off and no extinction results in $\sim 2\times$ higher number densities than observed (even with a standard IMF), and the empirical \citet{behroozi15} model can match the data.  Hence the results do not necessarily challenge the $\Lambda$CDM paradigm, but rather our understanding of the physics of early galaxy evolution within that paradigm.

\section{Summary and Conclusions}
We have presented the results from a study of galaxies at $z >$ 9 in the first epoch of NIRCam imaging from the Cosmic Evolution Early Release Science program.  These imaging cover $\sim$35 arcmin$^2$ with seven photometric filters (six broadband and one medium band) covering $\sim$1--5 $\mu$m, reaching m $\gtrsim$ 29.  In addition to being some of the widest moderate-depth imaging available early in Cycle 1, these data probe a parameter space in wavelength and depth optimal for studying these early redshifts.

Following a detailed reduction of the data (with full details available in \citealt{bagley22b}) we measure photometry from all sources in the field.  Using a combination of the F277W and F356W image as the detection images, we create a 13-band photometric catalog, inclusive of imaging in six {\it HST} filters in this field.  We emphasize the calculation of robust colors, total fluxes, and flux uncertainties, using simulations to validate our choices in cataloging.  We explore any systematic offsets in our photometry by comparing to the best-fitting model templates for $\sim$800 spectroscopically-confirmed sources in our field, finding that the large zeropoint offsets that affected early {\it JWST}/NIRCam studies have been resolved.  We do measure (and apply) offsets from $\sim$1--5\%, which are reflective of potential residual zeropoint corrections in the NIRCam imaging, adjustments to our estimations of the total fluxes, and potential mis-matches between the used template set and the SEDs of the real galaxies.

We estimate photometric redshifts for all sources in our catalog using  \texttt{\textsc{EAZY}}.  In addition to the standard set of 12 FSPS templates, we add six new templates designed to span the blue colors expected in very early galaxies, as well as very strong emission lines expected at low metallicities.  We designed a set of selection criteria to robustly select galaxies at $z >$ 9, balancing our desire to maximize completeness with our need to minimize contamination.  These selection criteria demand both robust detection significance (signal-to-noise ratio $>$ 5.5 in at least two filters, measured in 0.2\arcs\ diameter apertures), as well as photometric redshift probability distribution functions that are strongly constrained to $z >$ 8.5.

Following visual inspection to remove any remaining spurious sources (shown in the Appendix), our selection resulted in a sample of 26 robust candidate galaxies at $z >$ 8.5.  The majority of the sample (15 galaxies) lies at 8.5 $< z <$ 10, while nine galaxies lie at 10 $< z <$ 12.  Two candidates lie at higher redshifts -- the previously published $z \sim$ 16.5 source \citep{donnan22}, and a new candidate at $z \sim$ 13.  While a full analysis of the properties of these galaxies is reserved for future work, their rest-frame UV SEDs are fairly uniformly blue, and the galaxies are compact, with a median half-light radius of 3.6 pixels (0.11\arcs, or 0.46 kpc).  We explore information from multi-wavelength constraints on these galaxies, and find no significant detections in either X-ray or submm/mm wavelengths, strengthening the conclusion that these sources reside at $z >$ 9.  We compare to the few previous searches for $z >$ 9 sources in this field, and find that for sources in common between studies, the photometric redshifts agree quite well.  Of the two known {\it HST} sources previously published in this field, we very robustly constrain the SED of one, placing it at the slightly lower redshift of $z \sim$ 8.1.  The other is completely absent in our NIRCam data, and we conclude it was a likely supernova serendipitously captured by the {\it HST} data taken in 2013.

We estimate our sample completeness as a function of redshift and magnitude using source-injection simulations, and present an early look at the $z \sim$ 11 rest-UV luminosity function.  We find that the abundance of modestly bright ($M_{UV} \sim -$20) galaxies at $z \sim$ 11 does not appear to be evolving from $z \sim$9 -- 11, which is unexpected given the steady decline in the abundance of such galaxies from lower redshifts to $z =$ 8, though such non evolution at the very bright end from $z =$ 8--10 had previously been discussed by \citet{bowler20}.  We then compare the surface density of our sources to a variety of model predictions, finding that even after accounting for several sources of systematic and random uncertainty, the observed abundance of galaxies is in significant excess of these predictions.

We explore several potential explanations for these unexpected results.  While not the most exciting, significant sample contamination cannot be conclusively ruled out.  These data represent our first foray into a new cosmic epoch, and spectroscopic confirmation of the redshifts to at least a subset of these ultra-high-redshift sources are necessary to gain confidence in our sample selection processes.  However, such data will begin to flow soon, with CEERS scheduled to spectroscopically observe $\sim$10 of these sources in late 2022 (though these high redshifts may necessitate longer exposure times for future cycle programs).

Should these high abundances of $z =$ 9--13 galaxies be confirmed, we explore what possible changes in the models could bring their predictions into agreement with observations.  One very exciting possibility is that we are beginning to probe an era where star-formation in galaxies is dominated by a top-heavy IMF due to the presence of very low metallicities, which could increase the ratio of UV luminosity per unit halo mass.  We explore the ``excess" UV luminosity from our observations, both by comparing to the expected UV luminosity function based on extrapolations from lower redshift and through an abundance matching exercise, and find that our UV luminosities may be enhanced by 1.8--2.5$\times$.  This is very similar to the predicted excess UV emission from a low-metallicity stellar population where the IMF peaks at 10--40 M\sol\ \citep{tumlinson06} of a factor of $\sim$2.5$\times$, implying a top-heavy IMF is a plausible physical explanation.  
We also discuss how potential changes to the dust attenuation, star formation law, galactic feedback, and  resolution of numerical simulations could collectively contribute to reconciling our observations with model predictions.

These possibilities are exciting, and while one might expect that at $z >$ 10 we should expect to see changes in star-formation physics such as a top-heavy IMF, our results are just a first glimpse, and the data available in the near future will provide much stronger constraints.  Specifically, the remainder of Cycle 1 will see the creation of high-redshift galaxy samples orders of magnitude larger than we present here from the combination of the full CEERS survey, COSMOS-Web (PIs Casey \& Kartaltepe), PRIMER (PI Dunlop) and NGDEEP (PIs Finkelstein, Papovich \& Pirzkal), along with JADES (PI Rieke).  Additionally, Cycle 1 will also see spectroscopic followup with NIRSpec of NIRCam identified sources from CEERS and JADES, and several Cycle 2 programs will certainly target these sources with deep observations.  While it is early days with {\it JWST}, our first-look CEERS results provide an enthralling glimpse of the potential secrets the early universe has which our observations can unlock.

\facility{HST (ACS, WFC3)}
\facility{JWST (NIRCam)}

\begin{acknowledgements}
We acknowledge that the location where this work took place, the University of Texas at Austin, that sits on indigenous land. The Tonkawa lived in central Texas and the Comanche and Apache moved through this area. We pay our respects to all the American Indian and Indigenous Peoples and communities who have been or have become a part of these lands and territories in Texas, on this piece of Turtle Island. We acknowledge support from NASA through STScI ERS award JWST-ERS-1345. We thank Marcia Rieke, Daniel Schaerer, Volker Bromm and Mike Boylan-Kolchin for helpful conversations.
\end{acknowledgements}

\appendix

\section{Full Photometry}
In Tables~\ref{tab:nrcphot} and \ref{tab:hstphot} we list the measured photometry for our final sample.

\begin{deluxetable*}{cccccccc}
\vspace{2mm}
\tablecaption{NIRCam Photometry for $z >$ 8.5 Galaxy Sample}
\tablewidth{\textwidth}
\tablehead{\multicolumn{1}{c}{ID} & \multicolumn{1}{c}{F115W} & \multicolumn{1}{c}{F150W} & \multicolumn{1}{c}{F200W} & \multicolumn{1}{c}{F277W} & \multicolumn{1}{c}{F356W} & \multicolumn{1}{c}{F410M} & \multicolumn{1}{c}{F444W}}
\startdata
CEERS2\_2159&$-$4.1$\pm$5.9&7.0$\pm$6.7&22.5$\pm$4.9&94.2$\pm$4.6&96.3$\pm$3.7&104.4$\pm$7.3&89.7$\pm$5.4\\
CEERS1\_1730&7.7$\pm$5.3&6.0$\pm$5.7&33.3$\pm$4.4&30.1$\pm$3.4&33.3$\pm$3.3&26.1$\pm$6.8&39.4$\pm$4.6\\
\hline
CEERS2\_588&$-$5.3$\pm$8.3&28.7$\pm$8.5&49.8$\pm$7.2&63.6$\pm$5.1&58.5$\pm$5.1&59.4$\pm$10.8&68.8$\pm$7.3\\
CEERS1\_8817&$-$0.4$\pm$4.7&21.3$\pm$5.3&25.9$\pm$4.3&20.7$\pm$3.0&31.9$\pm$2.9&25.3$\pm$5.9&32.8$\pm$4.6\\
CEERS2\_7929&$-$6.8$\pm$5.1&17.0$\pm$5.8&25.0$\pm$4.2&20.5$\pm$2.9&18.0$\pm$2.9&22.9$\pm$6.2&23.4$\pm$4.6\\
CEERS6\_7641&4.2$\pm$3.8&18.5$\pm$4.2&27.7$\pm$3.8&20.3$\pm$2.7&26.8$\pm$2.8&17.3$\pm$5.0&41.0$\pm$3.6\\
CEERS2\_5429&$-$7.0$\pm$3.2&11.7$\pm$3.7&26.9$\pm$2.7&18.0$\pm$2.2&18.2$\pm$1.9&17.1$\pm$4.0&19.1$\pm$2.8\\
CEERS1\_7227&5.5$\pm$3.7&13.7$\pm$4.1&23.8$\pm$3.3&17.6$\pm$2.3&17.9$\pm$2.0&17.4$\pm$4.8&14.4$\pm$3.0\\
CEERS6\_7603&1.9$\pm$2.7&9.9$\pm$2.9&11.9$\pm$2.3&9.8$\pm$1.6&9.4$\pm$1.6&12.6$\pm$3.4&18.7$\pm$2.3\\
CEERS6\_4407&$-$0.5$\pm$2.3&11.2$\pm$3.0&18.1$\pm$2.8&8.9$\pm$2.0&11.1$\pm$1.9&11.6$\pm$4.8&8.6$\pm$2.5\\
CEERS6\_8056&$-$1.5$\pm$1.8&7.7$\pm$2.2&10.5$\pm$1.8&8.3$\pm$1.5&7.5$\pm$1.4&12.0$\pm$2.5&10.5$\pm$2.3\\
\hline
CEERS2\_2402&25.4$\pm$4.3&72.2$\pm$5.1&62.8$\pm$3.9&76.3$\pm$4.0&89.1$\pm$3.6&121.7$\pm$6.5&205.0$\pm$5.8\\
CEERS1\_6059&21.5$\pm$3.8&87.5$\pm$4.5&80.9$\pm$3.7&58.7$\pm$2.6&58.1$\pm$2.4&49.6$\pm$4.7&61.6$\pm$3.4\\
CEERS1\_1875&10.3$\pm$5.5&38.7$\pm$6.4&37.5$\pm$4.9&54.8$\pm$4.6&70.1$\pm$4.1&64.9$\pm$7.1&109.9$\pm$6.2\\
CEERS1\_3858&21.3$\pm$4.3&66.5$\pm$5.2&59.1$\pm$4.1&46.7$\pm$3.5&54.7$\pm$2.9&52.1$\pm$5.7&50.2$\pm$4.2\\
CEERS2\_7534&12.4$\pm$3.3&31.8$\pm$4.3&32.3$\pm$3.0&40.3$\pm$2.3&57.8$\pm$2.1&107.5$\pm$3.9&163.0$\pm$3.1\\
CEERS1\_3908&$-$1.8$\pm$6.9&47.2$\pm$7.9&43.0$\pm$6.9&44.2$\pm$4.7&54.0$\pm$4.9&41.1$\pm$8.9&73.2$\pm$7.8\\
CEERS6\_4012&9.6$\pm$7.3&30.7$\pm$8.4&36.4$\pm$6.7&32.3$\pm$5.9&31.5$\pm$4.8&27.7$\pm$8.9&29.8$\pm$7.6\\
CEERS2\_2324&$-$9.4$\pm$8.0&45.7$\pm$8.7&32.6$\pm$7.9&32.2$\pm$7.0&28.1$\pm$5.1&24.1$\pm$10.8&18.3$\pm$8.6\\
CEERS1\_3910&$-$8.8$\pm$7.0&30.4$\pm$9.7&26.5$\pm$6.3&23.1$\pm$4.7&49.6$\pm$4.9&71.6$\pm$11.3&76.9$\pm$6.9\\
CEERS1\_5534&14.8$\pm$5.2&35.4$\pm$5.9&24.5$\pm$4.6&25.7$\pm$3.8&37.5$\pm$3.3&25.4$\pm$6.7&66.0$\pm$5.3\\
CEERS1\_4143&5.0$\pm$5.3&26.2$\pm$6.2&18.0$\pm$4.9&20.6$\pm$4.0&16.3$\pm$3.7&7.1$\pm$6.4&16.7$\pm$5.4\\
CEERS3\_1748&3.2$\pm$2.8&13.9$\pm$3.2&11.4$\pm$3.0&14.9$\pm$2.1&34.6$\pm$2.1&62.7$\pm$4.3&78.0$\pm$3.3\\
CEERS2\_1298&6.5$\pm$2.2&11.9$\pm$2.5&15.8$\pm$2.1&13.6$\pm$1.6&30.6$\pm$1.5&41.9$\pm$3.1&68.5$\pm$2.2\\
CEERS2\_2274&5.5$\pm$3.3&9.6$\pm$3.8&14.6$\pm$2.9&8.4$\pm$2.8&17.1$\pm$2.7&11.5$\pm$5.3&22.9$\pm$5.1\\
CEERS2\_1075&5.4$\pm$2.5&12.1$\pm$2.9&9.9$\pm$2.1&8.2$\pm$1.8&9.5$\pm$1.6&8.0$\pm$3.6&21.2$\pm$2.2
\enddata
\tablecomments{All fluxes are in nJy.  The horizontal lines distinguish our $z >$ 12, $z =$ 10--12 and $z =$ 8.5--10 samples.}
\end{deluxetable*} \label{tab:nrcphot}

\begin{deluxetable*}{cccccc}
\vspace{2mm}
\tablecaption{{\it HST} ACS and WFC3 Photometry for $z >$ 8.5 Galaxy Sample}
\tablewidth{\textwidth}
\tablehead{\multicolumn{1}{c}{ID} & \multicolumn{1}{c}{F606W} & \multicolumn{1}{c}{F814W} & \multicolumn{1}{c}{F125W} & \multicolumn{1}{c}{F140W} & \multicolumn{1}{c}{F160W}}
\startdata
CEERS2\_2159&2.7$\pm$8.8&3.9$\pm$10.5&2.3$\pm$14.0&-23.6$\pm$27.6&6.4$\pm$12.7\\
CEERS1\_1730&15.0$\pm$8.7&-8.0$\pm$10.7&-13.6$\pm$12.8&33.1$\pm$22.4&-13.5$\pm$11.1\\
\hline
CEERS2\_588&-5.2$\pm$12.5&-20.2$\pm$18.3&5.8$\pm$21.0&-13.0$\pm$36.6&68.5$\pm$18.2\\
CEERS1\_8817&-2.0$\pm$6.6&-7.4$\pm$6.8&-24.9$\pm$10.3&---&34.6$\pm$10.2\\
CEERS2\_7929&-6.6$\pm$7.2&0.1$\pm$8.0&2.6$\pm$9.2&-7.9$\pm$24.7&15.9$\pm$8.8\\
CEERS6\_7641&2.8$\pm$7.7&1.6$\pm$8.0&10.8$\pm$9.4&28.7$\pm$13.6&23.3$\pm$10.7\\
CEERS2\_5429&8.1$\pm$5.2&1.4$\pm$6.5&-8.1$\pm$7.8&-31.8$\pm$14.0&29.6$\pm$7.1\\
CEERS1\_7227&1.5$\pm$5.8&1.6$\pm$6.5&-6.5$\pm$7.9&---&13.1$\pm$7.4\\
CEERS6\_7603&-2.2$\pm$4.7&-0.6$\pm$4.9&-0.1$\pm$5.1&5.9$\pm$9.8&1.1$\pm$4.6\\
CEERS6\_4407&1.0$\pm$4.7&-1.5$\pm$5.0&0.8$\pm$6.4&---&6.8$\pm$5.9\\
CEERS6\_8056&-0.1$\pm$3.2&-0.5$\pm$3.8&3.7$\pm$4.9&2.3$\pm$9.0&6.4$\pm$3.7\\
\hline
CEERS2\_2402&6.2$\pm$7.1&-6.0$\pm$8.9&50.5$\pm$11.3&70.3$\pm$13.3&65.0$\pm$9.5\\
CEERS1\_6059&-2.5$\pm$5.5&1.6$\pm$5.5&38.3$\pm$9.0&90.5$\pm$16.7&44.2$\pm$7.9\\
CEERS1\_1875&6.5$\pm$9.2&16.9$\pm$13.1&17.2$\pm$14.5&65.0$\pm$24.9&39.0$\pm$13.1\\
CEERS1\_3858&2.5$\pm$7.0&4.6$\pm$8.6&35.8$\pm$10.2&46.9$\pm$18.9&50.7$\pm$8.7\\
CEERS2\_7534&-0.3$\pm$4.8&0.8$\pm$5.1&21.7$\pm$7.0&20.3$\pm$9.1&24.0$\pm$6.1\\
CEERS1\_3908&14.7$\pm$11.5&19.2$\pm$13.9&46.7$\pm$18.7&14.1$\pm$33.7&53.8$\pm$16.8\\
CEERS6\_4012&4.9$\pm$12.6&-4.0$\pm$17.1&29.9$\pm$19.4&144.9$\pm$26.2&71.0$\pm$18.6\\
CEERS2\_2324&5.7$\pm$16.9&-0.2$\pm$16.9&29.8$\pm$20.6&-23.1$\pm$32.6&39.3$\pm$17.2\\
CEERS1\_3910&8.6$\pm$10.9&10.5$\pm$12.5&-2.8$\pm$17.8&10.1$\pm$31.3&24.2$\pm$16.0\\
CEERS1\_5534&10.9$\pm$7.0&-2.7$\pm$8.3&0.2$\pm$11.6&26.7$\pm$21.3&37.2$\pm$10.4\\
CEERS1\_4143&9.1$\pm$9.4&1.4$\pm$12.2&24.6$\pm$13.7&44.4$\pm$23.6&30.5$\pm$11.7\\
CEERS3\_1748&-5.6$\pm$4.4&5.7$\pm$5.7&18.5$\pm$7.9&-1.6$\pm$13.9&1.4$\pm$6.9\\
CEERS2\_1298&0.3$\pm$4.2&-0.3$\pm$4.6&6.4$\pm$5.6&7.0$\pm$9.7&17.5$\pm$4.9\\
CEERS2\_2274&-10.3$\pm$6.6&0.9$\pm$8.0&1.1$\pm$8.5&22.4$\pm$10.6&18.6$\pm$7.8\\
CEERS2\_1075&-4.8$\pm$4.0&-6.2$\pm$5.7&3.0$\pm$5.8&18.3$\pm$10.4&8.6$\pm$5.2
\enddata
\tablecomments{All fluxes are in nJy.  The horizontal lines distinguish our $z >$ 12, $z =$ 10--12 and $z =$ 8.5--10 samples.  We do not include a column for F105W as none of these candidate galaxies were covered by the sparse amount of F105W imaging in this field.}
\end{deluxetable*} \label{tab:hstphot}

\section{$z \sim$ 9 Sample Plots}
\begin{figure*}[!t]
\epsscale{0.9}
\plotone{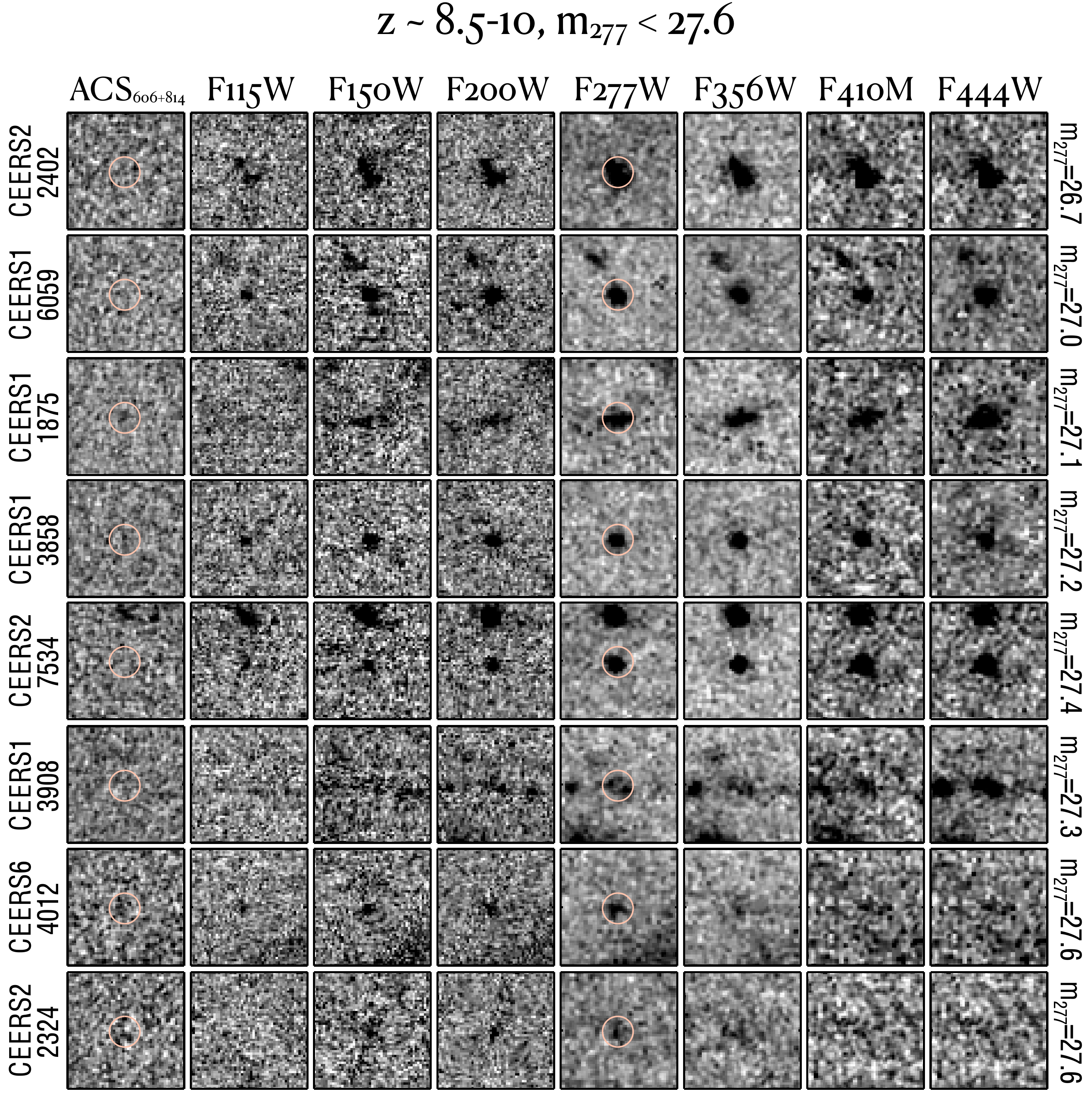}
\caption{Similar to Figure~\ref{fig:z12}, showing the eight candidates with best-fit photometric redshifts of $z \sim$ 9 and a F277W magnitude brighter than 27.6.}
\label{fig:z9stamps-bright}
\end{figure*}

\begin{figure*}[!t]
\epsscale{1.0}
\plotone{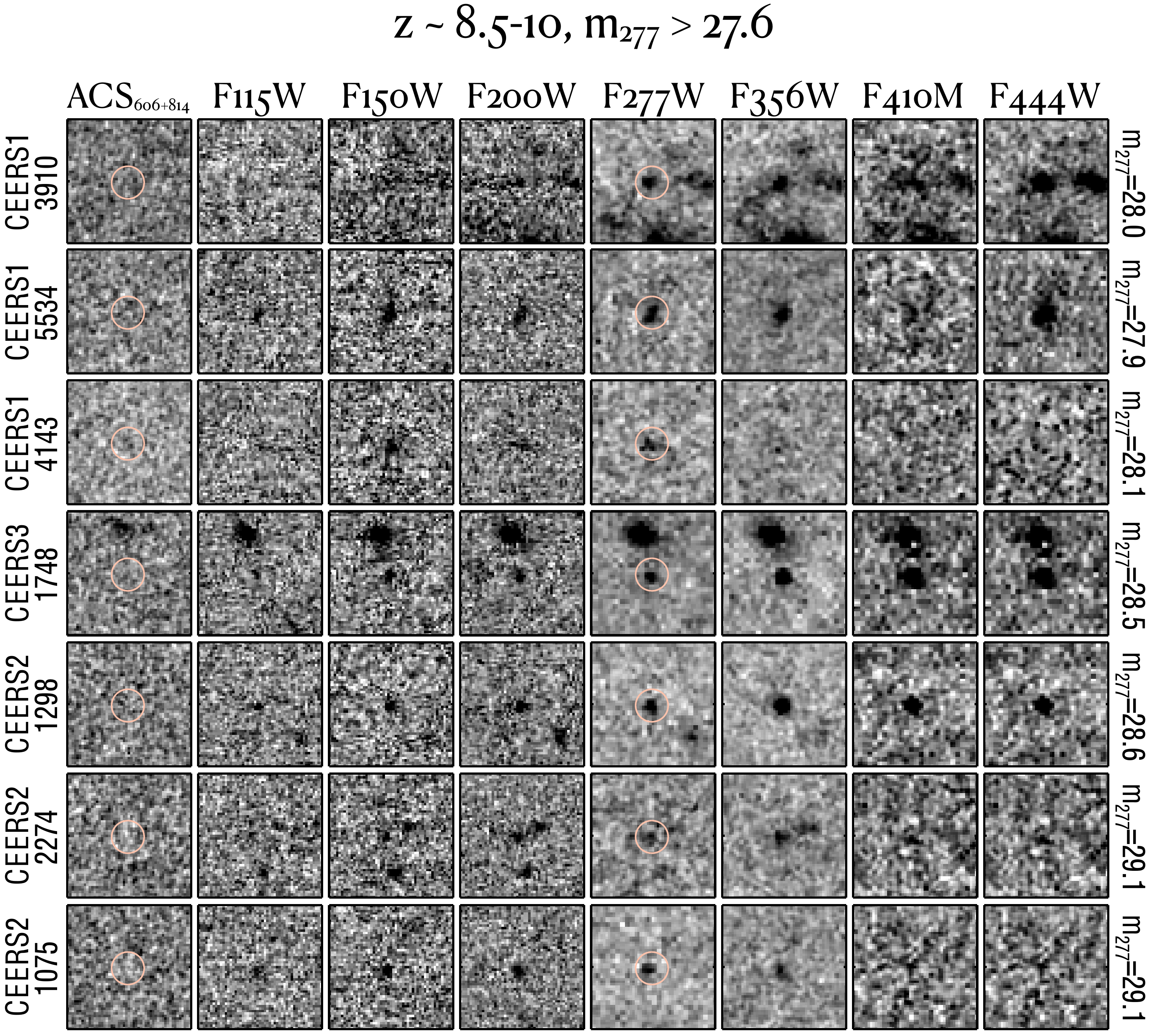}
\caption{Similar to Figures~\ref{fig:z12}, showing the nine candidates with best-fit photometric redshifts of $z \sim$ 9 and a F277W magnitude fainter than 27.6.}
\label{fig:z9stamps-faint}
\end{figure*}

Here we show the cutout images (Figure~\ref{fig:z9stamps-bright} and ~\ref{fig:z9stamps-faint}), and SED plots (Figure~\ref{fig:z9sed}) for the $z \sim$ 9 sample as described in \S 5.3, presented here in the Appendix for clarity in the main text.

\section{Sources Removed from Sample}

Figure~\ref{fig:largekron} shows two objects removed from our sample after re-measuring colors in smaller apertures, due to the stretching of their Kron apertures by nearby bright sources.  Figure~\ref{fig:spurious} shows cutout images for the 36 sources removed as spurious sources following visual inspection described in \S 4.3.  We list the coordinates for all removed sources in Table~\ref{tab:spurious}.

\begin{figure*}[!t]
\epsscale{1.15}
\plotone{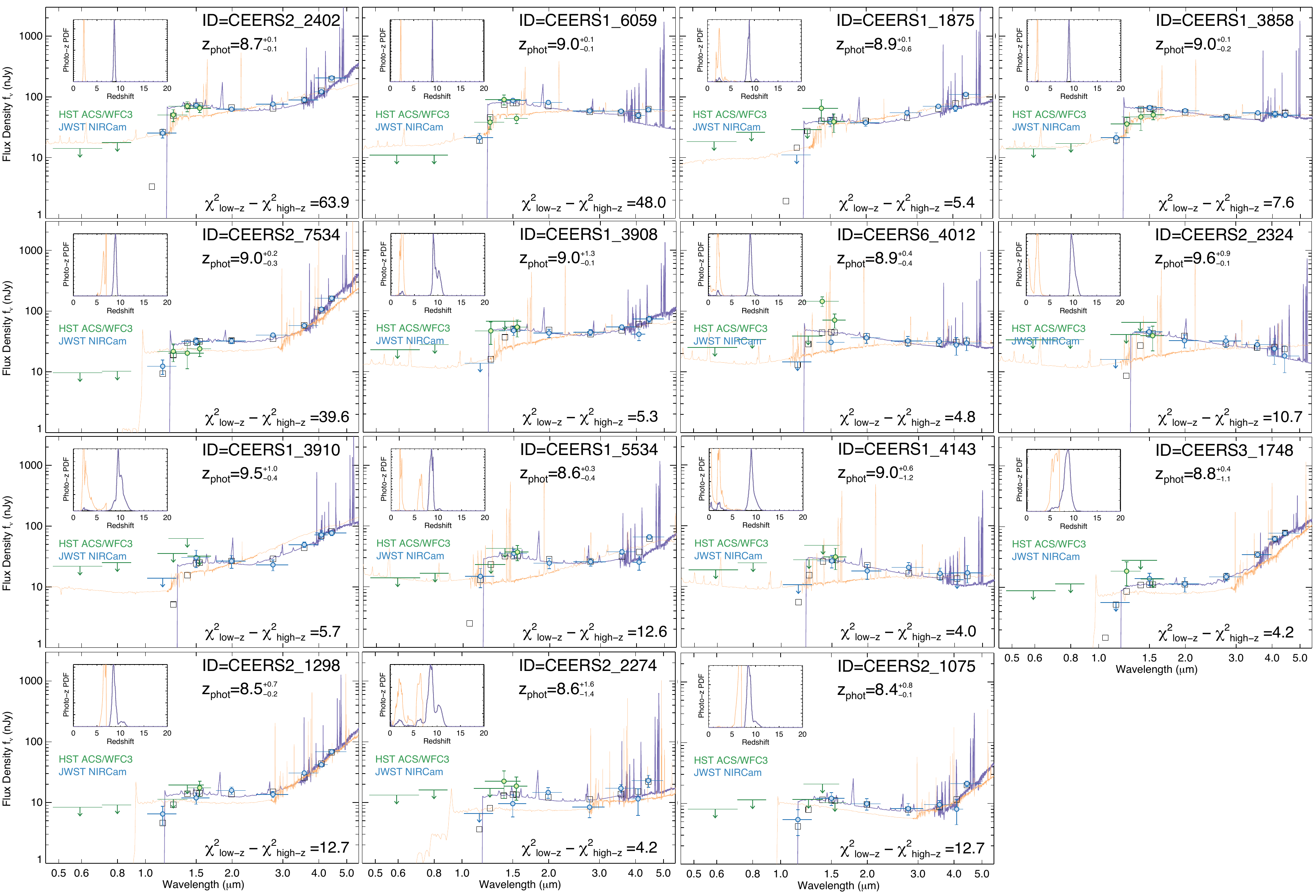}
\caption{The SEDs and $\mathcal{P}(z)$'s of the $z \sim$ 9 sample.}
\label{fig:z9sed}
\end{figure*}

\begin{figure*}[!t]
\epsscale{1.15}
\plotone{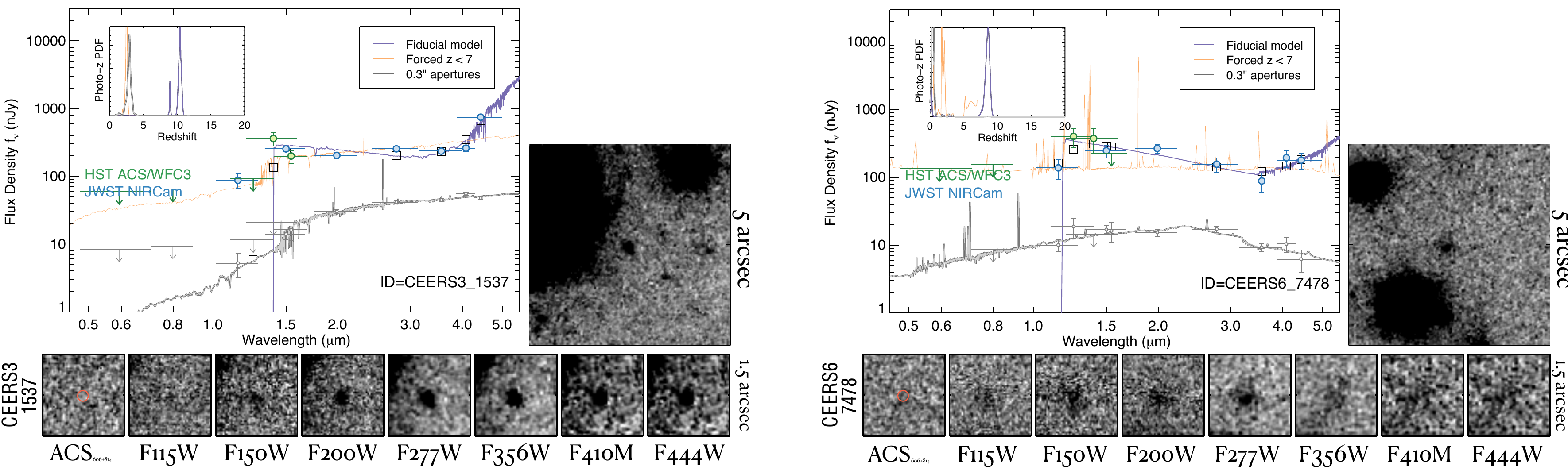}
\caption{Two sources which had incorrectly drawn Kron apertures due to the presence of nearby bright sources.  The large plot shows the SED, with the Kron NIRCam (HST) photometry in blue (green).  The best-fit \texttt{\textsc{EAZY}}model to these data is shown in purple (with squares denoting the model bandpass-averaged fluxes), with the best-fit $z <$ 7 solution shown in orange.  The small gray circles show the fluxes measured in 0.3\arcs\ diameter apertures, with the gray line showing the \texttt{\textsc{EAZY}}fit to these small-aperture fluxes.  The P(z) for all three \texttt{\textsc{EAZY}}runs are shown in the top-left.  The large image shows a 5\arcs\ $\times$ 5\arcs\ cutout around each source, highlighting the nearby neighbor responsible for stretching the Kron aperture leading to much brighter Kron fluxes.  The small images show a 1.5\arcs\ region around each source in the seven NIRCam bands (and the stacked ACS dropout bands), with the red circle denoting a 0.3\arcs\ diameter region around the source.  The P(z) from the colors measured in this small circular aperture prefers lower redshifts in both cases, thus these sources were removed from the sample.}
\label{fig:largekron}
\end{figure*}

\begin{figure*}[!t]
\epsscale{1.15}
\plotone{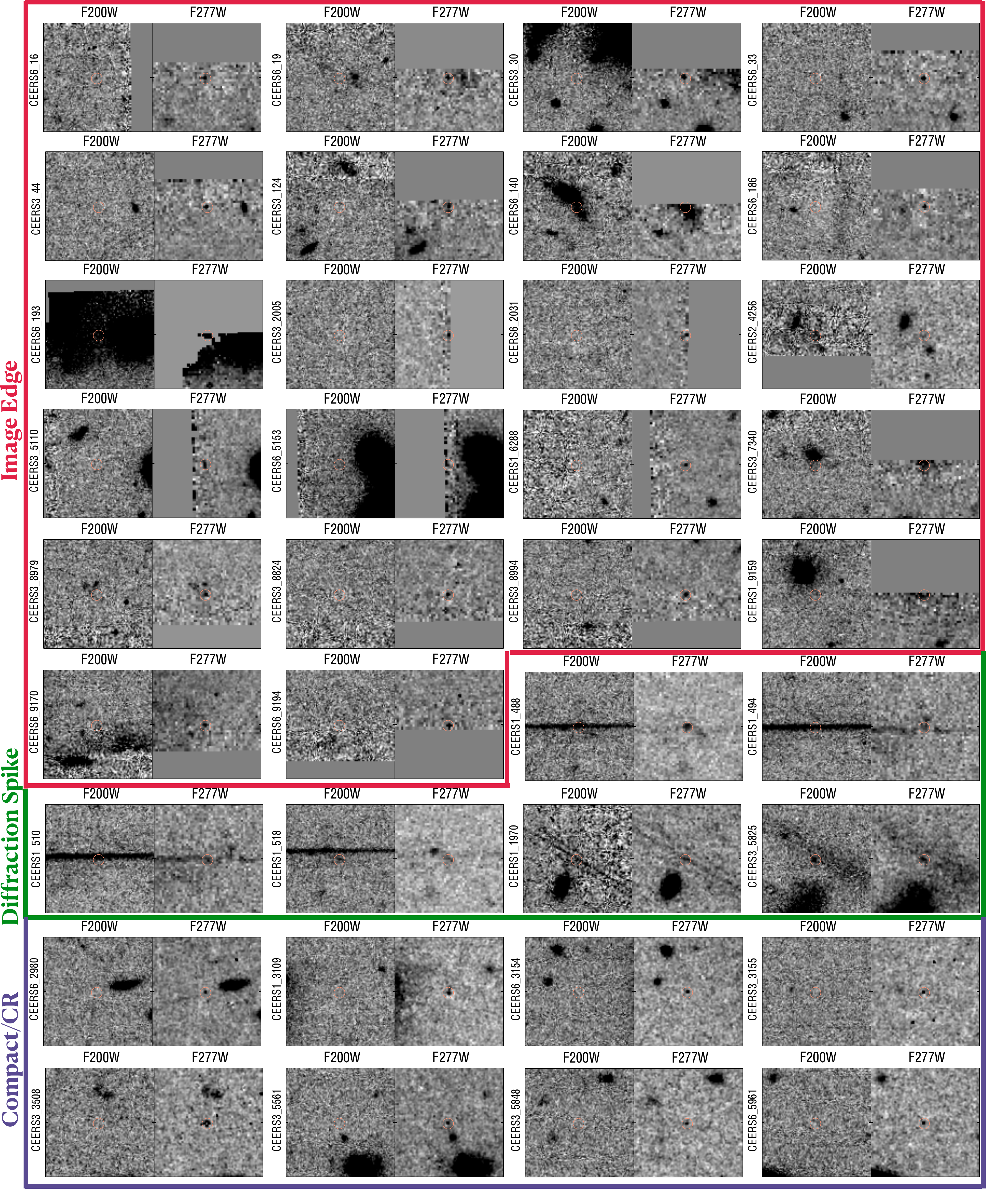}
\caption{This compilation shows 3\arcs\ $\times$ 3\arcs\ cutout images of the 36 sources identified as spurious from our visual inspection of the initial list of candidates.  The majority of these spurious detections come from very near to image edges, which are easily identifiable (and can in the future be automated).  Six objects are obvious diffraction spikes.  The remaining eight sources are very compact and boxy, and visible only in the LW channels. We conclude these are highly likely to be unflagged cosmic rays, which will be better flagged in future reductions \citep{bagley22c}.  We note that 7/8 are in the CEERS3 or 6 pointings, which had fewer images with longer exposure times than CEERS1 and 2, leading to more numerous cosmic ray hits.}
\label{fig:spurious}
\end{figure*}

\begin{deluxetable*}{ccccc}
\vspace{2mm}
\tablecaption{List of Removed Sources}
\tablewidth{\textwidth}
\tablehead{\multicolumn{1}{c}{ID} & \multicolumn{1}{c}{RA} & \multicolumn{1}{c}{Dec} & \multicolumn{1}{c}{Field} & \multicolumn{1}{c}{Reason}}
\startdata
488&215.005247&53.017772&1&Diffraction Spike\\
494&215.006368&53.018553&1&Diffraction Spike\\
510&215.006574&53.018658&1&Diffraction Spike\\
518&215.005202&53.017651&1&Diffraction Spike\\
1970&214.945032&52.966214&1&Diffraction Spike\\
3109&214.937432&52.953783&1&CR Residual\\
6288&214.985433&52.968347&1&Image Edge\\
9159&214.947769&52.980439&1&Image Edge\\
4256&214.878918&52.904385&2&Image Edge\\
30&214.735947&52.832000&3&Image Edge\\
44&214.768557&52.855112&3&Image Edge\\
124&214.798821&52.875759&3&Image Edge\\
1537&214.750580&52.829452&3&Bad Kron\\
2005&214.797283&52.862516&3&Image Edge\\
3155&214.758502&52.829340&3&CR Residual\\
3508&214.775758&52.838390&3&CR Residual\\
5110&214.798858&52.844780&3&Image Edge\\
5561&214.779455&52.827964&3&CR Residual\\
5825&214.775453&52.823445&3&Diffraction Spike\\
5848&214.780253&52.826894&3&CR Residual\\
7340&214.749582&52.841873&3&Image Edge\\
8824&214.811070&52.829804&3&CR Residual\\
8979&214.834667&52.845385&3&CR Residual\\
8994&214.861852&52.864635&3&CR Residual\\
16&214.806895&52.826457&6&Image Edge\\
19&214.788359&52.813215&6&Image Edge\\
33&214.820973&52.836319&6&Image Edge\\
140&214.864192&52.866051&6&Image Edge\\
186&214.871337&52.870820&6&Image Edge\\
193&214.878573&52.875993&6&Image Edge\\
2031&214.849667&52.843711&6&Image Edge\\
2980&214.811286&52.810462&6&CR Residual\\
3154&214.811199&52.809295&6&CR Residual\\
5153&214.851208&52.826106&6&Image Edge\\
5961&214.832642&52.808095&6&CR Residual\\
7478&214.851426&52.812204&6&Bad Kron\\
9170&214.914199&52.845844&6&Image Edge\\
9194&214.897088&52.833510&6&Image Edge
\enddata
\tablecomments{ID and coordinates of sources shown in Figures~\ref{fig:largekron} and ~\ref{fig:spurious}.}
\end{deluxetable*} \label{tab:spurious}


\end{document}